\documentclass[superscriptaddress,prx,twocolumn,nofootinbib,citeautoscript,longbibliography,notitlepage]{revtex4-1}

\usepackage{graphicx}
\usepackage{dcolumn}
\usepackage{bm}
\usepackage{color}
\usepackage{amsmath}
\usepackage{tabularx,graphicx}
\usepackage{epstopdf}
\usepackage{latexsym}
\usepackage{amssymb}
\usepackage{amsmath}
\usepackage{color, colortbl}
\usepackage{psfrag}
\usepackage{bbm}
\usepackage{booktabs}
\usepackage{bm}
\usepackage{titlesec}
\usepackage{dsfont}
\usepackage{feynmp}
\usepackage{slashed}
\usepackage{multirow}
\usepackage[tight]{subfigure}
\usepackage{comment}
\usepackage{float}
\usepackage{diagbox} 
\usepackage{makecell}
\usepackage{array}
\usepackage[papersize={8.5in,11in}]{geometry}
\usepackage{color}
\definecolor{darkblue}{rgb}{0.,0.,0.4}
\definecolor{darkred}{rgb}{0.5,0.,0.}
\definecolor{BlueViolet}{RGB}{138,43,226}
\definecolor{SkyBlue}{RGB}{30,144,255}
\definecolor{DarkGreen}{RGB}{0,100,0}
\definecolor{Green}{RGB}{0, 255, 0}
\usepackage[pdftex,colorlinks=true,linkcolor=darkblue,citecolor=blue,urlcolor=darkred]{hyperref}

\renewcommand{\epsilon}{\varepsilon}

\allowdisplaybreaks[3]

\newcommand{\red}[1]{\textcolor{red}{#1}}

 \UseRawInputEncoding

\usepackage{cases}

\allowdisplaybreaks[4]

\begin{document}



\title{Interaction and disorder effects on Cooper instability in two-dimensional fractional Dirac semimetals}

\author{Hua Zang}
\affiliation{Department of Physics, Tianjin University, Tianjin 300072, P.R. China}
\affiliation{Institute of Physics, Chinese Academy of Sciences, Beijing, 100080, P.R. China}

\author{Jing Wang}
\altaffiliation{Corresponding author: jing$\textunderscore$wang@tju.edu.cn}
\affiliation{Department of Physics, Tianjin University, Tianjin 300072, P.R. China}
\affiliation{Tianjin Key Laboratory of Low Dimensional Materials Physics and
Preparing Technology, Tianjin University, Tianjin 300072, P.R. China}

\date{\today}


\begin{abstract}

The fate of superconductivity in two-dimensional fractional Dirac semimetals featured
by a unique fractional energy dispersion $\alpha$ remains an open question. To address this, we
construct an effective low-energy theory that incorporates both the Cooper-pairing interactions
generated via a projection of attractive fermion-fermion couplings and multiple fermion-disorder scatterings
dubbed by $\Delta_{0,1,2,3}$. Employing a renormalization group analysis that allows for an unbiased
treatment of competing physical ingredients, we systematically trace how the interplay between Cooper pairing
and disorder scatterings governs the emergence or suppression of Cooper instability in the low-energy regime
of fractional Dirac semimetals. In the clean limit, we find that the emergence of Cooper instability requires
surpassing a finite interaction threshold $|\lambda_c|$, and depends sensitively on both the fractional exponent
$\alpha$ and the transfer momentum $\mathbf{Q}=(Q,\phi)$. Specifically, bigger values of $\alpha$ enhance the
tendency toward BCS instability. For $\alpha\in(0.001,0.61)$, the $(Q,\phi)$ parameter space separates into two
distinct regions: Zone-\uppercase\expandafter{\romannumeral1}, where Cooper instability is suppressed, and Zone-\uppercase\expandafter{\romannumeral2}, where it is allowed. In the presence of disorders, we demonstrate
that they can either promote or suppress Cooper instability. Disorder of type $\Delta_1$ or $\Delta_2$
enhances superconductivity by reducing the critical interaction threshold $|\lambda_c|$ and expanding the
superconducting phase space (Zone-\uppercase\expandafter{\romannumeral2}). In sharp contrast, either $\Delta_0$
or $\Delta_3$ suppresses Cooper pairing by increasing $|\lambda_c|$ and shrinking the available phase space (Zone-\uppercase\expandafter{\romannumeral1}). Although Cooper instability can be enhanced when promotive
disorders ($\Delta_1$, $\Delta_2$) coexist with a single suppressive disorder ($\Delta_0$ or $\Delta_3$),
the suppressive influence of $\Delta_{0,3}$ generally dominates the promotive effects of $\Delta_{1,2}$ in
the presence of all sorts of disorders. Our results provide rich information for Cooper instability governed
by the competition between Cooper pairing interaction and distinct types of disorders, which are helpful for
further studies of fractional Dirac semimetals and alike materials.
\end{abstract}


\maketitle

%
%
%
%
%
%
%
%
%



\section{Introduction}

Research on Dirac fermions in graphene and related two-dimensional crystals has fundamentally reshaped our understanding of quantum matter by unveiling a plethora of exotic quantum phenomena~\cite{Novoselov2005Nature,Neto2009RMP,Kane2007PRL,Roy2009PRB,Moore2010Nature,Hasan2010RMP,
Qi2011RMP,Sheng2012Book,Bernevig2013Book,Burkov2011PRL,Yang2011PRB,Wan2011PRB,
Huang2015PRX,Weng2015PRX,Hasan2015Science,Hasan2015NPhys,Ding2015NPhys,
WangFang2012PRB,Young2012PRL,Steinberg2014PRL,Hussain2014NMat,
LiuChen2014Science,Ong2015Science,Montambaux-Fuchs-PB2012,Banerjee2012PRB}.
A hallmark of these materials is the presence of symmetry-protected, discrete band-touching
points that generate gapless quasiparticle excitations, which are equipped with either a
linear energy dispersion in Dirac and Weyl semimetals~\cite{WangFang2012PRB,Young2012PRL,Steinberg2014PRL,Hussain2014NMat,LiuChen2014Science,
Ong2015Science,Neto2009RMP,Burkov2011PRL,Yang2011PRB,
Wan2011PRB,Huang2015PRX,Hasan2015Science,Hasan2015NPhys,Ding2015NPhys,Weng2015PRX,
Novoselov2005Nature,Kane2007PRL,Roy2009PRB,Moore2010Nature,Hasan2010RMP,
Qi2011RMP,Korshunov2014PRB,Hung2016PRB,Sondhi2013PRB,Sondhi2014PRB,Wang2017PRB_BCS,Wang2019JPCM,
Nandkishore2017PRB,Roy-Saram2016PRB,Herbut2018Science,WGB2020PRL} or
parabolic dispersion where conduction and valence bands cross parabolically in
quadratic-band-crossing materials~\cite{Chong2008PRB,Fradkin2008PRB,
Fradkin2009PRL,Vafek2012PRB,Vafek2014PRB,Herbut2012PRB,Mandal2019CMP,Zhu2016PRL,
Vafek2010PRB,Yang2010PRB,Wang2017PRB_QBCP,DZZW2020PRB,Roy2020-arxiv,Janssen2020PRB,Shah2011.00249,
Luttinger1956PR,Murakami2004PRB,Janssen2015PRB,Boettcher2016PRB,Janssen2017PRB,Boettcher2017PRB,
Mandal2018PRB,Lin2018PRB,Savary2014PRX,Savary2017PRB,Vojta1810.07695,Lai2014arXiv,Goswami2017PRB,
Szabo2018arXiv,Foster2019PRB,Wang1911.09654,Wang2303.10163}.
Unlike conventional Dirac semimetals, whose energy dispersion is characterized by an integer index, the
fractional Dirac semimetals (FDSMs) in two and three dimensions have been proposed as a distinct
quantum state~\cite{Roy2023PRR,Lin-Fruger2026PRB,MGSB2016PRB,HTN2019PRR} in which the energy
dispersion obeys an anomalous fractional momentum
scaling $E \propto k^{n/m}$ with $m,n$ being integer exponents and $m>n$. Quantum Monte Carlo
simulations~\cite{Garttner2015PRB,Shang2015NC,Kempkes2019NP} have succeeded in realizing these systems
with non-trivial Berry connections. The resulting gapless quasiparticles with non-integer dispersion
render FDSMs an ideal platform for exploring novel quantum criticality.

Understanding the low-energy physics of two-dimensional (2D) FDSMs is therefore of central importance.
Among their emergent phenomena, superconductivity emerges as a particularly emergent phenomenon in these materials.
The famous Bardeen-Cooper-Schrieffer (BCS) theory~\cite{BCS1957PR} tells us that an arbitrarily weak attractive interaction can bind a
pair of electrons and trigger a Cooper instability in conventional metals.
In Dirac systems, however, the linear dispersion and vanishing density of states at the nodal points~\cite{Neto2009RMP,Zhao2006PRL,Honerkamp2008PRL,Roy-Herbut2010PRB,Roy-Herbut2013PRB,Roy-Jurici2014PRB,
Ponte-Lee2014NJP,Yao2015PRL,Maciejko2016PRL,Nandkishore2012NP,Sondhi2013PRB,Sondhi2014PRB} impose
a finite strength of attraction interaction~\cite{Sondhi2013PRB,Sondhi2014PRB}. In other words,
the pairing occurs only when the interaction strength exceeds a threshold~\cite{MDR2025PRB},
rendering the interaction itself a control parameter for the quantum phase transition to
the superconducting state~\cite{Zhao2006PRL,Honerkamp2008PRL,Sondhi2013PRB,Sondhi2014PRB}.
Compared to the Dirac materials, the 2D FDSMs exhibit profoundly unconventional
characteristics due to their fractional dispersion. This fundamentally reshapes low-energy
quasiparticle physics in several key aspects. At first, it modifies the density of states to
renormalize quasiparticle properties~\cite{Neto2009RMP,Altland2006Book,MSBJ2026PRB}
and thus alters quasiparticle interactions via changing the renormalized scalings
~\cite{Altland2006Book,Coleman2015Book,Nandkishore2012NP,Roy2018PRX,Vafek2012PRB,Vafek2014PRB}.
In addition, these together can influence the quasiparticle-disorder scattering effects~\cite{Edwards1975JPF,Ramakrishnan1985RMP,Lerner0307471,Nersesyan1995NPB,Stauber2005PRB,Wang2011PRB, Mirlin2008RMP,Coleman2015Book,Roy2018PRX,Wang2015PLA},
which are of close relevance to transport quantities~\cite{Sachdev1999Book,Altland2002PR,
Lee2006RMP,Neto2009RMP,Fradkin2010ARCMP,Hasan2010RMP,Sarma2011RMP,Qi2011RMP,Kotov2012RMP}.
In particular, disorder creates two competing effects by simultaneously enhancing the density
of states and shortening quasiparticle lifetimes. These considerations consequently raise intriguing questions:
Can a Cooper instability associated with the superconsudting state survive in 2D FDSMs?
What is the critical interaction strength required to overcome the fractional-statistics barrier?
How do various disorder types reshape the superconducting phase boundary?

To address these questions, we construct an effective low-energy theory that incorporates Cooper-pairing interactions
derived through a projection procedure of attractive fermion-fermion couplings alongside the non-interacting Hamiltonian and the fermion-impurity scatterings. For an unbiased treatment of these competing physical ingredients, we employ the renormalization group (RG)
approach~\cite{Wilson1975RMP,Polchinski9210046,Shankar1994RMP}. Within the RG formalism, the onset of Cooper instability manifests as a (marginally) relevant flow of an attractive interaction, which inevitably evolves toward strong coupling and signals the emergence of superconductivity~\cite{Shankar1994RMP}. Elucidating these inquiries would be helpful to deepen our understanding
of the properties of 2D FDSM materials, and provide clues for the study of other Dirac-like materials~\cite{LCJS2009PRL,Beenakker2009PRL,Rosenberg2010PRL,
Hasan2011Science,Bahramy2012NC,Viyuela2012PRB,Bardyn2012PRL,Garate2003PRL,
Oka2009PRB,Lindner2011NP,Gedik2013Science,CBHR2016PRL,Slager1802}.

Employing the RG analysis, we systematically examine how these interactions combined with disorder
effects govern the emergence of Cooper instability in the low-energy regime of 2D FDSMs.

At first, we consider the clean limit adopting a combined analytical and numerical approach.
At tree level, the Cooper-pairing coupling strength $\lambda$ flows to zero as the energy scale
decreases, indicating the absence of instability. However, beyond the tree level, we find that
Cooper instability arises only when the initial pairing strength $|\lambda_0|$ exceeds a critical
threshold $|\lambda_c|$. This critical value exhibits parametric dependence on the momentum-transfer
space $(Q, \phi)$, which is characterized by the magnitude $|\mathbf{Q}|$ and angular orientation $\phi$,
as well as on the fractional dispersion exponent $\alpha$ and the fermionic velocity $v_\alpha$.
In particular, the $(Q, \phi)$ space separates into two distinct zones: Zone-\uppercase\expandafter{\romannumeral1},
where $|\lambda_c|$ diverges and Cooper instability is prohibited, and Zone-\uppercase\expandafter{\romannumeral2},
where $|\lambda_c|$ remains finite and instability is allowed. There exist two critical values of $\alpha$
dubbed $\alpha_{c1}$ and $\alpha_{c2}$. For $\alpha \in (\alpha_{c1}, \alpha_{c2})$, both
Zone-\uppercase\expandafter{\romannumeral1} and Zone-\uppercase\expandafter{\romannumeral2} coexist,
with their areas depending explicitly on $\alpha$. Outside this region, i.e., when $\alpha \notin (\alpha_{c1}, \alpha_{c2})$,
only Zone-\uppercase\expandafter{\romannumeral2} persists across the entire parameter space. Additionally,
we find that lower values of $v_\alpha$ and higher values of $\alpha$ systematically reduce the critical
coupling strength $|\lambda_c|$, thereby favoring the emergence of Cooper instability.

Next, we examine the effects of distinct types of disorders including $\Delta_0$, $\Delta_1$,
$\Delta_2$, and $\Delta_3$ as detailed in Sec.~\ref{Sec_effective_theory}. Our results indicate that
the sole presence of either $\Delta_0$ or $\Delta_3$ (classified as suppressive disorders)
inhibits Cooper instability, while either $\Delta_1$ or $\Delta_2$ (promotive disorders) enhances it.
Turning to the presence of multiple types of disorders, they compete with each other. Specifically,
the simultaneous presence of both promotive disorders ($\Delta_1$, $\Delta_2$) together with a single
suppressive disorder ($\Delta_0$ or $\Delta_3$) can still enhance Cooper instability. However, when all
types of disorders are present, the suppressive influence of $\Delta_{0,3}$ generally dominates over
the promotive effects of $\Delta_{1,2}$. Among the promotive disorders, $\Delta_1$ exhibits a stronger
effect than $\Delta_2$, while the combined suppressive impact of $\Delta_0$ and $\Delta_3$ systematically
outweighs the promotive contribution of $\Delta_1$ and $\Delta_2$.

The remainder of this paper is organized as follows. In Sec.~\ref{Sec_effective_theory},
we construct the low-energy effective theory, which includes the non-interacting Hamiltonian
with the fractional dispersion, the projection procedure for attractive fermion-fermion interactions,
and the classification of disorder scatterings.  Subsequently, in Sec.~\ref{Sec_RG_analysis}, we perform the RG analysis
and derive the coupled flow equations for the interaction and disorder parameters. Our central results
are presented in Sec.~\ref{Sec_results_clean} and Sec.~\ref{Sec_dis_effects}.
In Sec.~\ref{Sec_results_clean}, we conduct both analytical and numerical studies of the clean limit,
revealing the existence of a critical interaction strength for Cooper instability in 2D FDSMs, which
depends sensitively on the fractional dispersion exponent and transfer momentum. Sec.~\ref{Sec_dis_effects}
examines the pronounced effects of individual and combined disorder scatterings,
quantifying their influence on the critical interaction strength and the stability of Cooper pairing
in the low-energy regime. Finally, a brief summary is provided in Sec.~\ref{Sec_summary}.

\begin{figure}[h]
\centering
\subfigure[]{
\includegraphics[width=1in]{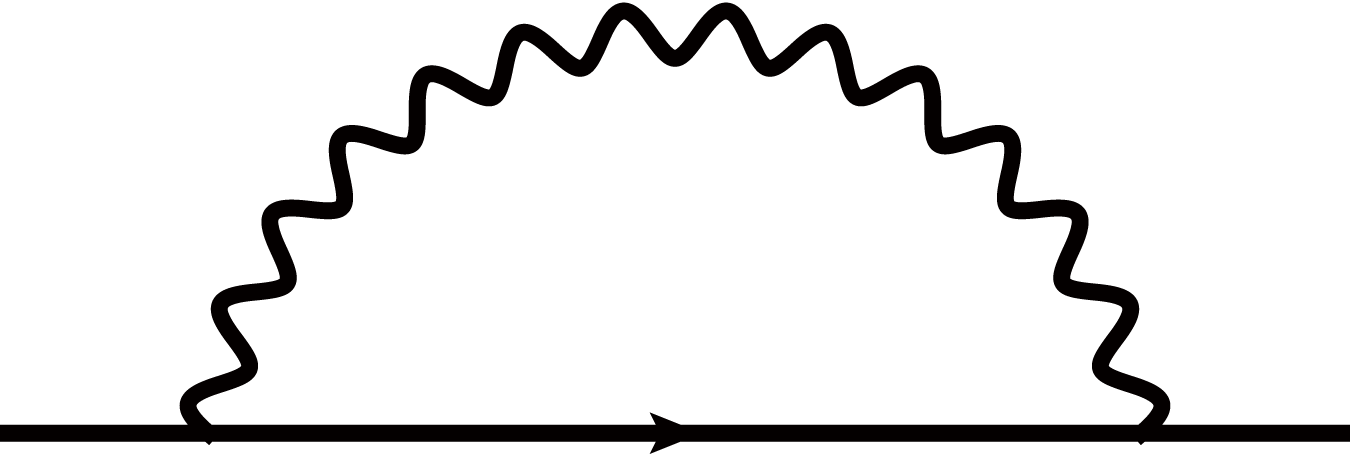}
}\hspace{0.25cm}
\subfigure[]{
\includegraphics[width=1in]{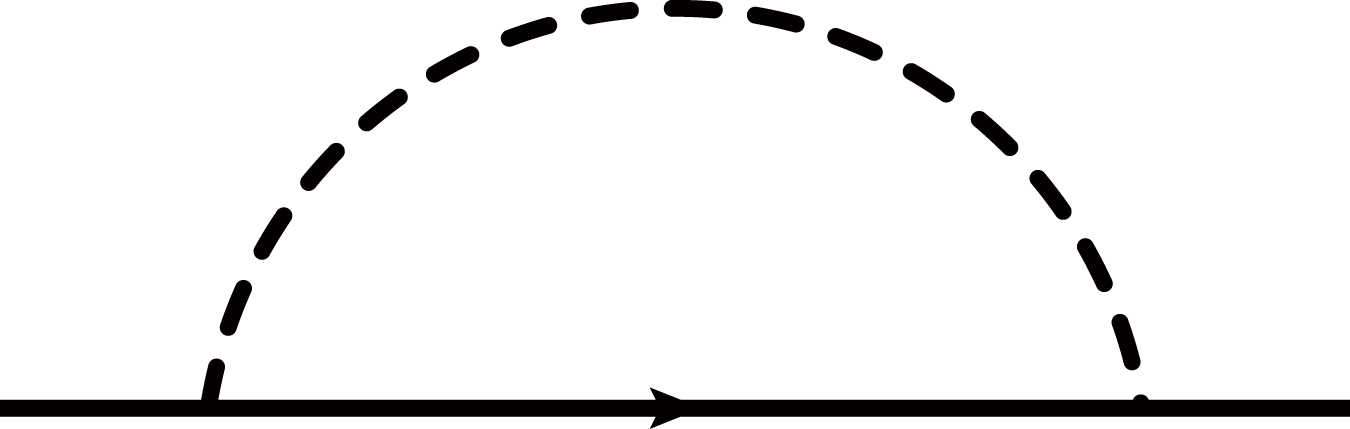}
}
\\
\vspace{-0.3cm}
\caption{One-loop corrections to the fermionic propagator due to (a) the
Cooper-pairing interaction and (b) the fermion-disorder interaction.
The wave and dashed lines denote the Cooper-pairing interaction and the
fermion-disorder interaction, respectively.}\label{fig_one_loop_fermion}
\end{figure}

\section{Model and effective theory}\label{Sec_effective_theory}

The effective Hamiltonian density for a $d$-dimensional fractional Dirac semimetal (FDSM) of order $\alpha$ is
expressed as~\cite{Roy2023PRR},
\begin{equation}
H_{\mathrm{FD}}(\mathbf{k})
=\sum_{j=1}^{d} v_\alpha |k_j|^\alpha \mathrm{sign}(k_j) \Gamma_j,\label{Hamiltonian}
\end{equation}
with $0 < \alpha < 1$ and $j$ being a component index. Here, $k_j$ denotes the $j$-component momentum
and $v_\alpha$ serves as the effective velocity~\cite{Roy2023PRR}. In addition, the $\Gamma_j$ matrices satisfy the Clifford algebra anticommutation relations $\left\{ \Gamma_i, \Gamma_j \right\} = 2\delta_{ij}\mathrm{I}_d$
where $\mathrm{I}_d$ denotes the $d$-dimensional identity matrix. For convenience, we within this work focus on the $d=2$ case. For two-dimensional systems, the $\Gamma_j$ can be represented by Pauli matrices $\sigma_i$ ($i = 1,2,3$), i.e., $\Gamma_1 = \sigma_1$ and $\Gamma_2 = \sigma_2$. This Hamiltonian density~(\ref{Hamiltonian})
gives rise to the following non-interacting effective action~\cite{Roy2023PRR,Roy2018PRX,Villegas2025PRB},
\begin{eqnarray}
S_{0} & = & \int \frac{\mathrm{d}\omega  }{2\pi } \int \frac{\mathrm{d^2}\mathbf{k}  }{\left ( 2\pi  \right )^2 } \psi ^{\dagger } \left [ -i\omega +v_\alpha \left | k_1 \right | ^\alpha \mathrm{sign} \left ( k_1 \right ) \sigma _1 \right. \nonumber \\
 & & +\left . v_\alpha \left | k_2 \right | ^\alpha \mathrm{sign} \left ( k_2 \right ) \sigma _2\right ]  \psi,
\label{Eq_S0}
\end{eqnarray}
where the spinors $\psi^\dagger(i\omega,\mathbf{k} ) $ and $\psi(i\omega,\mathbf{k} ) $ denote the low-energy excitations of fermionic degrees from the Dirac point. In consequence, the free propagator for these fermionic excitations can be obtained as,
\begin{eqnarray}
G_{0}= \frac{1}{-i\omega + \sum_{i=1}^{2}v_\alpha \left | k_i \right | ^\alpha \mathrm{sign} \left ( k_i \right ) \sigma _i}.
\end{eqnarray}

To proceed, let us take into account an attractive fermion-fermion interaction
$\mathcal{H}_{\mathrm{ff}} = \int d^2\mathbf{r}  \lambda/4 \left[ \psi^\dagger(\mathbf{r}) \psi(\mathbf{r}) \right]^2$~\cite{Sondhi2013PRB,Sondhi2014PRB,Wang2017PRB_BCS} where the coupling $\lambda$ is negative and
becomes energy-dependent after incorporating higher-order corrections.
By following Nandkishore \emph{et al}.~\cite{Sondhi2013PRB,Sondhi2014PRB}, we project $\mathcal{H}_{\mathrm{ff}}$ onto the Cooper channel where fermions with antiparallel spins and opposite momenta form bound states, and obtain the Cooper-channel interaction
with the restriction of singlet-pairing interaction as,
\begin{equation}
\mathcal{H}_{\mathrm{C}}
\!=\!\!\!\sum_{\mathbf{k}_1,\mathbf{k}_2}\!\!\frac{\lambda\Lambda^2}{4}
\psi^\dagger_{\mathbf{k}_1,\uparrow}(-i\Gamma_2)\psi^\dagger_{-\mathbf{k}_1,\downarrow}
\psi_{-\mathbf{k}_2,\downarrow}(i\Gamma_2)\psi_{\mathbf{k}_2,\uparrow},\label{Eq_H_int2}
\end{equation}
where a UV cutoff $\Lambda$ is introduced to maintain dimensional consistency. To simplify the analysis,
we redefine the coupling constant as $\lambda \Lambda^2 / 4 \rightarrow \lambda$ and subsequently
arrive at the Cooper-interaction action as~\cite{Sondhi2013PRB,Sondhi2014PRB,Wang2017PRB_BCS},
\begin{eqnarray}
 S_{\mathrm{C}}
\!&=& \!\!\int \frac{\mathrm{d}\omega _{1}\mathrm{d}\omega _{2}\mathrm{d}\omega _{3}  }{\left ( 2\pi \right )^3  } \int \frac{\mathrm{d^2}\mathbf{k}\mathrm{d^2}\mathbf{q}   }{\left ( 2\pi  \right ) ^4} \frac{\lambda \Lambda ^2}{4} \psi^{\dagger }_{\omega _{1},\mathbf{k}  ,\uparrow }\left ( -i\Gamma _{2}  \right )\nonumber \\
&& \times   \psi^{\dagger }_{\omega _{2},-\mathbf{k}  ,\downarrow }  \psi_{\omega _{3},\mathbf{q}  ,\downarrow }
\left ( i\Gamma_{2}  \right) \psi _{\omega _{1}+\omega _{2}-\omega _{3},\mathbf{q}  ,\uparrow }.
\label{eq:cooper-pairing interaction}
\end{eqnarray}

\begin{figure}[h]
\centering
\subfigure[]{
\includegraphics[width=1in]{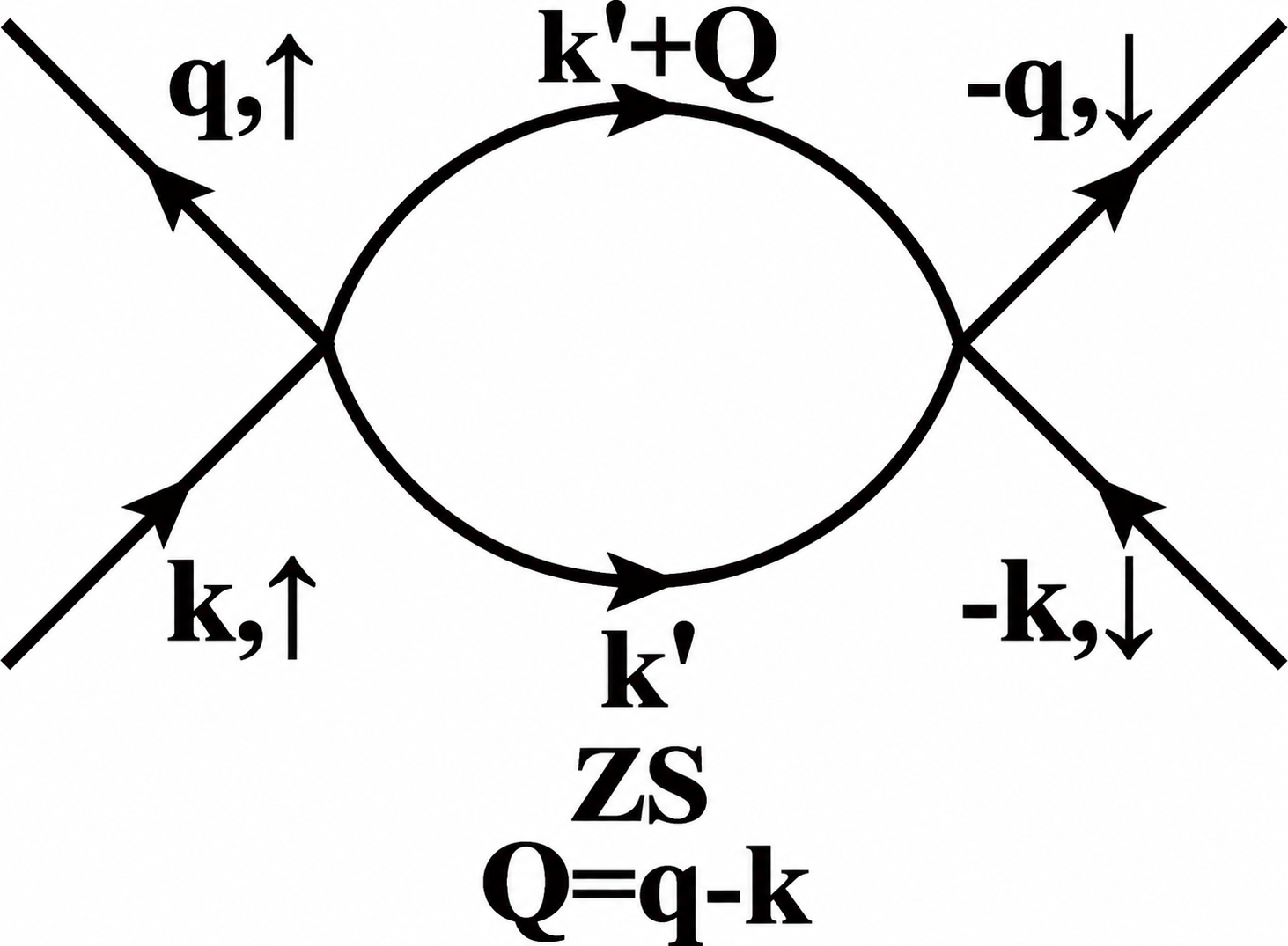}
}
\subfigure[]{
\includegraphics[width=1in]{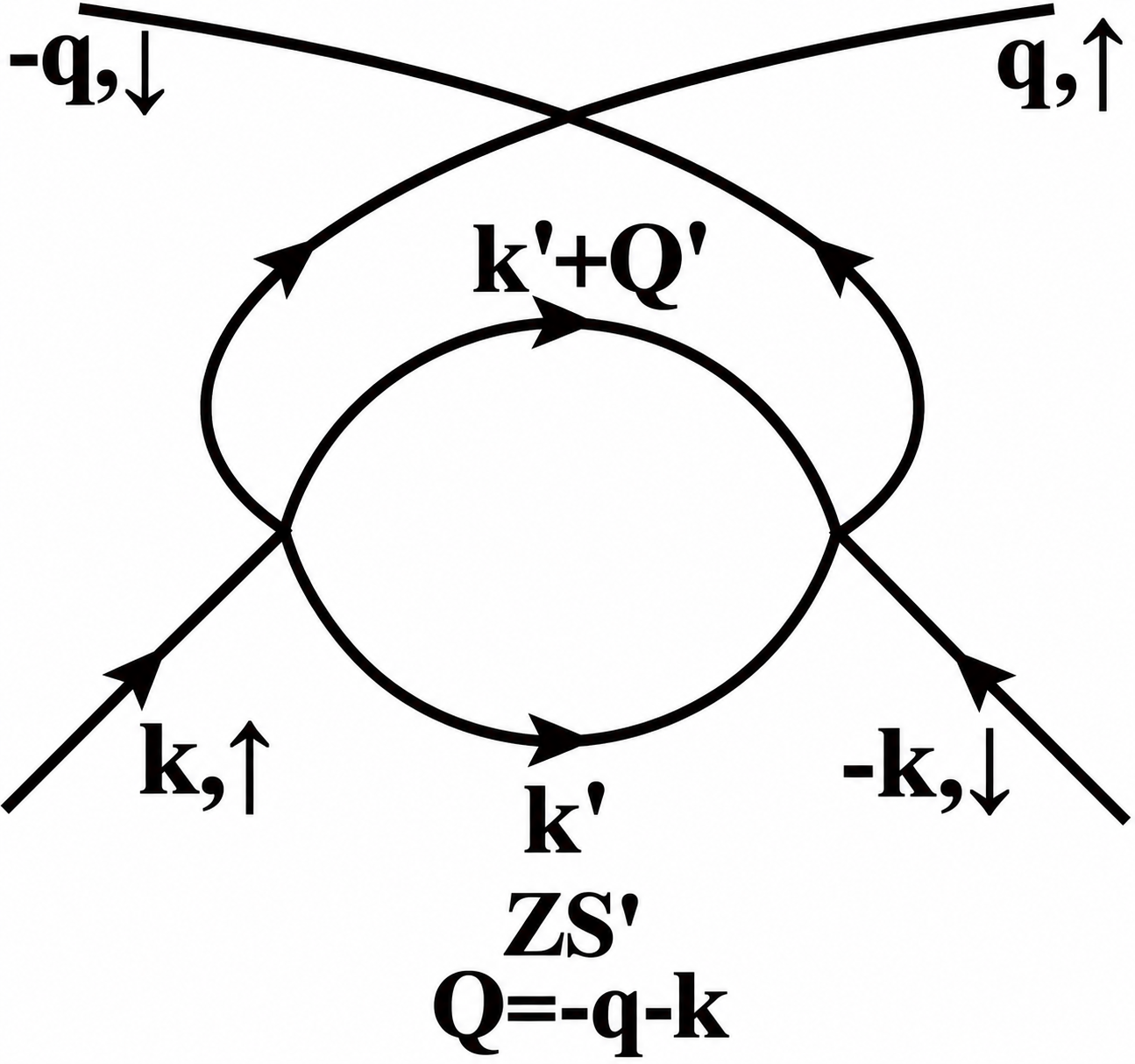}
}
\subfigure[]{
\includegraphics[width=0.65in]{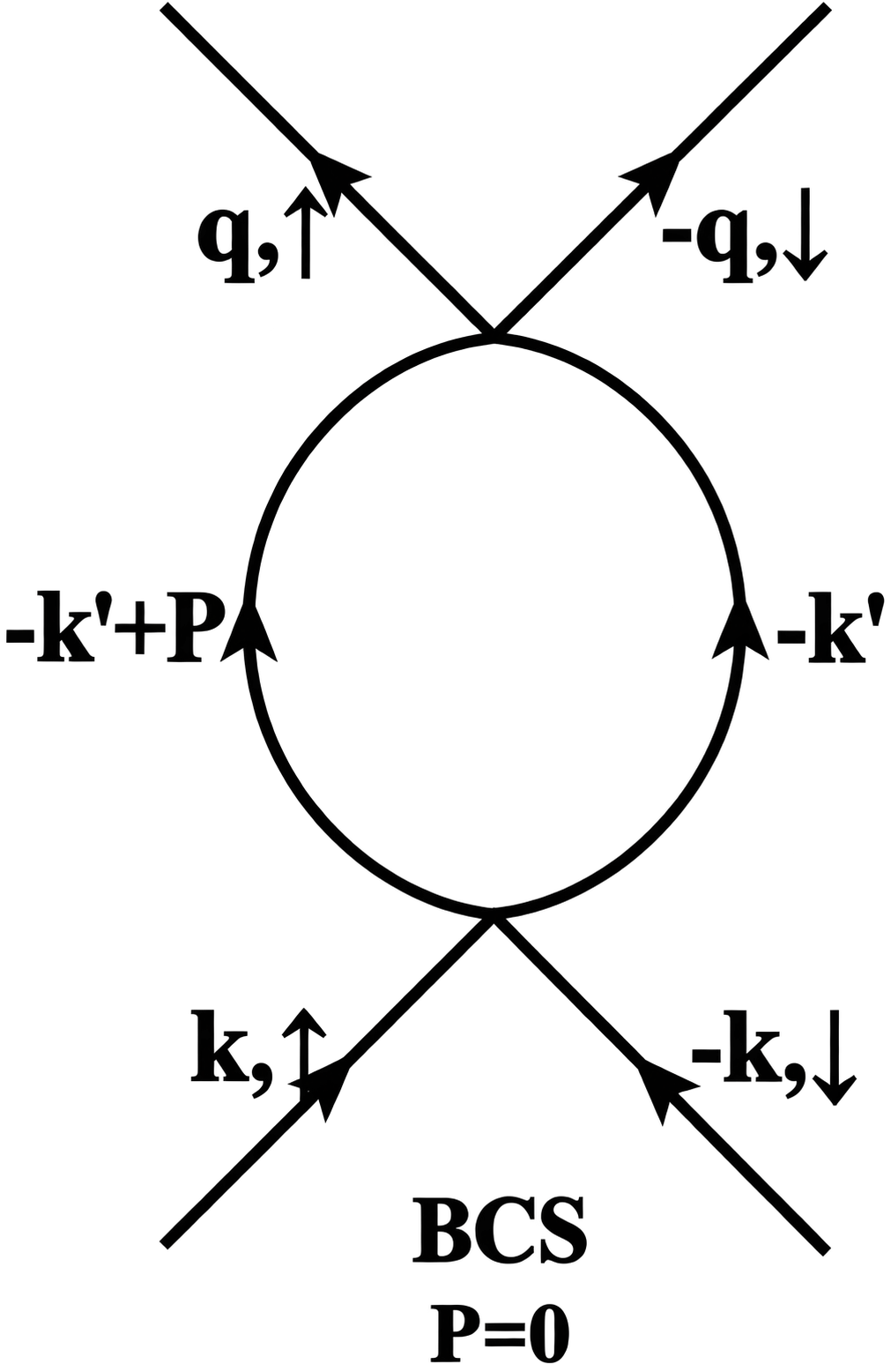}
}
\subfigure[]{
\includegraphics[width=1in]{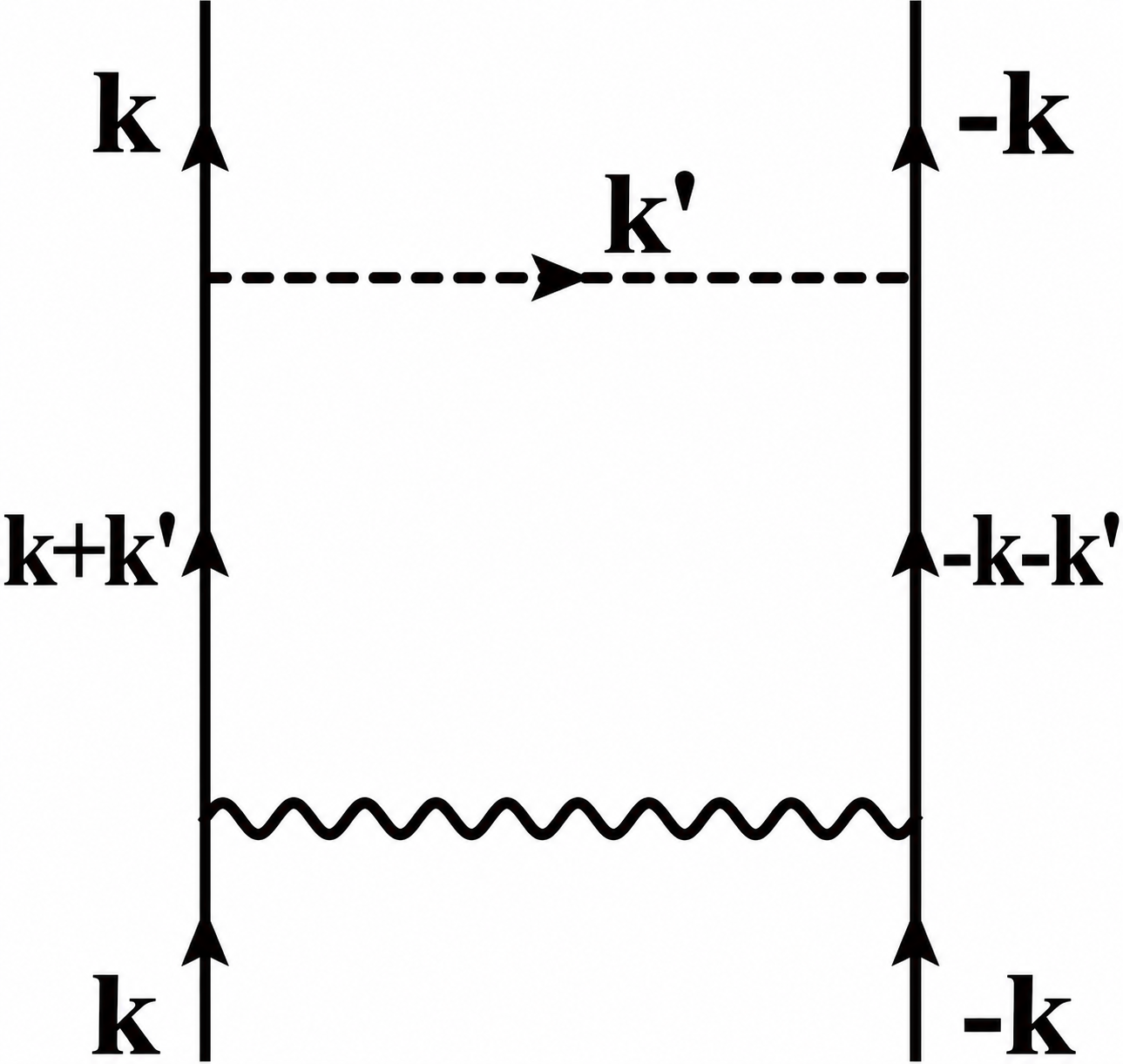}
}
\subfigure[]{
\includegraphics[width=1in]{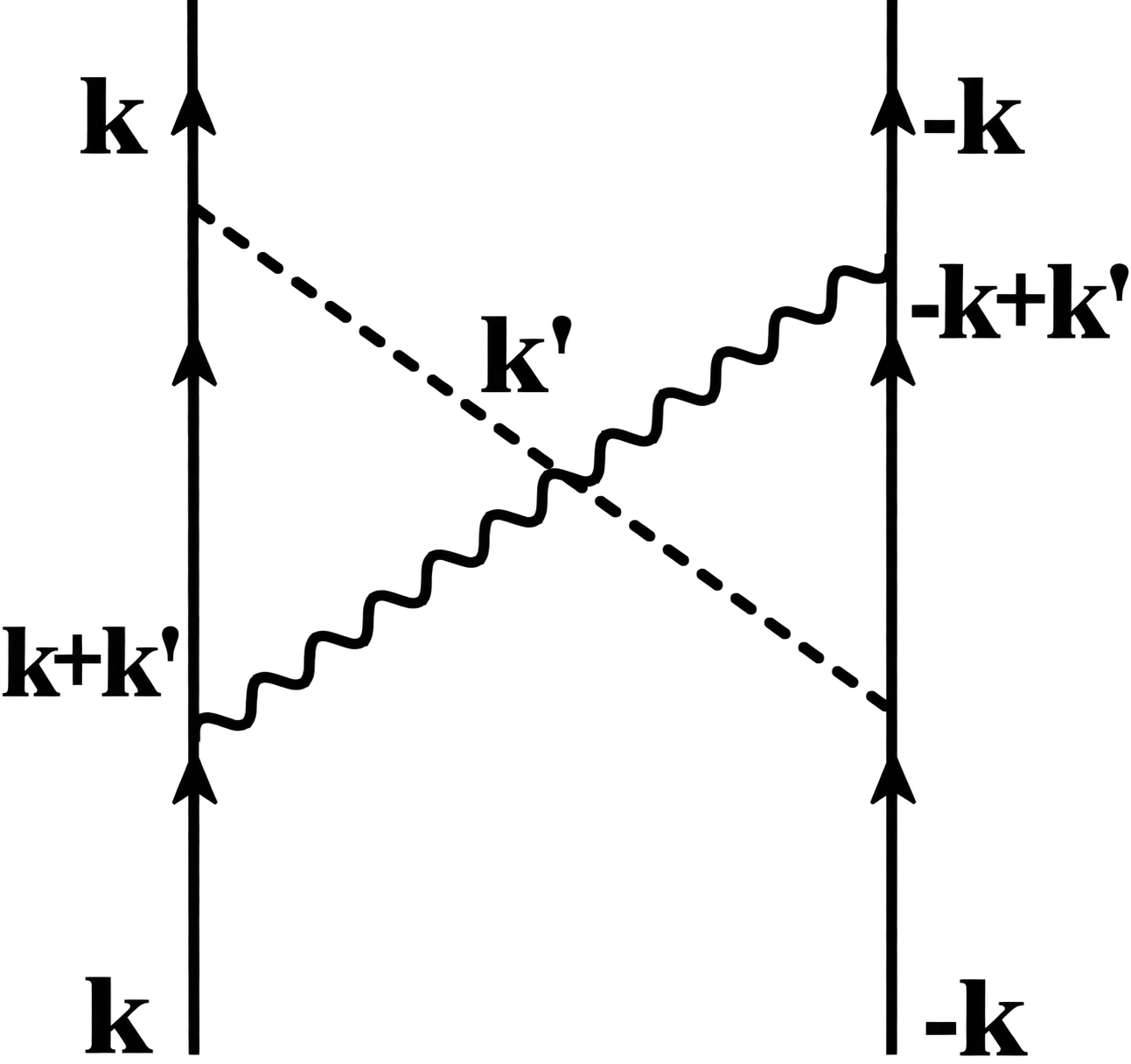}
}
\\
\vspace{-0.2cm}
\caption{One-loop corrections to the attractive Cooper-pairing coupling due to (a)-(c) the
Cooper-pairing interaction and (d)-(e) the fermion-disorder interaction.
The wave and dashed lines denote the Cooper-pairing interaction and the
fermion-disorder interaction, respectively.}
\label{fig_one_loop_cooper_pairing}
\end{figure}

\begin{figure}[h]
\centering
\subfigure[]{
\includegraphics[width=1in]{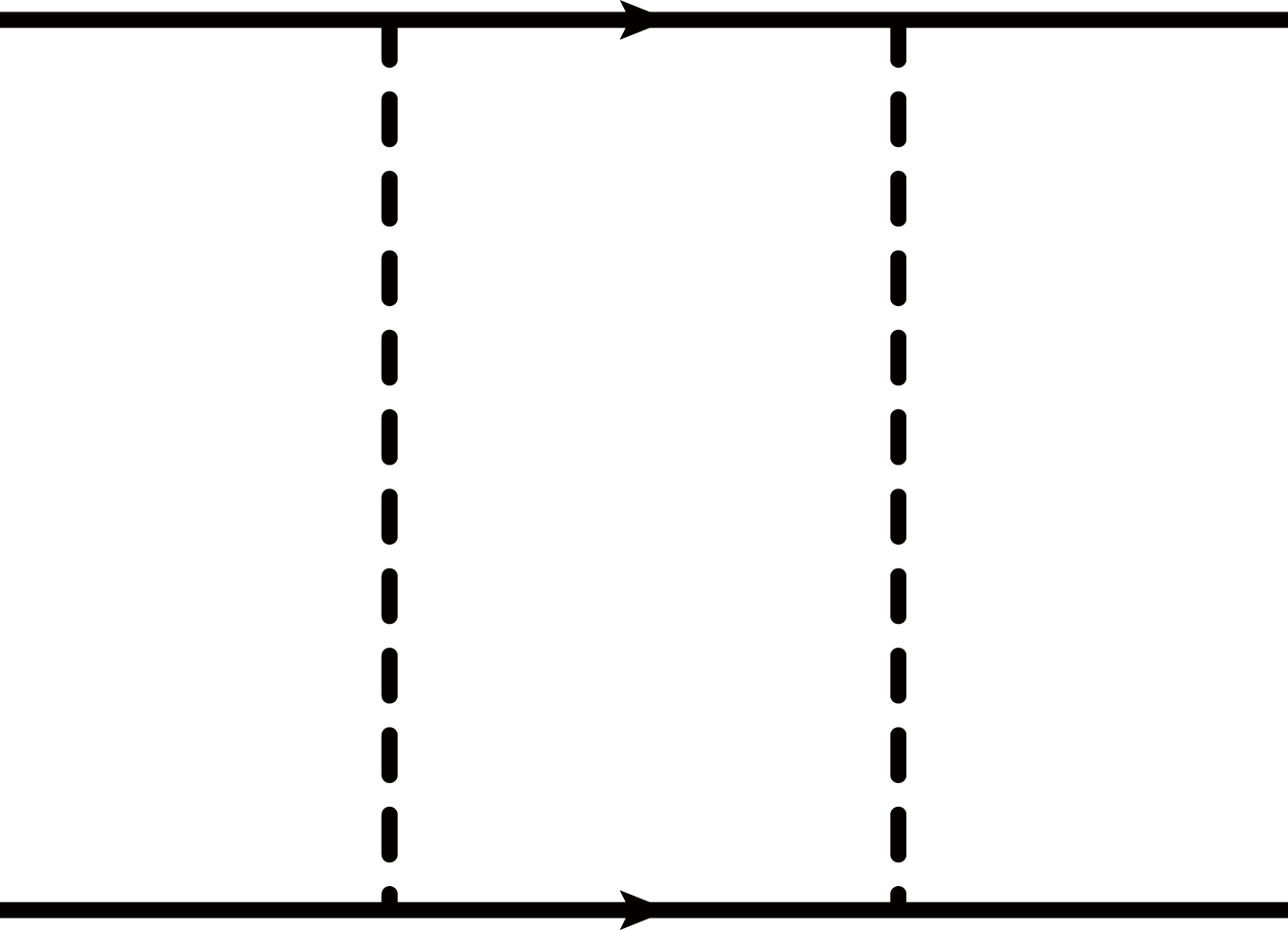}
}
\subfigure[]{
\includegraphics[width=1in]{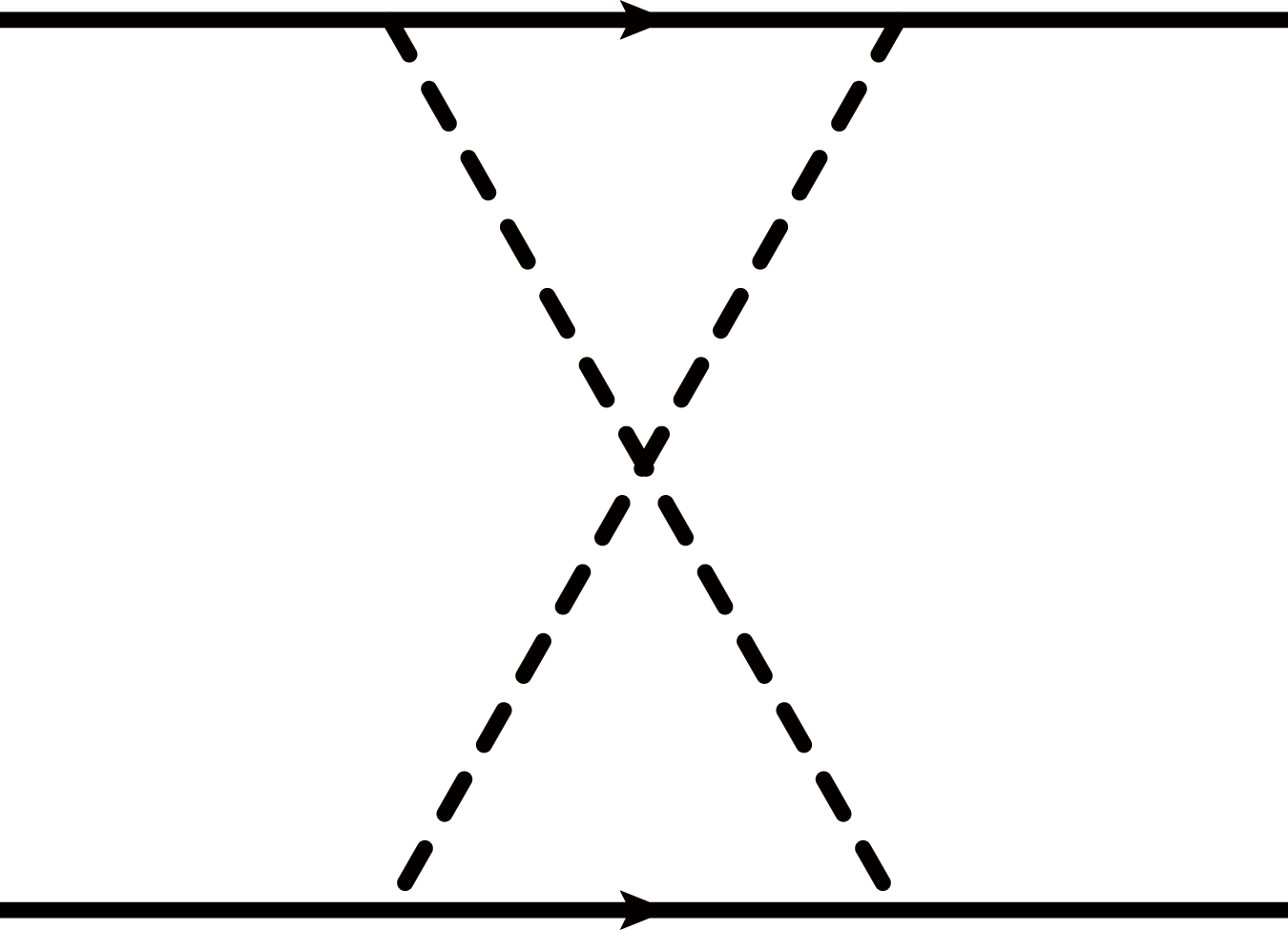}
}
\subfigure[]{
\includegraphics[width=1in]{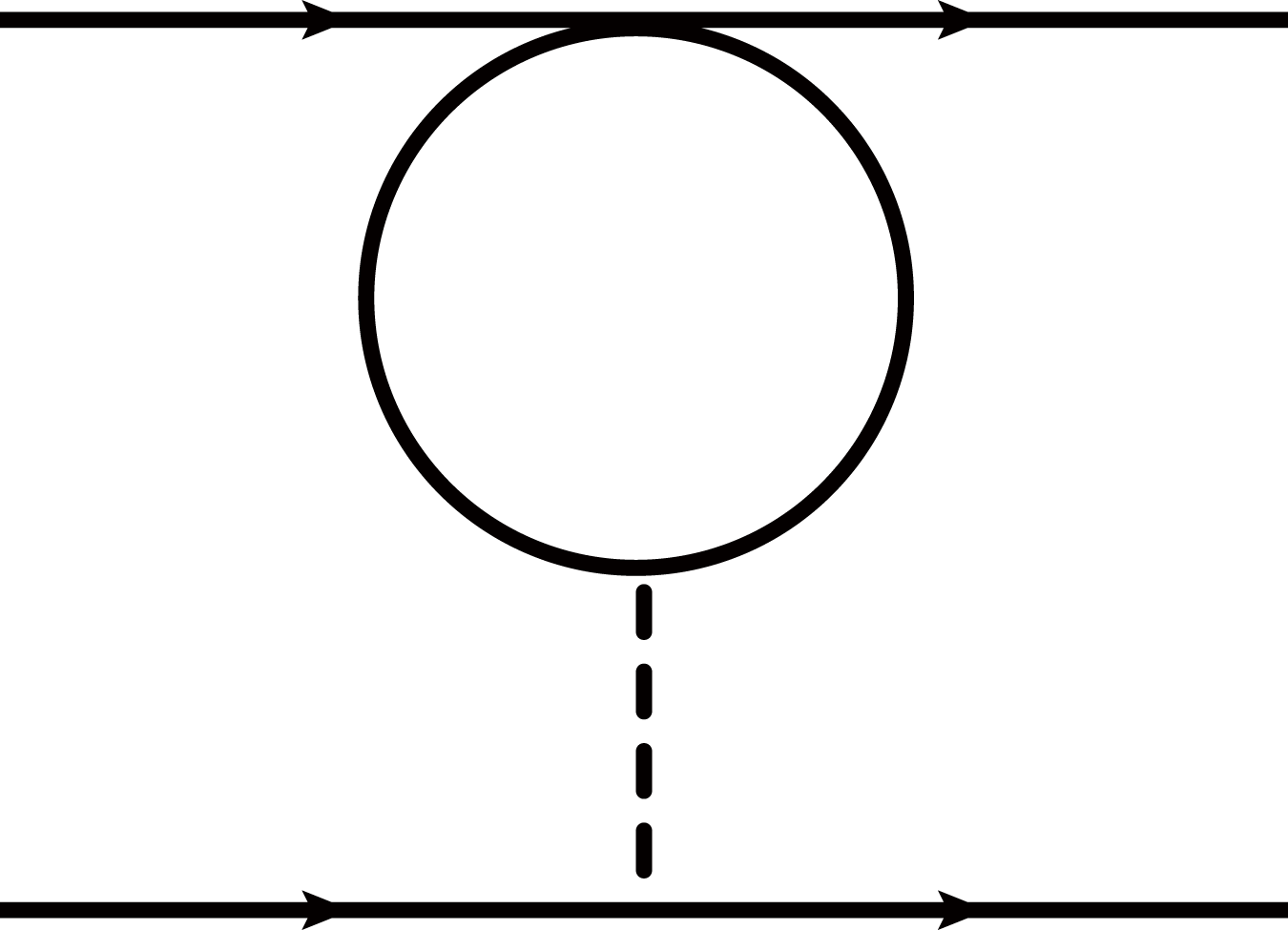}
}
\subfigure[]{
\includegraphics[width=1in]{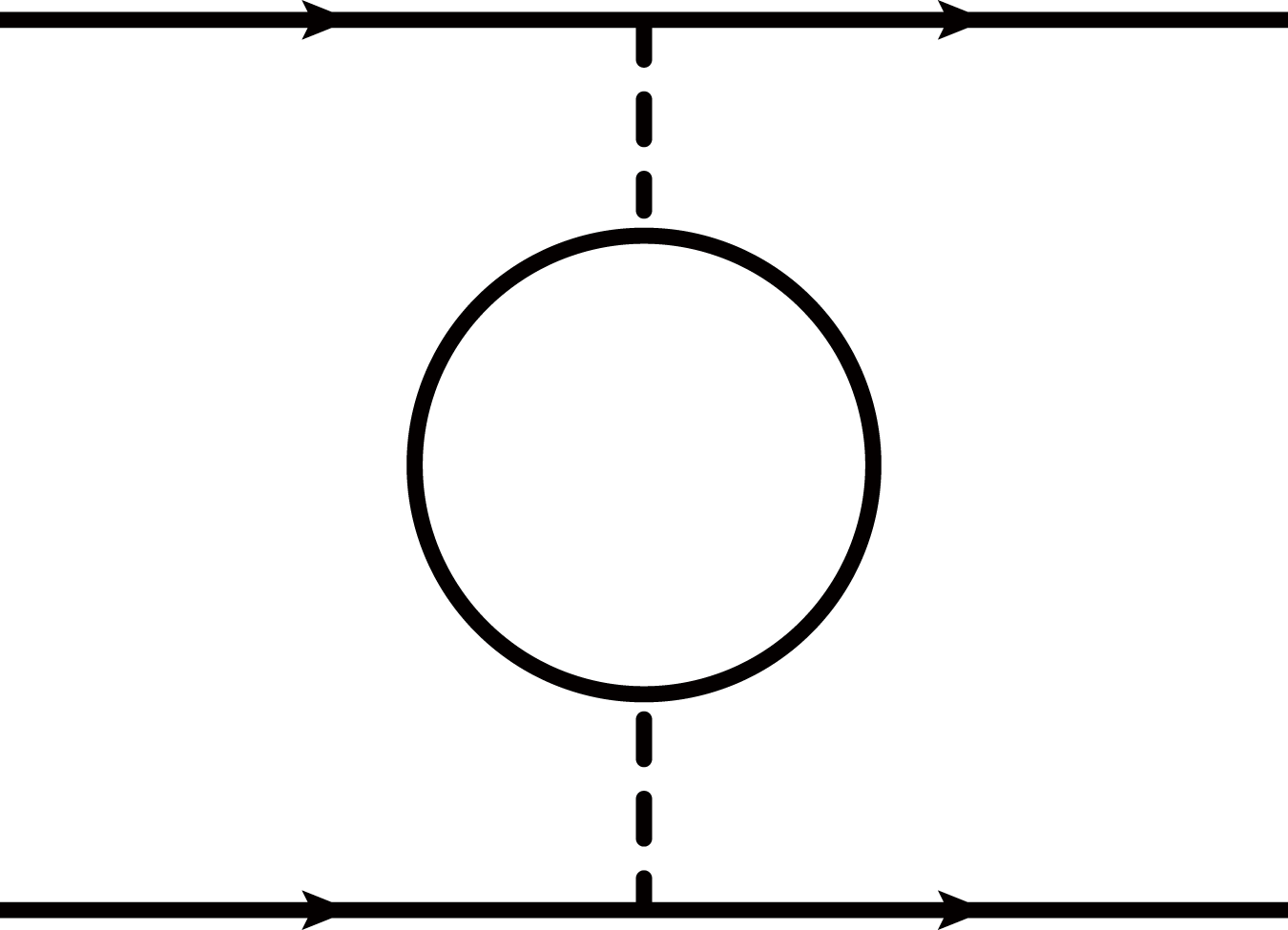}
}
\subfigure[]{
\includegraphics[width=1in]{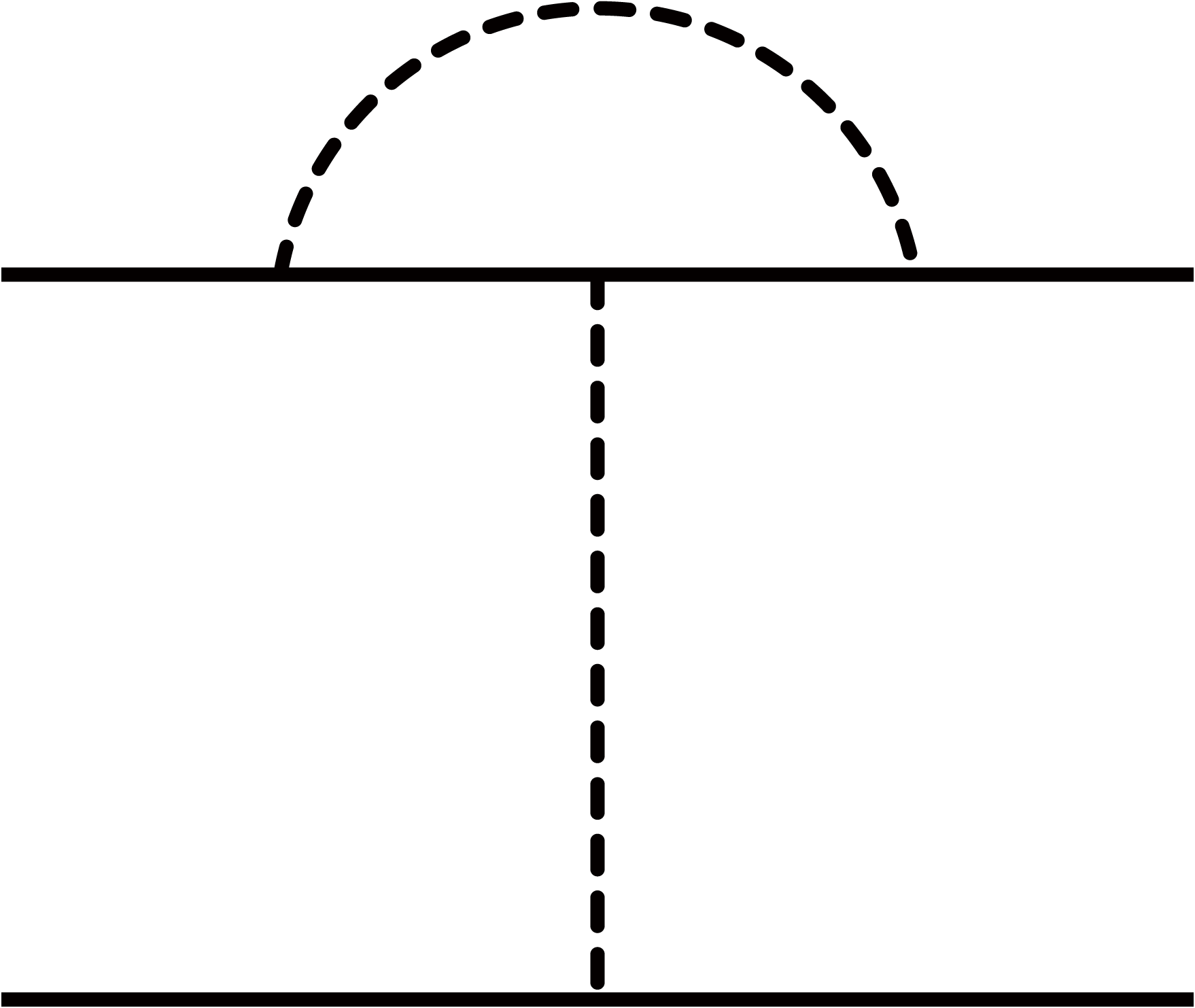}
}
\\
\vspace{-0.2cm}
\caption{(Color online) One-loop corrections to the fermion-disorder interaction due to (a)-(d) the fermion-disorder interaction
and (e) the Cooper interaction. The wave and dashed lines denote the Cooper-pairing interaction and the
fermion-disorder interaction, respectively.}\label{fig_one_loop_fermion_disorder}
\end{figure}

Further, we bring out the fermion-impurity interaction~(scattering) via adopting the
replica technique~\cite{Edwards1975JPF,Ramakrishnan1985RMP,Lerner0307471,
Mirlin2008RMP,Wang2015PLA,Roy2018PRX} to average over the random impurity potential $\mathcal{D}(\mathbf{x})$ which
satisfies $\langle \mathcal{D}(\mathbf{x})\rangle=0$ and $\langle \mathcal{D}(\mathbf{x})
\mathcal{D}(\mathbf{x}')\rangle=\Delta\delta(\mathbf{x}-\mathbf{x}')$~\cite{Nersesyan1995NPB,Stauber2005PRB,Wang2011PRB, Mirlin2008RMP,Coleman2015Book,Roy2018PRX} with $\mathcal{D}$ specifying the impurity field and the parameter
$\Delta$ denoting the concentration of the
impurity. This accordingly gives rise to the fermion-disorder interaction as~\cite{Roy-Saram2016PRB,Roy2018PRX},
\begin{eqnarray}
S_{\mathrm{dis}}
&=&\sum_{i=1} \frac{\Delta _i}{2} \int \prod_{\nu =1,\nu '=1 }^{\nu =2,\nu '=3} \frac{\mathrm{d}\omega _{\nu }\mathrm{d^2}\mathbf{k_\nu '}    }{\left ( 2\pi  \right )^8 }\psi_{m}^{\dagger }\left ( \omega _{1 } ,\mathbf{k_{1 }}   \right ) \mathcal{M}_i   \nonumber \\
&& \times  \psi_{m}\left ( \omega _{1} ,\mathbf{k_{2}}   \right )   \psi_n^{\dagger }\left( \omega _{2} ,\mathbf{k_{3}}   \right )
\mathcal{M}_i  \psi_n \left( \omega _{2} ,\mathbf{k_{4}}   \right ),
\label{eq:fermion-disorder interaction}
\end{eqnarray}
where $\Delta_i$ characterizes the strength of disorder and
$\mathcal{M}_i=\sigma_0$, $\sigma_{1,3}$, and $\sigma_2$ correspond to the random chemical potential,
the random gauge potential (two components), and the random mass disorders,
respectively~\cite{Nersesyan1995NPB,Stauber2005PRB,Wang2011PRB, Mirlin2008RMP,Coleman2015Book,Roy2018PRX}.
To wrap up, combining Eq.~(\ref{Eq_S0}) and Eq.~(\ref{eq:cooper-pairing interaction})
as well as Eq.~(\ref{eq:fermion-disorder interaction}), we arrive at the effective action as follows
\begin{equation}
S_{\mathrm{eff}}=S_{0}+S_{\mathrm{C}}+S_{\mathrm{dis}}.\label{Eq_S_eff}
\end{equation}
which serve as our starting point. This effective action contributes the one-loop corrections to
the fermionic propagator, the coupling $\lambda$, and the disorder strength as shown in
Fig.~\ref{fig_one_loop_fermion}, Fig.~\ref{fig_one_loop_cooper_pairing}, and
Fig.~\ref{fig_one_loop_fermion_disorder}, respectively. It is of particular importance to highlight that
the attractive Cooper-channel interaction~(\ref{Eq_H_int2}) produces three distinct classes of one-loop diagrams
including ZS, $\mathrm{ZS}'$, and BCS as shown in Fig.~\ref{fig_one_loop_cooper_pairing}~(a)-(c)~\cite{Shankar1994RMP}.
These collectively renormalize the coupling strength $\lambda$ and fundamentally govern low-energy physics~\cite{Shankar1994RMP,Sondhi2013PRB,Sondhi2014PRB} of the FDSM. We are going to examine
the potential emergence of Cooper instability in these fractional Dirac materials under the combined influence
of all interaction terms in $S_{\mathrm{eff}}$.

\section{RG analysis and coupled flow equations}\label{Sec_RG_analysis}

Within this section, we perform a one-loop renormalization
group (RG)~\cite{Wilson1975RMP,Polchinski9210046,Shankar1994RMP} analysis of the
effective theory~(\ref{Eq_S_eff}) to derive coupled flow equations for all relevant
parameters as the energy scale is lowered, which are generally crucial
to dictate the low-energy physical behavior. To this end, following the spirit of RG approach~\cite{Wilson1975RMP,Polchinski9210046,Shankar1994RMP},
one integrates out the fast-mode fermionic fields within the momentum shell $b\Lambda < |k| < \Lambda$,
where $b \equiv e^{-l} < 1$ and $l$ is the running energy scale, and then incorporates their corrections
into the slow modes, and finally rescales the slow modes to new ``fast modes"~\cite{Huh2008PRB,Kim2008PRB,
Maiti2010PRB,She2010PRB,She2015PRB,Vafek2012PRB,Vafek2014PRB,
Roy2016PRB,Wang2011PRB,Wang2013PRB,Wang2017PRB_QBCP,MDR2025PRB,Wang2025PRB_3D-tDSM}.

In order to connect two steps of RG processes, the non-interacting parts of effective action can be
considered as a fixed point that is invariant during the RG transformations. This yields the RG rescaling transformations of energies and momenta as well as fields in the momentum-frequency space~\cite{Shankar1994RMP,Huh2008PRB,Wang2011PRB},
\begin{eqnarray}
k_{x}&=&k'_{x}e^{-l},\label{Eq_rescaling_k_x}\\
k_{y}&=&k'_{y}e^{-l},\\
\omega&=&\omega'e^{-\alpha l},\\
\Psi_{\sigma}(i\omega,\mathbf{k})
&=&\Psi_{\sigma}'(i\omega',\mathbf{k}')e^{\left(\alpha+1-\eta_f\right)l},\label{Eq_rescaling_psi}
\end{eqnarray}
where $\eta_f$ is the anomalous dimension that captures the contribution from the one-loop corrections.

In particular, it is essential to reemphasize that the one-loop corrections to the interaction parameter $\lambda$ decompose into three distinct diagrammatic channels~\cite{Shankar1994RMP}, i.e. ZS, $\mathrm{ZS}'$, and BCS subchannels as depicted in Fig.~\ref{fig_one_loop_cooper_pairing}~(a)-(c). While both ZS and $\mathrm{ZS}'$ diagrams feature finite transfer momenta $\mathbf{Q} = \mathbf{q} - \mathbf{k}$ and $\mathbf{Q}' = -\mathbf{q} - \mathbf{k}$ respectively, under the Cooper interaction, it is restricted with $|\mathbf{Q}|\ll|\mathbf{Q}'|$ once two external momenta $\mathbf{q}$ and $\mathbf{k}$ possess the same sign
or $|\mathbf{Q}'|\ll|\mathbf{Q}|$ if they own opposite signs~\cite{Shankar1994RMP}. In this sense, we adopt the approximation
$\mathbf{Q} \approx 0$ while retaining finite $\mathbf{Q}'$ or vice versa~\cite{Shankar1994RMP,Wang2017PRB_BCS}.
Hereby, the finite transfer momentum can be parameterized as $\mathbf{Q} = Q(\cos\phi, \sin\phi)$
where $Q$ and $\phi$ denote magnitude and angular orientation, respectively.
Paralleling the analogous procedures in Refs.~\cite{Vafek2014PRB,Wang2017PRB_QBCP,DZZW2020PRB,Huh2008PRB,Maiti2010PRB,She2010PRB,
Wang2011PRB,Wang2013PRB,Kim2008PRB,She2015PRB,Roy2016PRB,Nandkishore2012NP} and carrying out lengthy but straightforward calculations,
we derive the anomalous dimension as
\begin{eqnarray}
\eta_f =\frac{\mathcal{F}_5}{16\pi^2v_\alpha^2} \sum_{j=0}^{3} \Delta_i,\label{Eq_1L_loop_corrections_1}
\end{eqnarray}
and all one-loop corrections for the Cooper pairing interaction
\begin{eqnarray}
\delta \lambda =\!\left\{\begin{matrix}
\frac{\lambda^2 \Lambda^4 l }{8\pi^2v_\alpha }\!\left [ \mathcal F_{4}-\frac{(\mathcal F_0+v_\alpha \mathcal F_{1}) }{2} -\frac{3}{8v_\alpha^4}  \mathcal F_{12} \right ]   \mathbf{Q} \neq 0,\mathbf{Q} '=0, \\
\\
\frac{\lambda^2 \Lambda^4 l }{8\pi^2v_\alpha }\!\left [ \mathcal F_{4}-\frac{(\mathcal F_0'+v_\alpha \mathcal F_{1}' ) }{2} -\frac{3}{8v_\alpha^4}  \mathcal F_{12}  \right ]  \mathbf{Q} = 0,\mathbf{Q} '\neq 0,
\end{matrix}\right.
\end{eqnarray}
as well as
\begin{eqnarray}
\delta \Delta _{0} & = &\frac{l}{4\pi^2} \left ( \frac{\Delta _0^2}{v_\alpha^2 } \mathcal F_{11}\right ),\\
\delta \Delta _{1} & = &\frac{l}{4\pi^2} \left ( \frac{\lambda }{16}\Delta_1  \mathcal F_3 -2\Delta_1^2\frac{1}{v_\alpha^2 } \mathcal F_2\right),\\
\delta \Delta _{2} & = &\frac{l}{4\pi^2} \left [ \frac{\lambda }{16}\Delta_2 \mathcal F_4 -\left ( \lambda  \Delta_2-2\Delta_2^2 \right ) \frac{1}{v_\alpha^2 } \mathcal F_2\right ],\\
\delta \Delta _{3} & = &\frac{l}{4\pi^2}\left [ \frac{\lambda }{16}\Delta_3\left ( 2\mathcal F_3+\mathcal F_4 \right )+2\Delta_3^2 \frac{1}{v_\alpha^2 } \mathcal F_{5} \right ],\label{Eq_1L_loop_corrections_2}
\end{eqnarray}
for the disorder strengths, where all the related coefficients are provided
in Appendix~\ref{Appendix_Related coefficients}.

\begin{figure}[h]
\hspace{-0.5cm}
\includegraphics [width=0.9\linewidth] {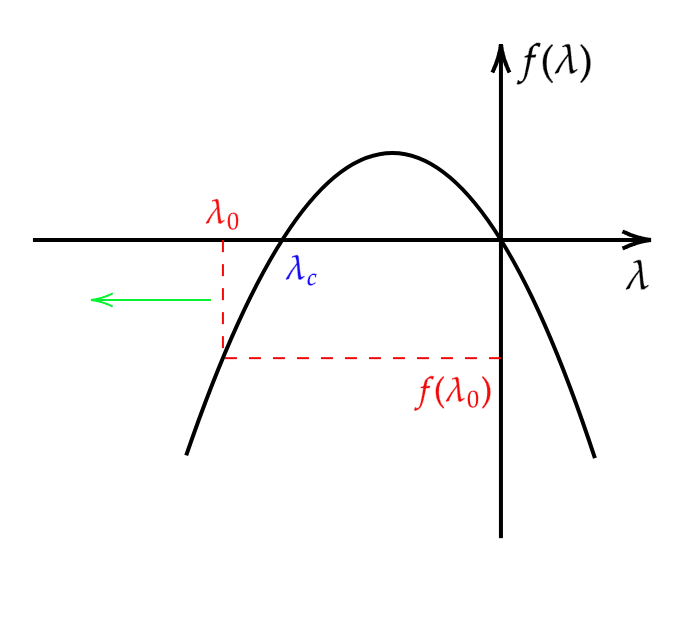}
\vspace{-1cm}
\caption{(Color online) Schematic illustration of the flow divergence for the Cooper channel coupling strength $\lambda$ when the initial coupling $|\lambda_0|$ exceeds the critical threshold $|\lambda_c|$.}
\label{sc}
\end{figure}

To proceed, following the standard procedures of RG approach~\cite{Shankar1994RMP} with the
help of the RG rescaling transformations~(\ref{Eq_rescaling_k_x})-(\ref{Eq_rescaling_psi}),  in tandem with
the one-loop corrections~(\ref{Eq_1L_loop_corrections_1})-(\ref{Eq_1L_loop_corrections_2}), the coupled RG equations of all interactions
in the effective action can be derived as,
\begin{widetext}
\begin{eqnarray}
\frac{dv_\alpha}{dl}
&=&-\frac{\mathcal F_{5}}{8\pi^2}\sum_{j=0}^{3}\frac{\Delta _j}{v_\alpha}
,\label{Eq_RG_v_alpha_Q}\\
\frac{d\lambda}{dl} &=& \lambda \left[ (\alpha - 2) - \frac{ \mathcal F_0+v_\alpha\mathcal{F}_1}{4\pi^2 v_\alpha}\lambda - \frac{\mathcal{F}_5}{4\pi^2} \sum_{j=0}^3 \Delta_j \right],
\end{eqnarray}
and
\begin{eqnarray}
\frac{d\Delta_0}{dl}
&=&\left[2(\alpha-1)+\frac{\mathcal F_{5}}{4\pi^2} \frac{\Delta _0-\Delta _1-\Delta _2-\Delta _3}{v_\alpha^2}
\right]\Delta _0,\\
\frac{d\Delta_1}{dl}
&=&\left[2(\alpha-1)
+  \frac{\mathcal F_3 }{32\pi^2}\lambda  -\frac{\mathcal F_2}{\pi^2}\frac{\Delta_1}{v_\alpha^2 } -\frac{\mathcal F_{5}}{4\pi^2}\sum_{j=0}^{3}\frac{\Delta _j}{v_\alpha^2}
\right]\Delta _1,\\
\frac{d\Delta_2}{dl}
&=&\left[2(\alpha-1)
+   \frac{\mathcal F_4 }{32\pi^2} \lambda -\frac{\mathcal F_2}{2\pi^2}\frac{\lambda  -2\Delta_2  }{v_\alpha^2 } -\frac{\mathcal F_{5}}{4\pi^2}\sum_{j=0}^{3}\frac{\Delta _j}{v_\alpha^2}\right]\Delta _2,\\
\frac{d\Delta_3}{dl}
&=&\left[2(\alpha-1)
+ \frac{2\mathcal F_3+\mathcal F_4  }{32\pi^2}\lambda+\frac{\mathcal F_{5} }{4\pi^2}\frac{3\Delta_3-\Delta _0-\Delta _1-\Delta _2}{v_\alpha^2 } \right]\Delta _3,\label{Eq_RG_Delta3_alpha_Q}
\end{eqnarray}
for the case $\mathbf{Q} \neq 0,\mathbf{Q} '=0$, and
\begin{eqnarray}
\frac{dv_\alpha}{dl}
&=&-\frac{\mathcal F_{5}}{8\pi^2}\sum_{j=0}^{3}\frac{\Delta _j}{v_\alpha}
\label{Eq_RG_v_alpha_Q_prime},\\
\frac{d\lambda}{dl} &=& \lambda \left[ (\alpha - 2) - \frac{ \mathcal F_0 '+v_\alpha\mathcal{F}_1'}{4\pi^2 v_\alpha}\lambda - \frac{\mathcal{F}_5}{4\pi^2} \sum_{j=0}^3 \Delta_j \right],
\end{eqnarray}
and
\begin{eqnarray}
\frac{d\Delta_0}{dl}
&=&\left[2(\alpha-1)+\frac{\mathcal F_{5}}{4\pi^2} \frac{\Delta _0-\Delta _1-\Delta _2-\Delta _3}{v_\alpha^2}
\right]\Delta _0,\\
\frac{d\Delta_1}{dl}
&=&\left[2(\alpha-1)
+  \frac{\mathcal F_3 }{32\pi^2}\lambda  -\frac{\mathcal F_2}{\pi^2}\frac{\Delta_1}{v_\alpha^2 } -\frac{\mathcal F_{5}}{4\pi^2}\sum_{j=0}^{3}\frac{\Delta _j}{v_\alpha^2}
\right]\Delta _1,\\
\frac{d\Delta_2}{dl}
&=&\left[2(\alpha-1)
+   \frac{\mathcal F_4 }{32\pi^2} \lambda -\frac{\mathcal F_2}{2\pi^2}\frac{\lambda  -2\Delta_2  }{v_\alpha^2 } -\frac{\mathcal F_{5}}{4\pi^2}\sum_{j=0}^{3}\frac{\Delta _j}{v_\alpha^2}\right]\Delta _2,\\
\frac{d\Delta_3}{dl}
&=&\left[2(\alpha-1)
+ \frac{2\mathcal F_3+\mathcal F_4  }{32\pi^2}\lambda+\frac{\mathcal F_{5} }{4\pi^2}\frac{3\Delta_3-\Delta _0-\Delta _1-\Delta _2}{v_\alpha^2 } \right]\Delta _3,\label{Eq_RG_v_Delta3_Q_prime}
\end{eqnarray}
for the case $\mathbf{Q} = 0,\mathbf{Q} '\neq 0$. All the related coefficients are presented in Appendix~\ref{Appendix_Related coefficients}.
\end{widetext}

Before proceeding, we highlight essential characteristics of above coupled RG evolutions for interaction parameters.
These equations render all couplings interdependent and thus their low-energy fates are intimately
associated with each other. Consequently, the low-energy behavior may deviate significantly from tree-level counterparts,
potentially undergoing either quantitative or qualitative modifications. Of particular importance is the renormalization of coupling
$\lambda$, which may flow toward strong-coupling regimes and potentially induce Cooper instability under certain conditions.
In the following sections, we are going to investigate whether Cooper instability can be generated
and how the system parameters govern its emergence.

\section{Potential Cooper instability at clean limit}\label{Sec_results_clean}

Using the coupled renormalization group (RG) equations~(\ref{Eq_RG_v_alpha_Q})-(\ref{Eq_RG_Delta3_alpha_Q}) for
$\mathbf{Q} \neq 0,\mathbf{Q} '=0$ and (\ref{Eq_RG_v_alpha_Q_prime})-(\ref{Eq_RG_v_Delta3_Q_prime})
for $\mathbf{Q} =0,\mathbf{Q} '\neq0$, we analyze
the low-energy evolution of the Cooper interaction strength $\lambda$. At tree level,
these RG equations reduce to
\begin{eqnarray}
\frac{dv}{dl}&=&0,\hspace{0.3cm}\frac{d\lambda}{dl}=(\alpha-2)\lambda,\label{Eq_RG_tree_lambda}
\end{eqnarray}
where $\alpha\in(0,1)$ characterizes the fractional dispersion in the FDSM~\cite{Roy2023PRR}. In this circumstance, $\lambda$ decreases monotonically as the energy scale decreases. Accordingly, the Cooper instability is strictly forbidden.
We subsequently turn attention to how one-loop corrections modify this behavior.
As mentioned in Sec.~\ref{Sec_RG_analysis}, the transfer momenta $\mathbf{Q}$ and $\mathbf{Q'}$ cluster into
two distinct scenarios, namely Scenario-A  and Scenario-B for  $\mathbf{Q} \neq 0$, $\mathbf{Q'} = 0$ and
$\mathbf{Q} = 0$, $\mathbf{Q'} \neq 0$, respectively. Without loss of generality, we primarily examine
the physical behavior in Scenario-A and address the discussion of Scenario-B for the end of this section.
For clarity, we restrict this section to the clean limit and defer disorder effects to Section~\ref{Sec_dis_effects}.

\subsection{Warm up: preliminary approximation analysis}

In the clean limit, the disorder parameters $\Delta_i$ (for $i=1,2,3$) vanish, reducing the RG equation for the Cooper interaction to
\begin{eqnarray}
\frac{d\lambda}{dl}
 & =  &\lambda\left[(\alpha-2)-\frac{\lambda}{4\pi^2v_\alpha }\left (  \mathcal F_0 + v_\alpha\mathcal F_1    \right) \right ],\label{eq:RG_theory_clean}
\end{eqnarray}
where the fermionic velocity $v_\alpha$ is a $l$-independent constant, and the coefficients
$\mathcal F_0$ and $\mathcal F_1$ are defined in Eq.~(\ref{Eq_F_0}) and Eq.~(\ref{Eq_F_1}), respectively.

\begin{figure}[h]
\centering
\subfigure[]{
\includegraphics[width=2.5in]{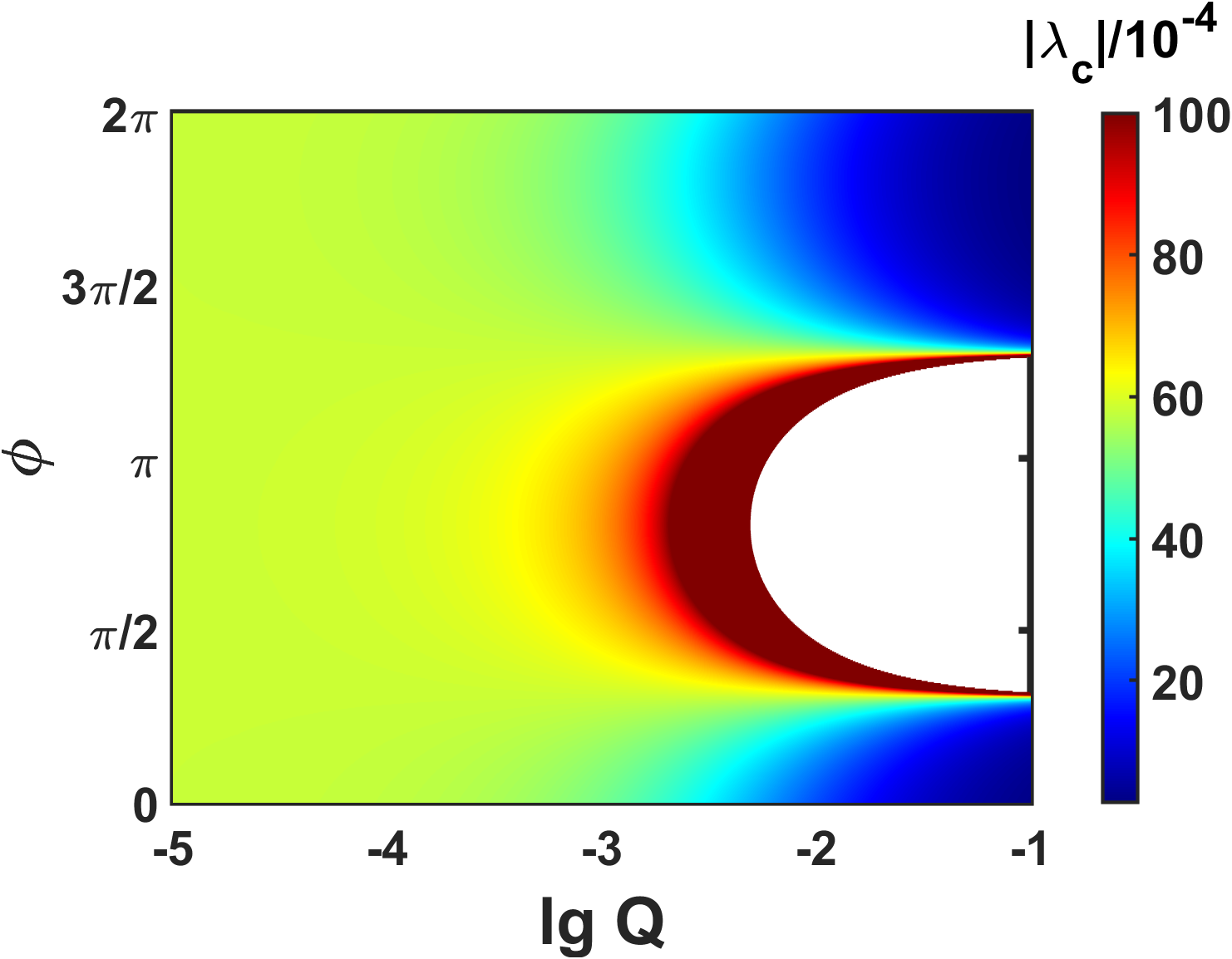}
}\\
\vspace{-0.15cm}
\subfigure[]{
\includegraphics[width=2.5in]{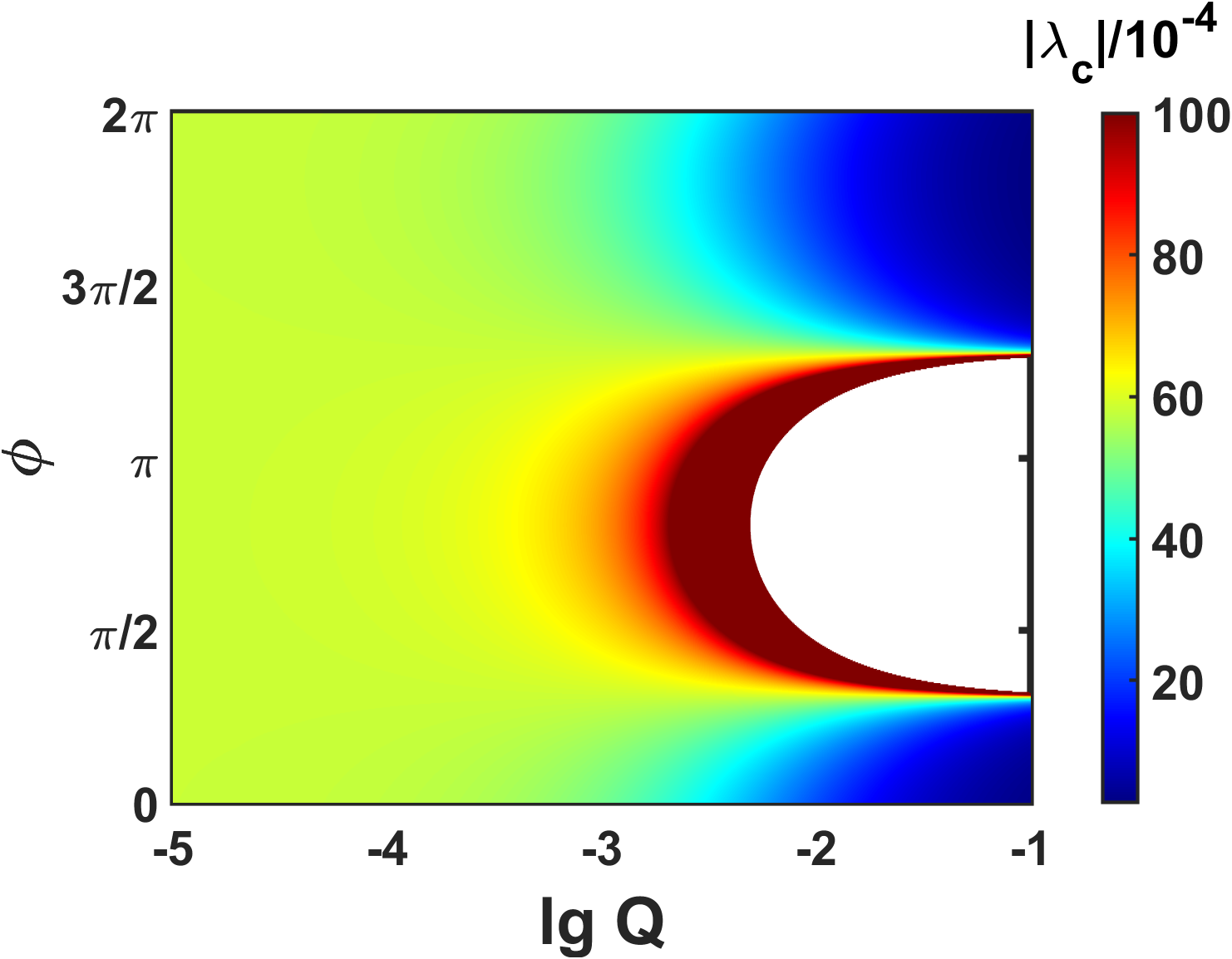}
}\\
\vspace{-0.15cm}
\subfigure[]{
\includegraphics[width=2.5in]{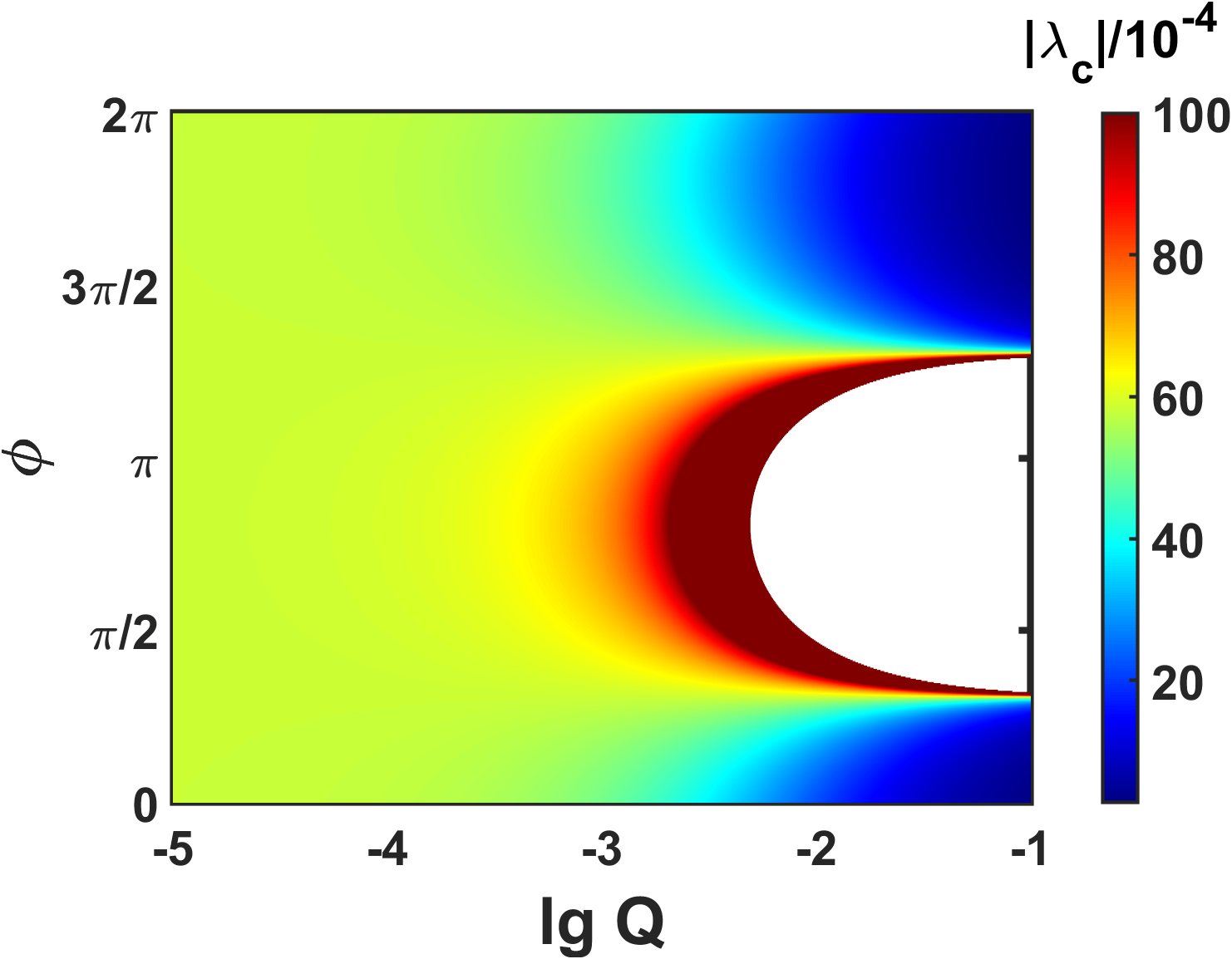}
}\\
\vspace{-0.2cm}
\caption{(Color online) The critical coupling strength $|\lambda_c|$ required for Cooper instability
with the fractional exponent $\alpha = 0.3$ and three fermionic velocities
(a) $v_\alpha = 10^{-3}$, (b) $10^{-4}$, and (c) $10^{-5}$.
The parameter space partitions into two distinct sectors: Zone-I (white) and Zone-II (colored)
where the Cooper instability is prohibited and  permitted at $|\lambda_0| > |\lambda_c|$.}
\label{fig_Q_phi_1}
\end{figure}


As a warm-up, let us assume $\mathcal F_0$ and $\mathcal F_1$ to be two constants and provide an attentive
and preliminary analysis of the behavior of Cooper interaction. For convenience, we introduce $a\equiv\alpha-2$ and $b\equiv-\frac{(\mathcal F_0+\mathcal F_1v_{\alpha})}{4\pi^2v_{\alpha}}$ and thus Eq.~(\ref{eq:RG_theory_clean})
is cast into a compact form
\begin{equation}
\frac{d\lambda}{dl} = a\lambda + b\lambda^2.
\label{eq:RG_transformed}
\end{equation}
It can be found that the evolution of $\lambda$ is governed by the sign of its slope.
On one hand, a positive slope with $d\lambda/dl>0$ indicates that $\lambda$ increases toward asymptotic stability at $\lambda=0$,
thus suppressing the BCS instability. On the other hand, a negative slope causes $\lambda$ to
toward strong-coupling divergence, which is an indicator of BCS instability~\cite{Shankar1994RMP,Zhao2006PRL,Honerkamp2008PRL,Sondhi2013PRB,Sondhi2014PRB}.


\begin{figure*}[htbp]
\centering
\subfigure[]{
\includegraphics[width=2.25in]{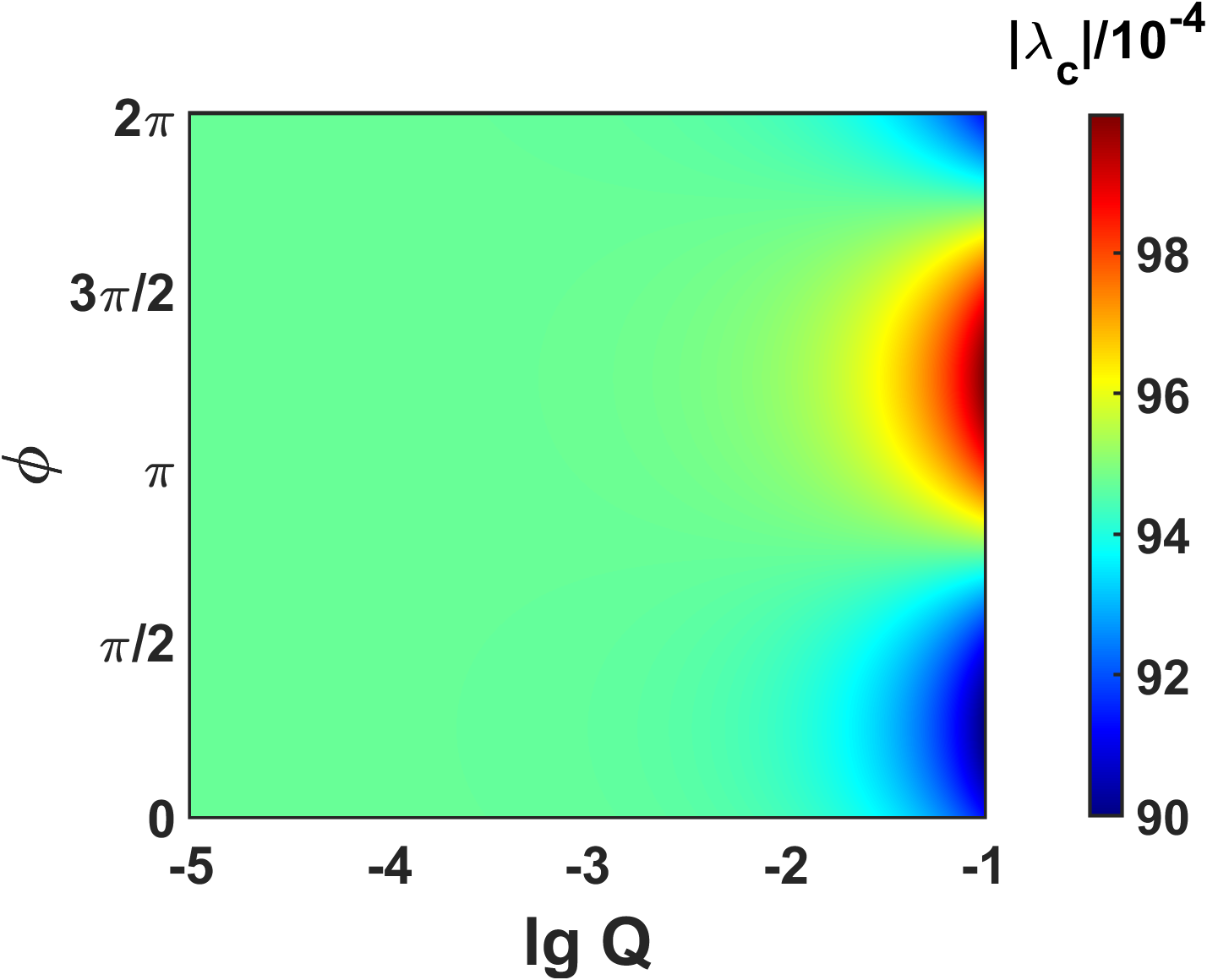}\hspace{-0.3cm}
}
\subfigure[]{
\includegraphics[width=2.25in]{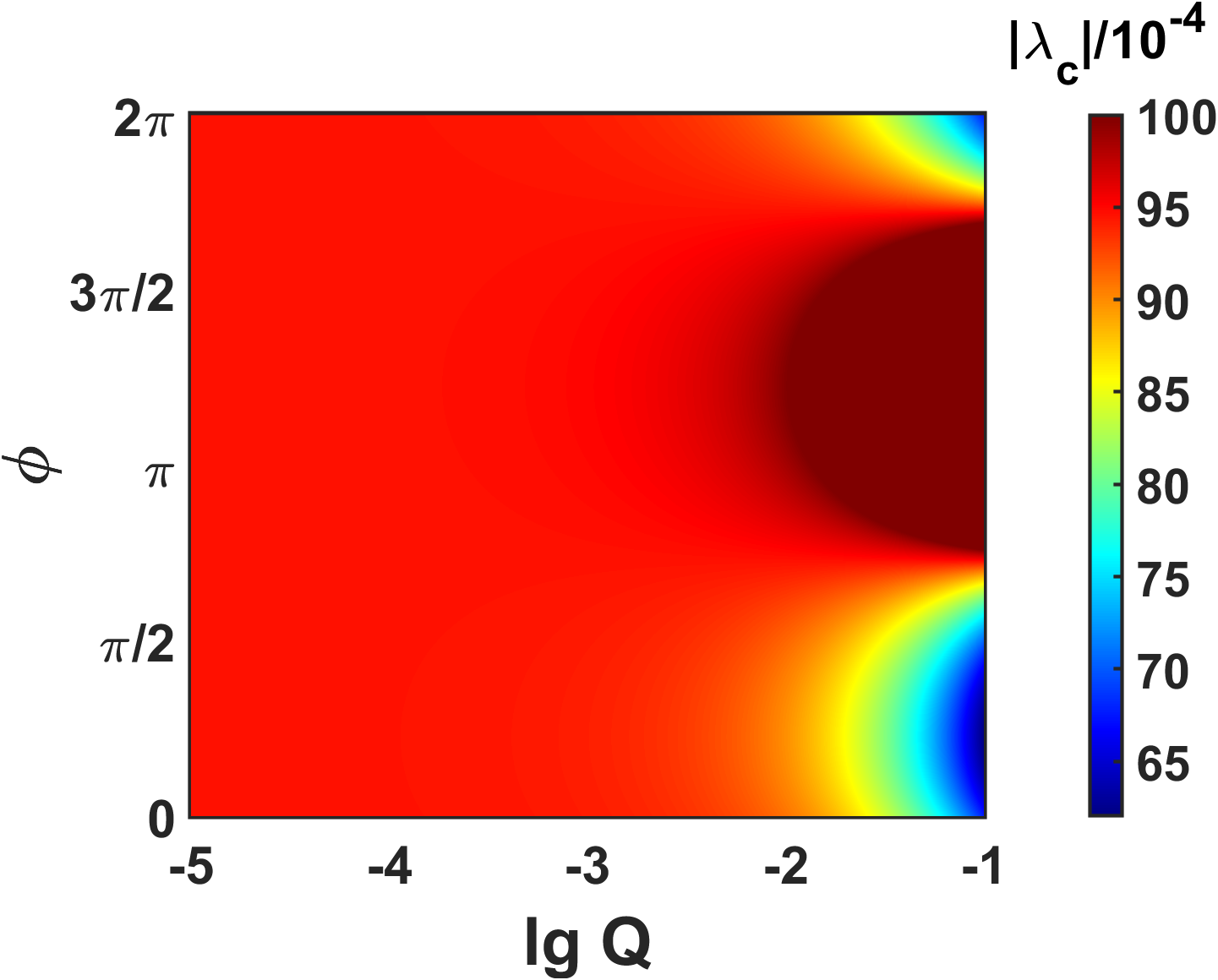}\hspace{-0.3cm}
}
\subfigure[]{
\includegraphics[width=2.25in]{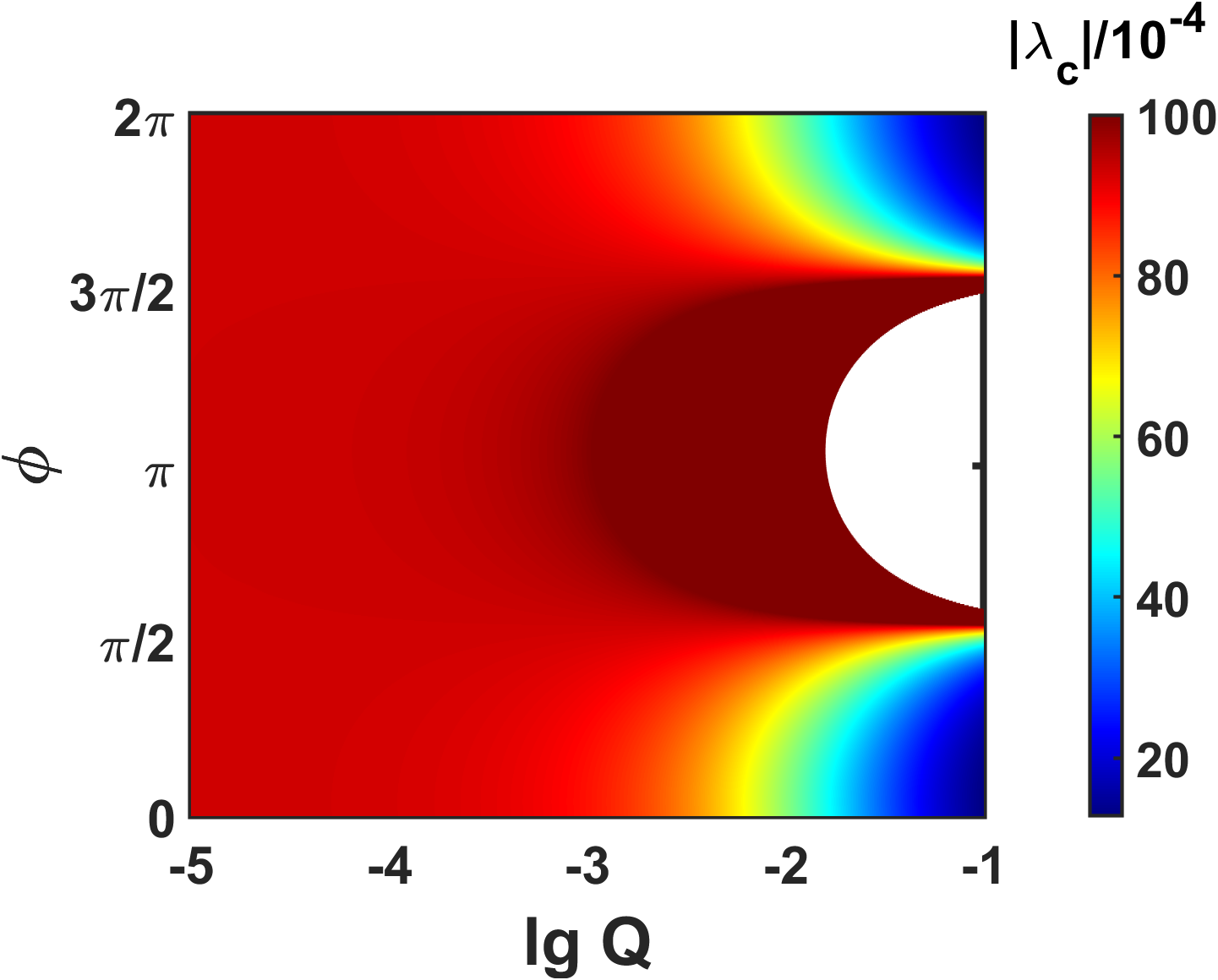}
}
\\
\subfigure[]{
\includegraphics[width=2.15in]{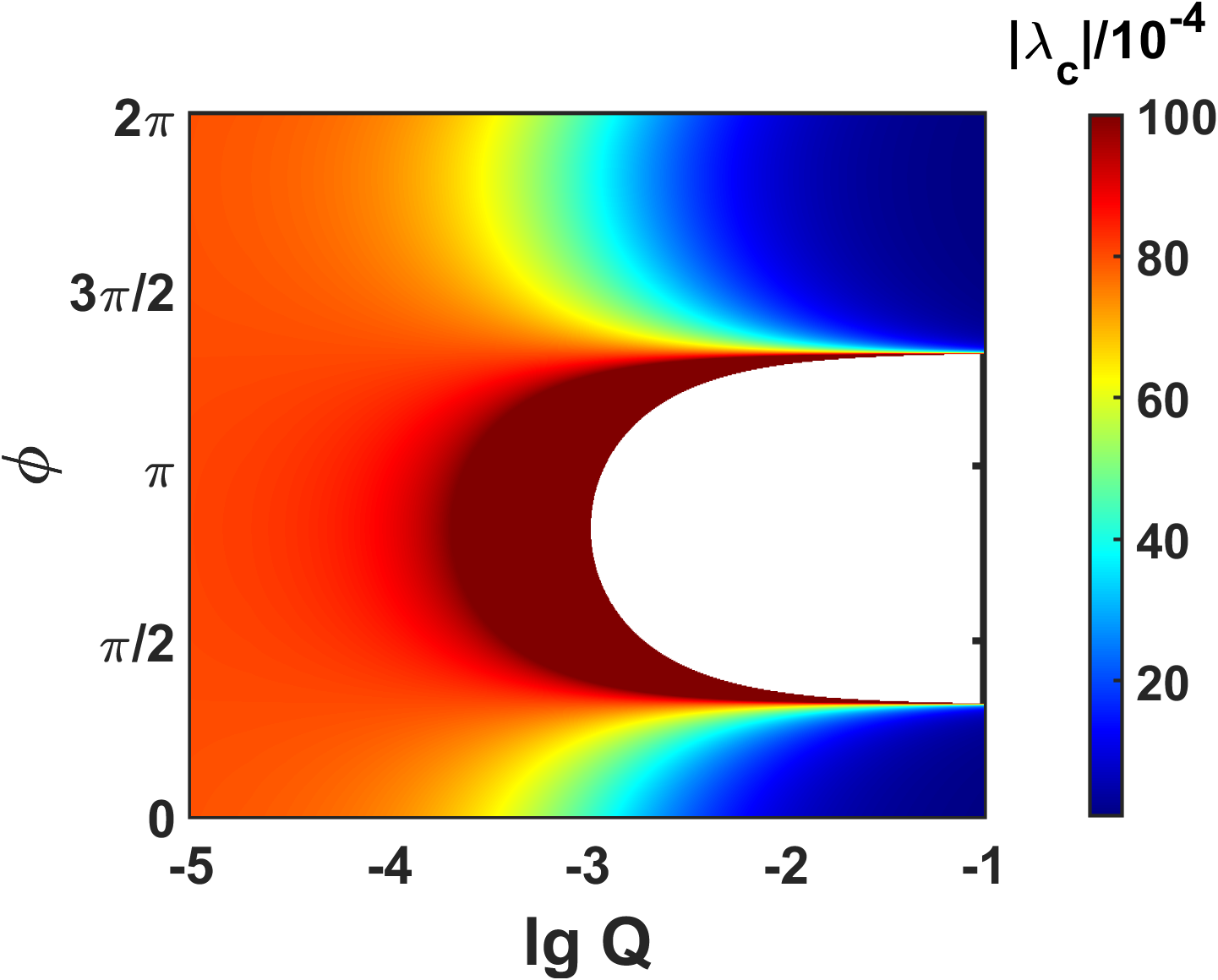}\hspace{-0.3cm}
}
\subfigure[]{
\includegraphics[width=2.15in]{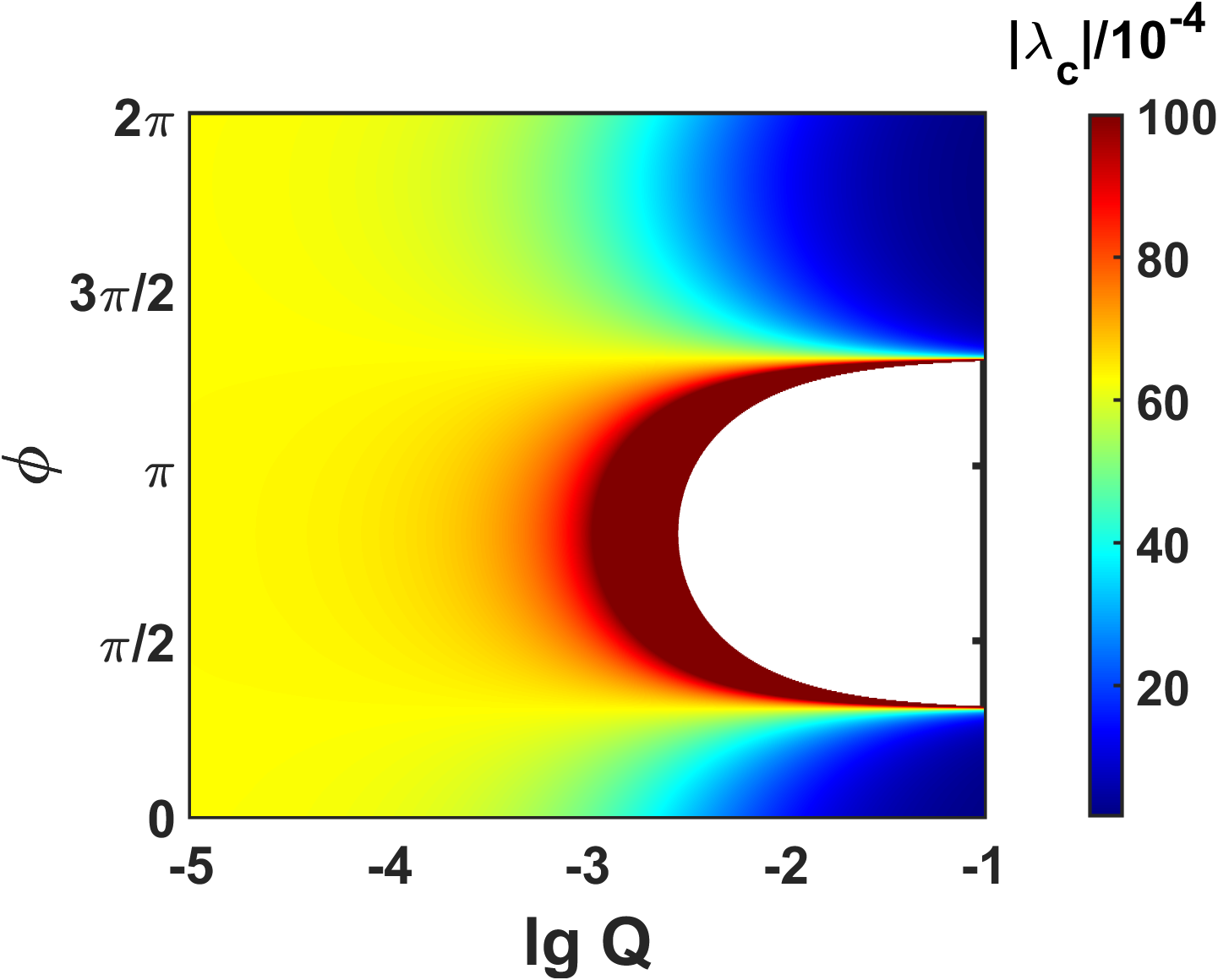}\hspace{-0.3cm}
}
\subfigure[]{
\includegraphics[width=2.15in]{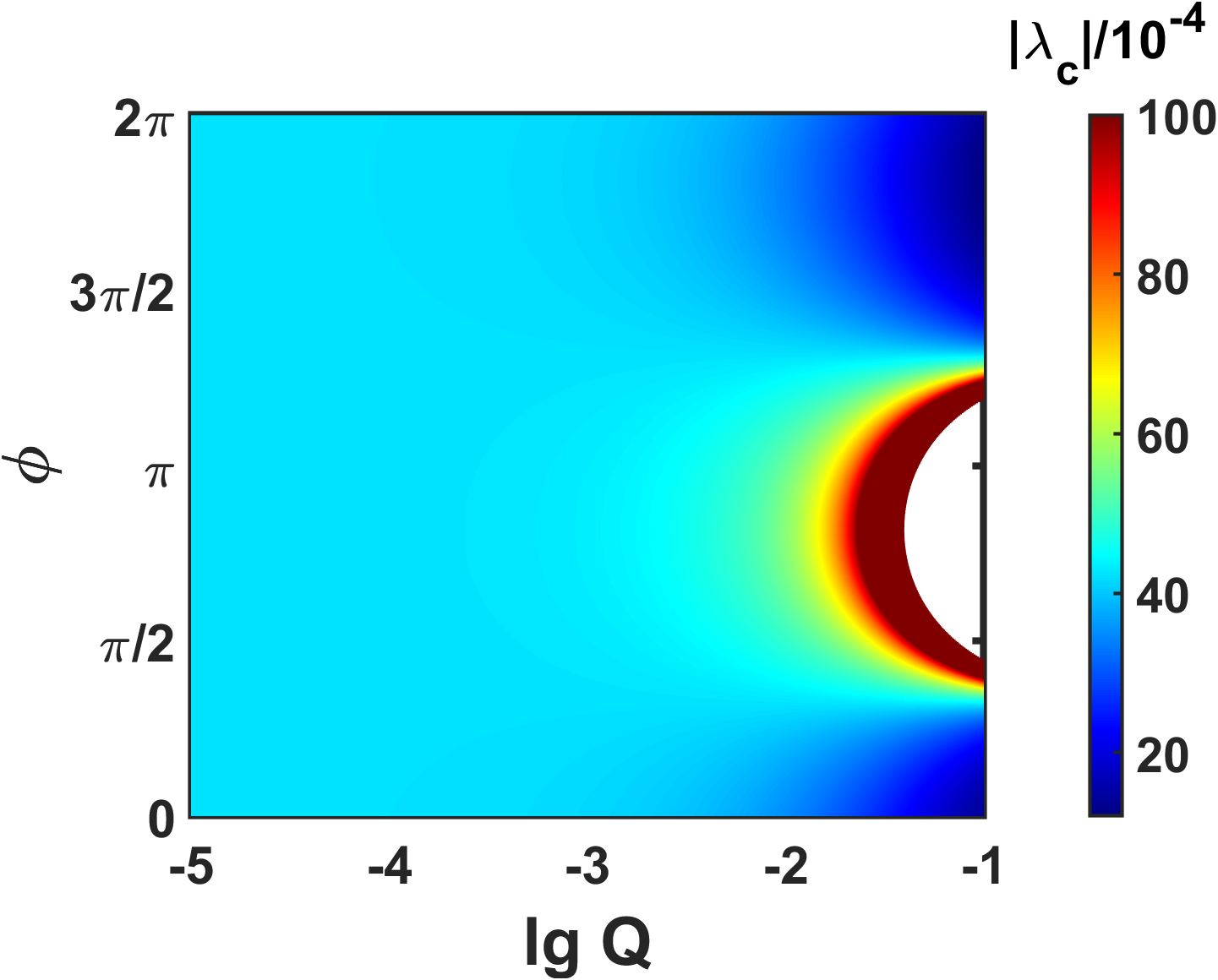}
}
\\
\subfigure[]{
\includegraphics[width=2.15in]{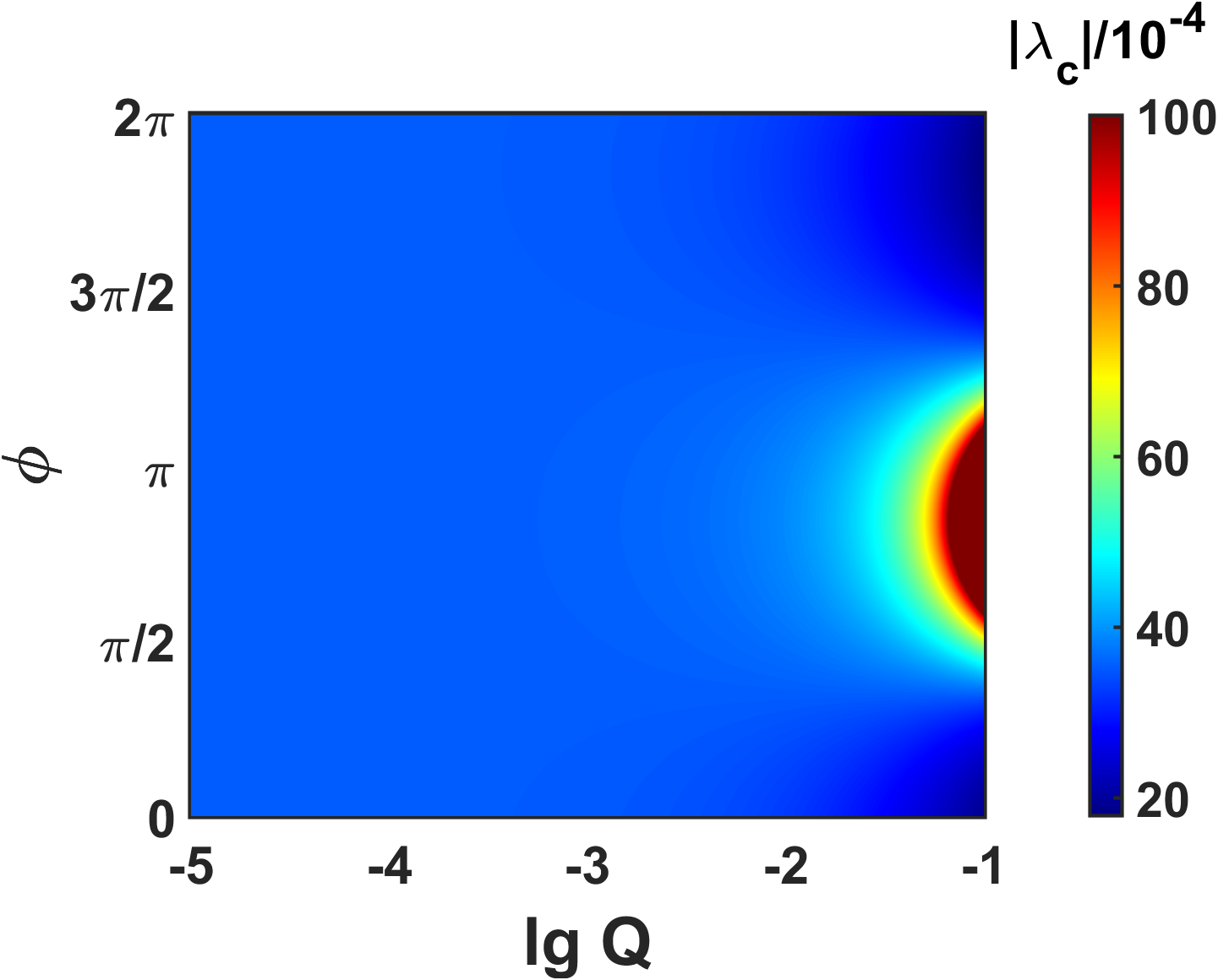}\hspace{-0.3cm}
}
\subfigure[]{
\includegraphics[width=2.15in]{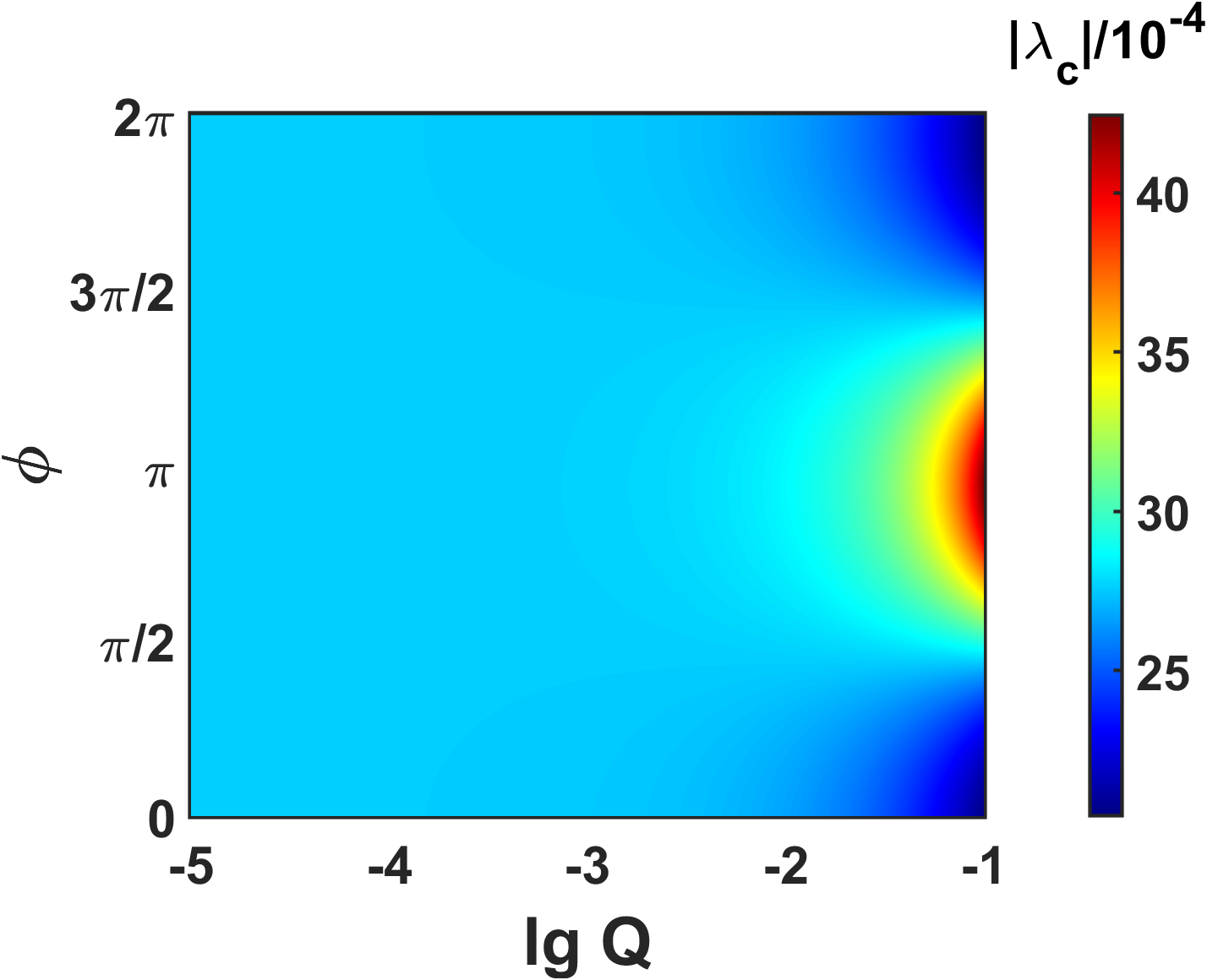}\hspace{-0.3cm}
}
\subfigure[]{
\includegraphics[width=2.15in]{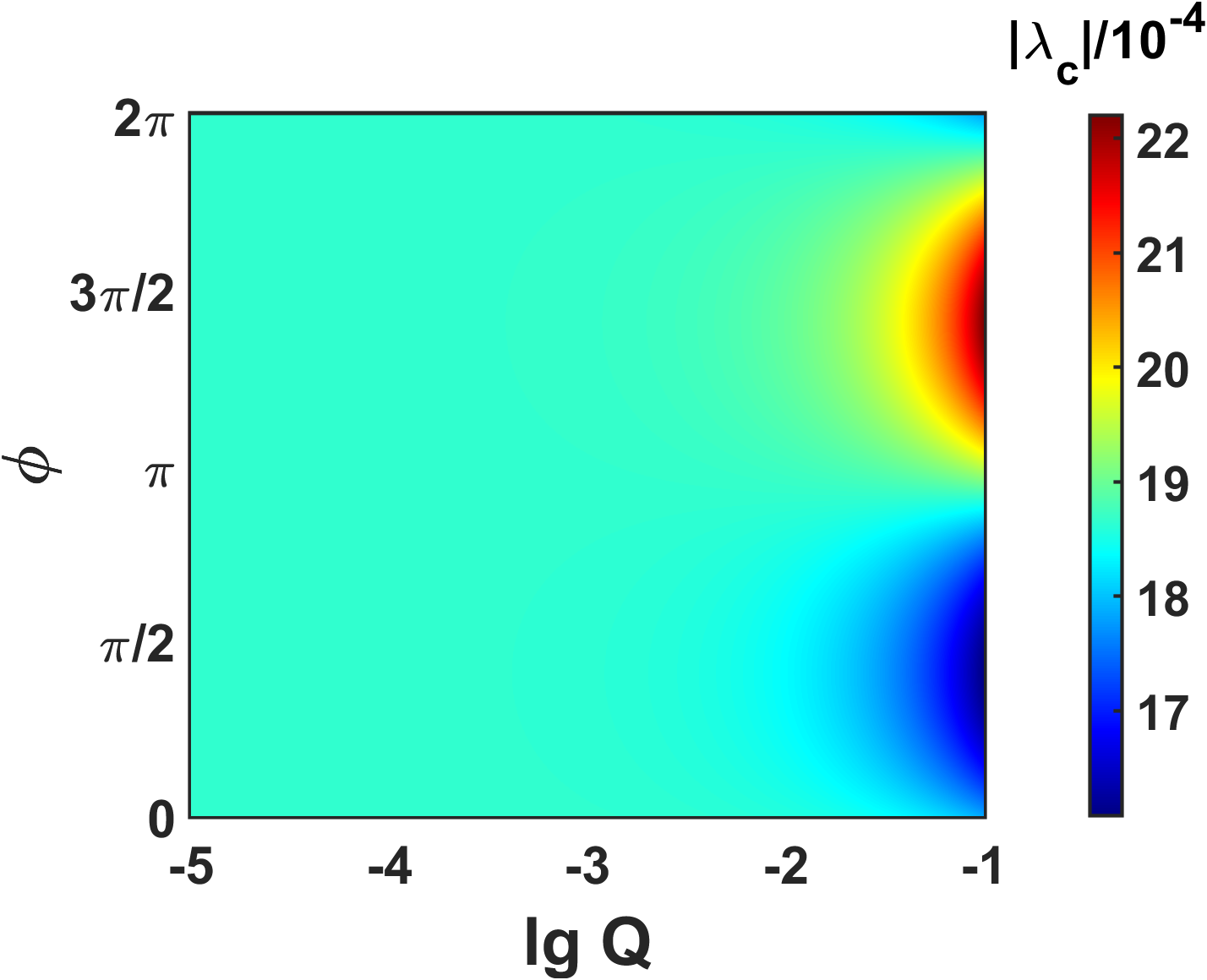}
}
\\
\vspace{-0.2cm}
\caption{(Color online) The critical interaction strength $|\lambda_c|$ required for Cooper instability
with fixing the fermionic velocity $v_\alpha = 10^{-3}$ and tuning the parameter $\alpha$ as
(a) $10^{-4}$, (b) $10^{-3}$, (c) $10^{-2}$, (d) $10^{-1}$, (e) 0.25, (f) 0.50, (g) 0.61, (h) 0.75 , and (i) 0.95.}
\label{fig_Q_phi}
\end{figure*}

In this sense, taking $d\lambda/dl=0$ yields two critical points $\lambda_c = 0$  and $\lambda_c = -a/b$.
Since $\lambda_c = 0$ is trivial, we put the focus on the latter. As illustrated in Fig.~\ref{sc},
the Cooper interaction $\lambda$ goes towards divergence as long as the following initial condition satisfies
\begin{eqnarray}
\left | \lambda_0 \right |>|\lambda_c|=\frac{4(2-\alpha)\pi^2v_\alpha}
{\mathcal F_0+\mathcal F_1 v_\alpha }.\label{eq:Superconductivity condition}
\end{eqnarray}
This inequality represents a preliminary criterion for Cooper instability with all related coefficients being constant.
However, both $\mathcal F_0$ and $\mathcal F_1$ are not constants but dependent on other parameters. It is necessary to
go beyond the coarse analytical analysis and examine this more systematically.

\subsection{Critical Cooper strength of Scenario-A}\label{Subsec_lambda_c}

Subsequently, let us examine the tendency of critical strength $|\lambda_c|$ at clean limit, which
is composed of $v_\alpha$ as well as the parameters $(\alpha, Q, \phi)$.

At the outset, we find from Fig.~\ref{fig_Q_phi_1}
that the variation of $v_\alpha$ can only provide quantitative effects on the value of
$\lambda_c$ but do not qualitatively modify the overall structure of dependence of other parameters.
It is of particular importance to highlight that there yield two distinct regions in the $Q$-$\phi$ space, which are
designated as $\text{Zone-}\mathrm{I}$ and $\text{Zone-}\mathrm{II}$, corresponding
to the white region and the colored region in Fig.~\ref{fig_Q_phi_1}, respectively.
As to the former, $\mathcal F_0+\mathcal F_1 v_\alpha <0$ implies that the BCS instability
is strictly forbidden, but instead $\frac{4(2-\alpha)\pi^2v_\alpha}{\mathcal F_0+\mathcal F_1 v_\alpha}>0$ for the
latter and thus the BCS instability is available at $|\lambda_0|>|\lambda_c|$. Reading off Fig.~\ref{fig_Q_phi_1}, it is clear that
the shapes of Zone-\uppercase\expandafter{\romannumeral1} and Zone-\uppercase\expandafter{\romannumeral2}
is insensitive to the variation of $v_\alpha$ which only causes $|\lambda_c|$ to change.
As a consequence, the $v_\alpha$ is subordinate to other parameters in impacting the BCS criterion that is primarily determined by the
$\text{Zone-}\mathrm{II}$. To simplify the analysis, we take $v_\alpha = 10^{-3}$ in the following discussions.

Next, let us consider the influence of the fractional exponent $\alpha$ on the Cooper instability.
Fig.~\ref{fig_Q_phi} illustrates that the areas of Zone-\uppercase\expandafter{\romannumeral1} and
Zone-\uppercase\expandafter{\romannumeral2} are highly sensitive to the value of $\alpha$.
To be concrete, Fig.~\ref{fig_Q_phi} presents that there exist two
critical values, $\alpha_{c1} \approx 10^{-3}$ and $\alpha_{c2} \approx 0.61$.
As $\alpha$ increases from a very small value, Zone-\uppercase\expandafter{\romannumeral1} emerges at $\alpha=\alpha_{c1}$,
then expands, contracts, and finally disappears beyond $\alpha=\alpha_{c2}$ as shown in Fig.~\ref{fig_Q_phi}(a)-(d).
This implies that within the region $\alpha \in (\alpha_{c1}, \alpha_{c2})$, Zone-\uppercase\expandafter{\romannumeral1} and
Zone-\uppercase\expandafter{\romannumeral2} can coexist and compete with each other, while Zone-II dominates when $\alpha < \alpha_{c1}$ or $\alpha > \alpha_{c2}$. Besides, Fig.~\ref{fig_Q_phi}(e)-(i) clearly show that the increase of $\alpha$ is helpful to enlarge the blue portion of Zone-\uppercase\expandafter{\romannumeral2}, where the critical strength of Cooper interaction $|\lambda|$
decreases. This indicates that a higher $\alpha$ favors the emergence of BCS instability.

Furthermore, we examine the effects of transfer momentum characterized by momentum magnitude $Q$
and angular orientation $\phi$ on the critical strength of Cooper interaction.
As depicted in Fig.~\ref{fig_Q_phi}, the response to $Q$-variation is strongly modulated by $\phi$.
For orientations near $\phi \approx 0$ or $2\pi$, the critical strength $|\lambda_c|$ decreases
monotonically with increasing $Q$. In sharp contrast, $|\lambda_c|$ exhibits the inverse behavior near
$\phi \approx \pi$. Specifically, at low momentum scales, $|\lambda_c|$ shows insensitive variation with $\phi$, reflecting weak directional dependence. Conversely, in the high-momentum regime, it displays angular dependence,
first increasing and then decreasing as $\phi$ is tuned from $0$ to $2\pi$. Additionally,
for $\alpha \in (\alpha_{c1}, \alpha_{c2})$, when $Q$ and $\phi$ take appropriate values, the
critical strength $|\lambda_c|$ goes toward infinity. This indicates that the Cooper instability is forbidden,
corresponding to the emergence of $\text{Zone-}\mathrm{I}$ which is shown as the white region in Fig.~\ref{fig_Q_phi}.

\subsection{Fate of Cooper interaction for Scenario-A}\label{Subsec_fate_Cooper_inter_Scenario_A}

\begin{figure}[h]
\hspace{-0.95cm}
\subfigure[]{
\includegraphics[width=1.9in]{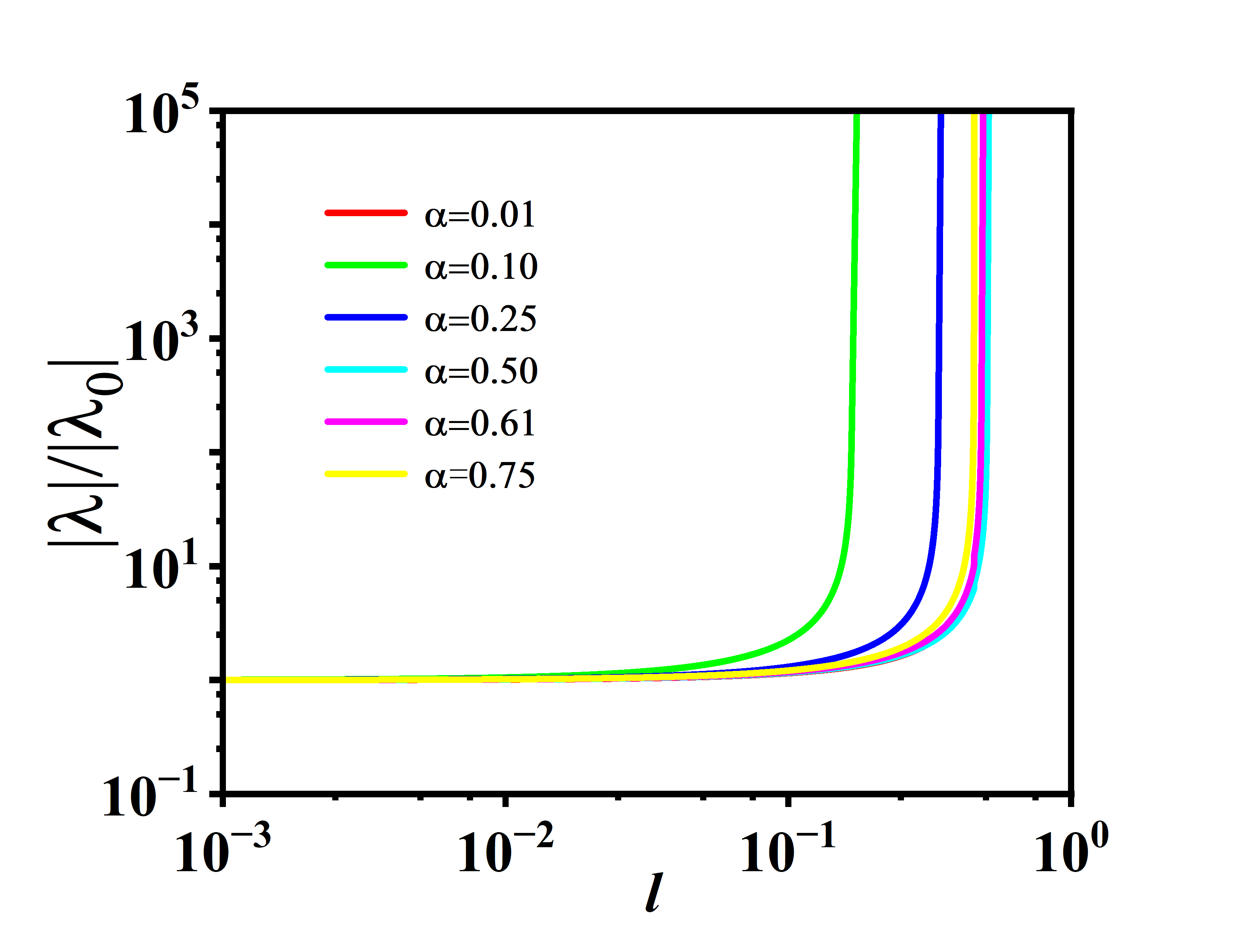}\hspace{-0.85cm}
}
\subfigure[]{
\includegraphics[width=1.9in]{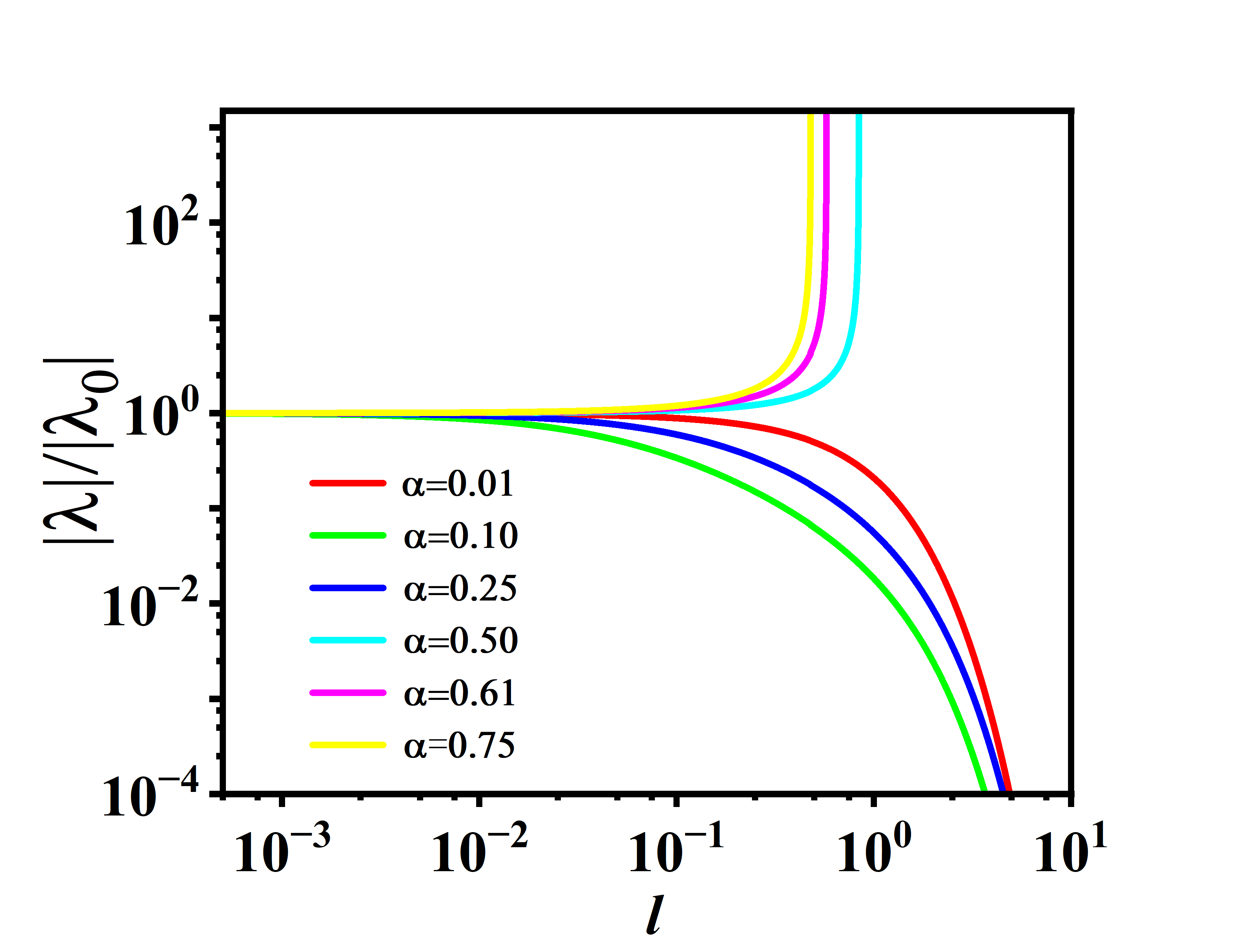}
}\\
\hspace{-0.95cm}
\subfigure[]{
\includegraphics[width=1.9in]{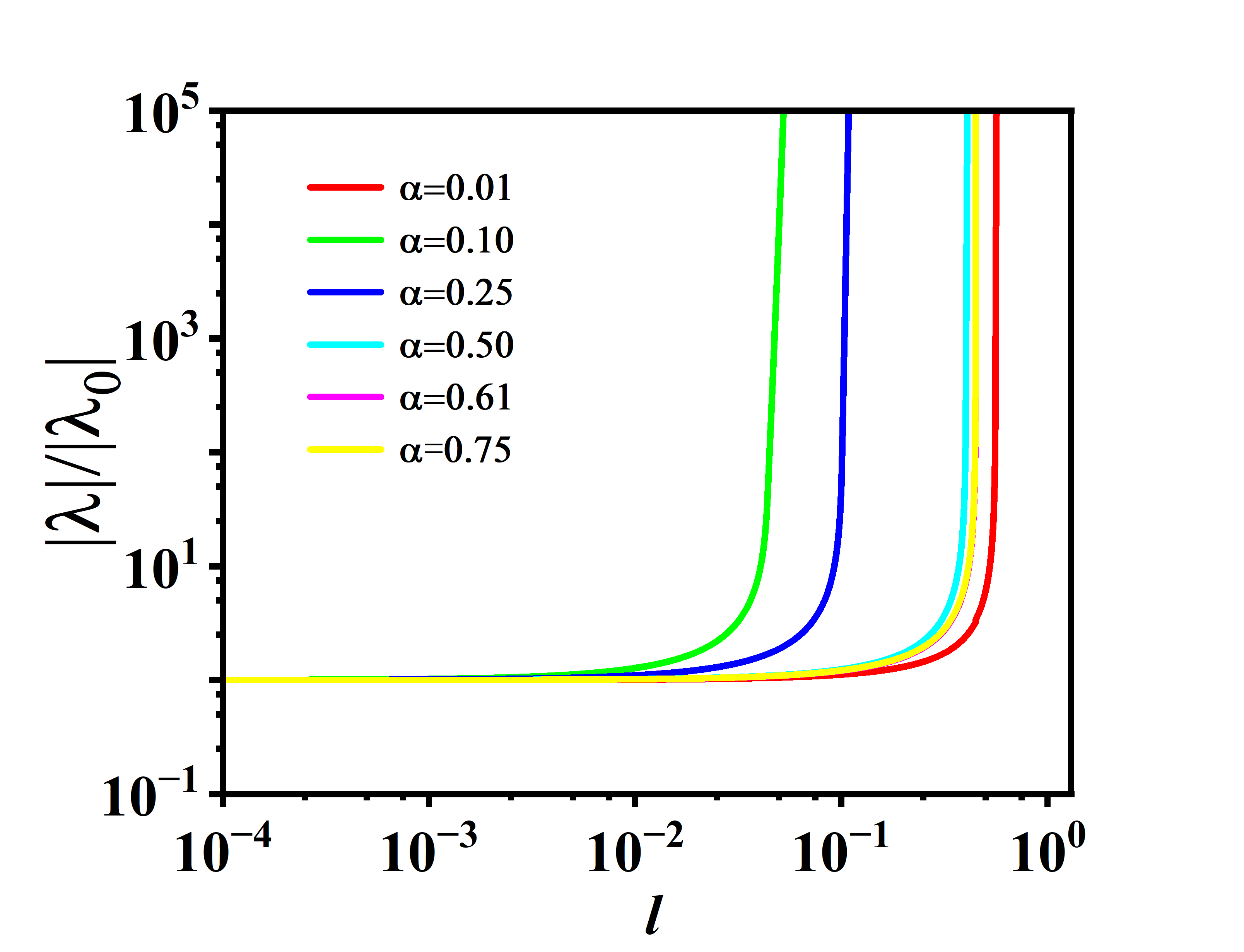}\hspace{-0.85cm}
}
\subfigure[]{
\includegraphics[width=1.9in]{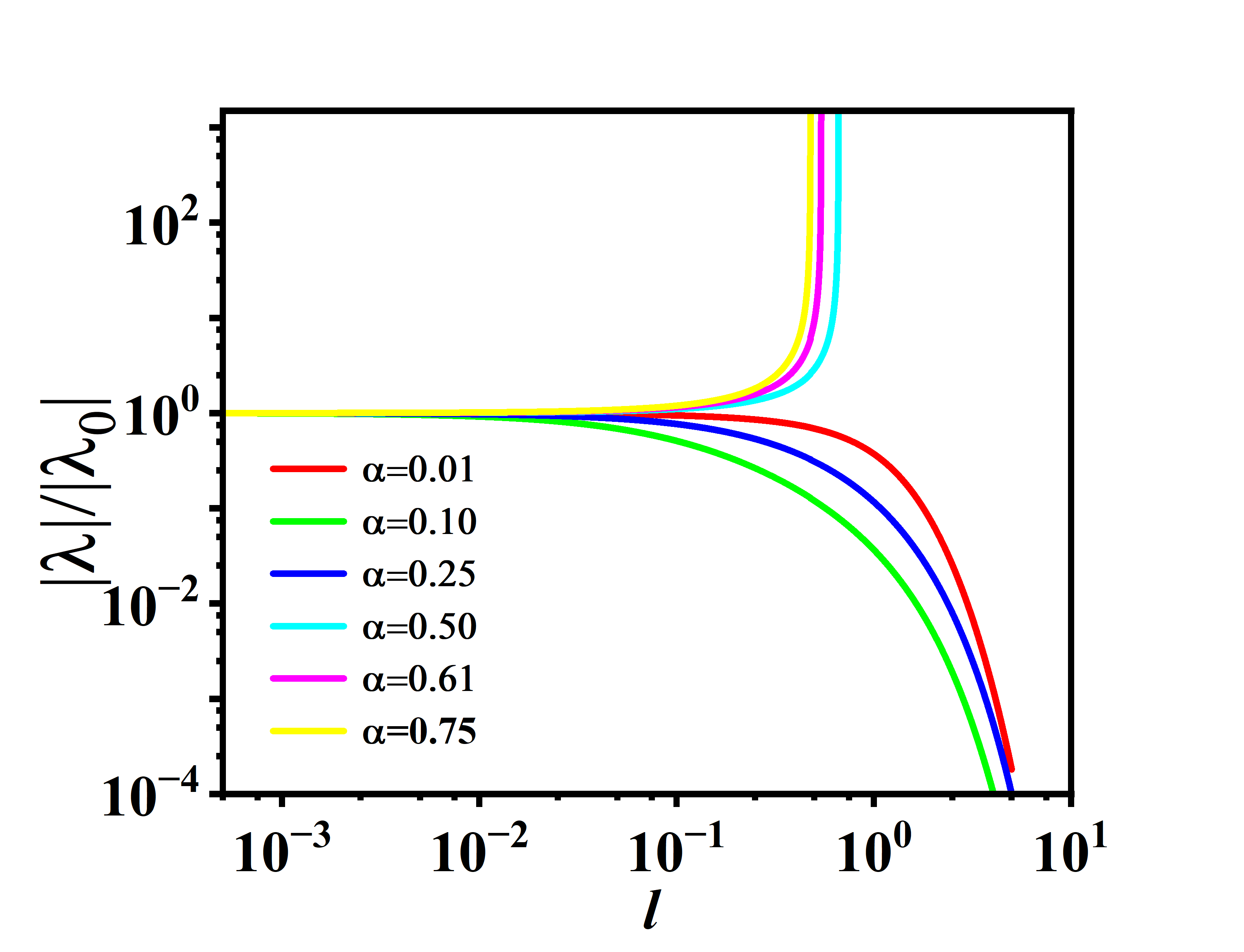}
}
\caption{(Color online) Energy-dependent evolutions of the relative interaction strength $|\lambda|/|\lambda_0|$
with variations of the parameter $\alpha$ at specific points in the $(Q, \phi)$ parameter space from Fig.~\ref{fig_Q_phi}:
(a) $Q=0.01, \phi=\pi/4$, (b) $Q=0.01, \phi=\pi$,
(c) $Q=0.01, \phi=7\pi/4$, and (d) $Q=0.005, \phi=\pi$.}
\label{fig_lambda}
\end{figure}

To move forward, let us combine the RG behavior of $\lambda$~(\ref{eq:RG_theory_clean})
and the constraints on critical strength $\lambda_c$ from previous subsections to examine the
energy-dependent behavior of $\lambda$. Fig.~\ref{fig_lambda} illustrates the energy-dependent
behavior of $\lambda$ under different conditions associated with Fig.~\ref{fig_Q_phi}.
We find that the RG analysis of $\lambda$ is well consistent with the study of critical Cooper
strength in Sec.~\ref{Subsec_lambda_c}.


\begin{figure*}[htbp]
\centering
\subfigure[]{
\includegraphics[width=2.45in]{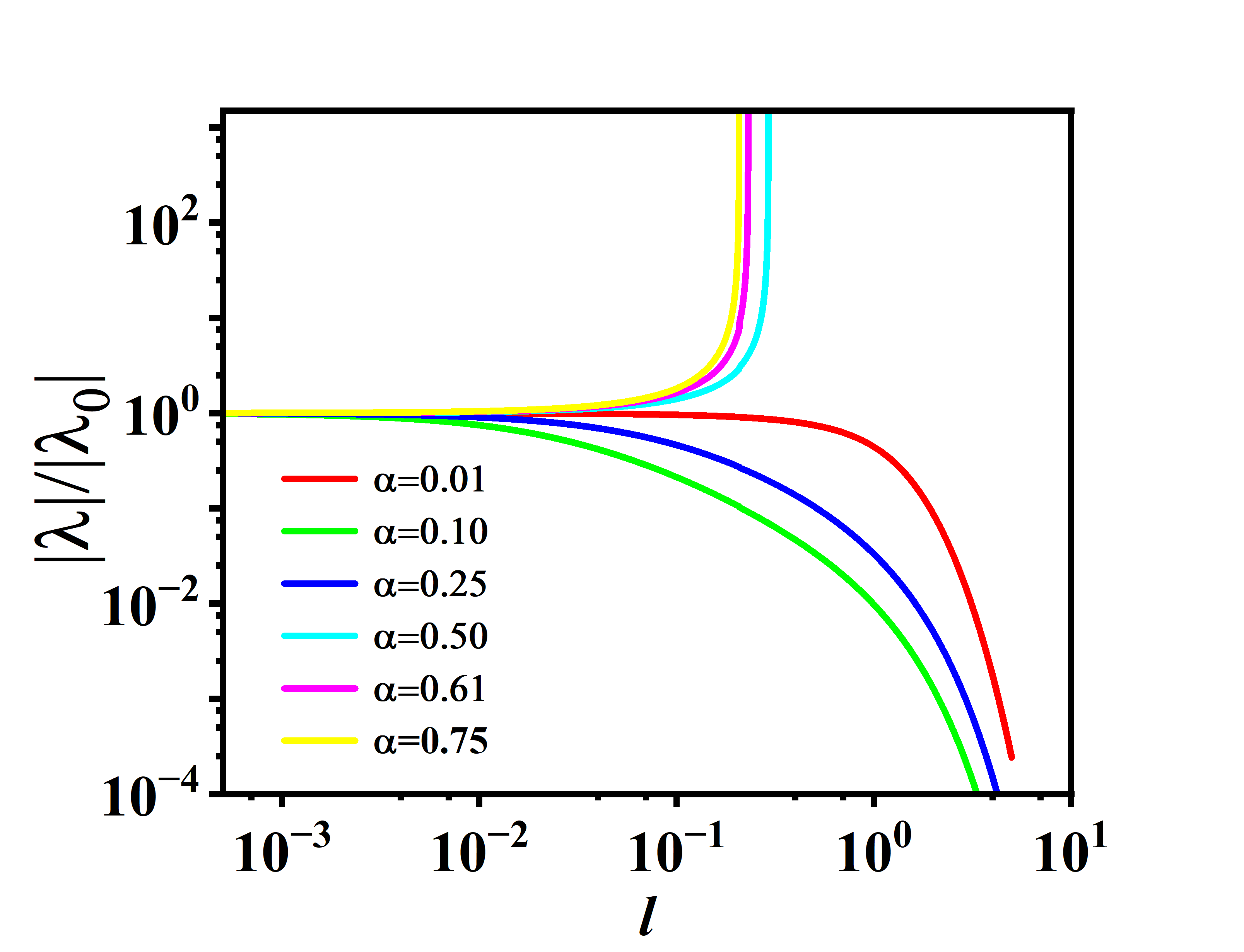}
}\hspace{-1.0cm}
\subfigure[]{
\includegraphics[width=2.45in]{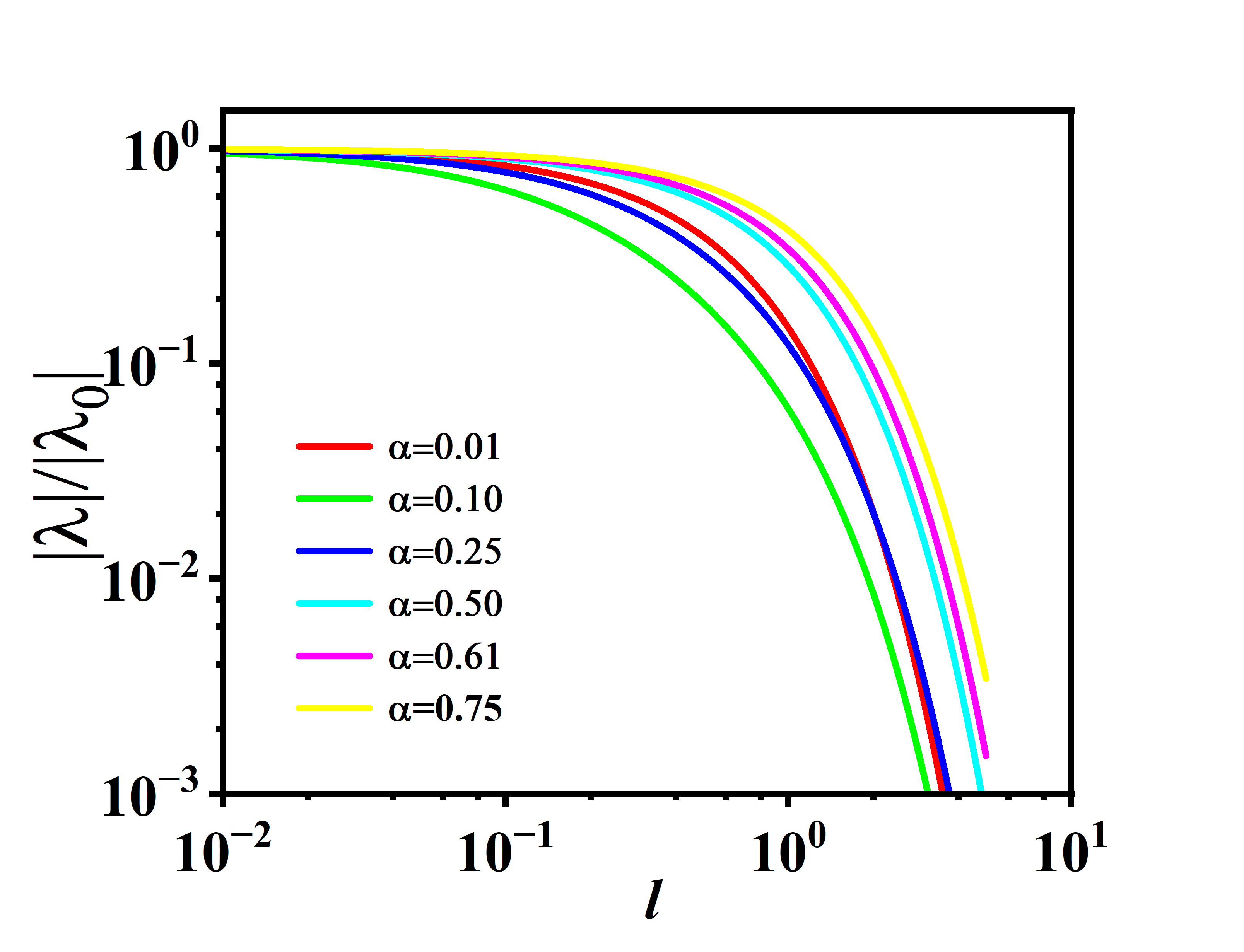}
}\hspace{-1.0cm}
\subfigure[]{
\includegraphics[width=2.45in]{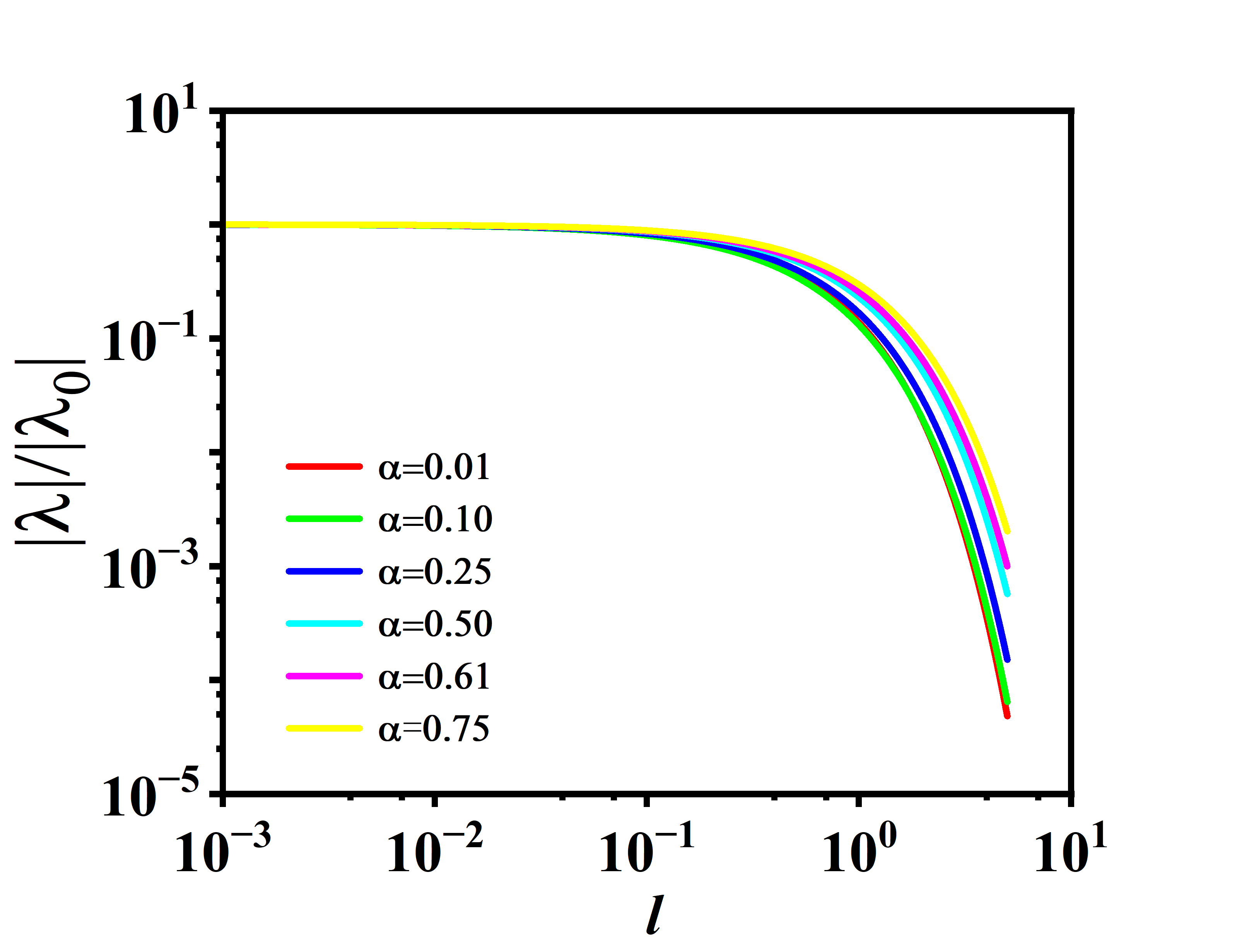}
}
\\
\hspace{-0.40cm}
\subfigure[]{
\includegraphics[width=2.45in]{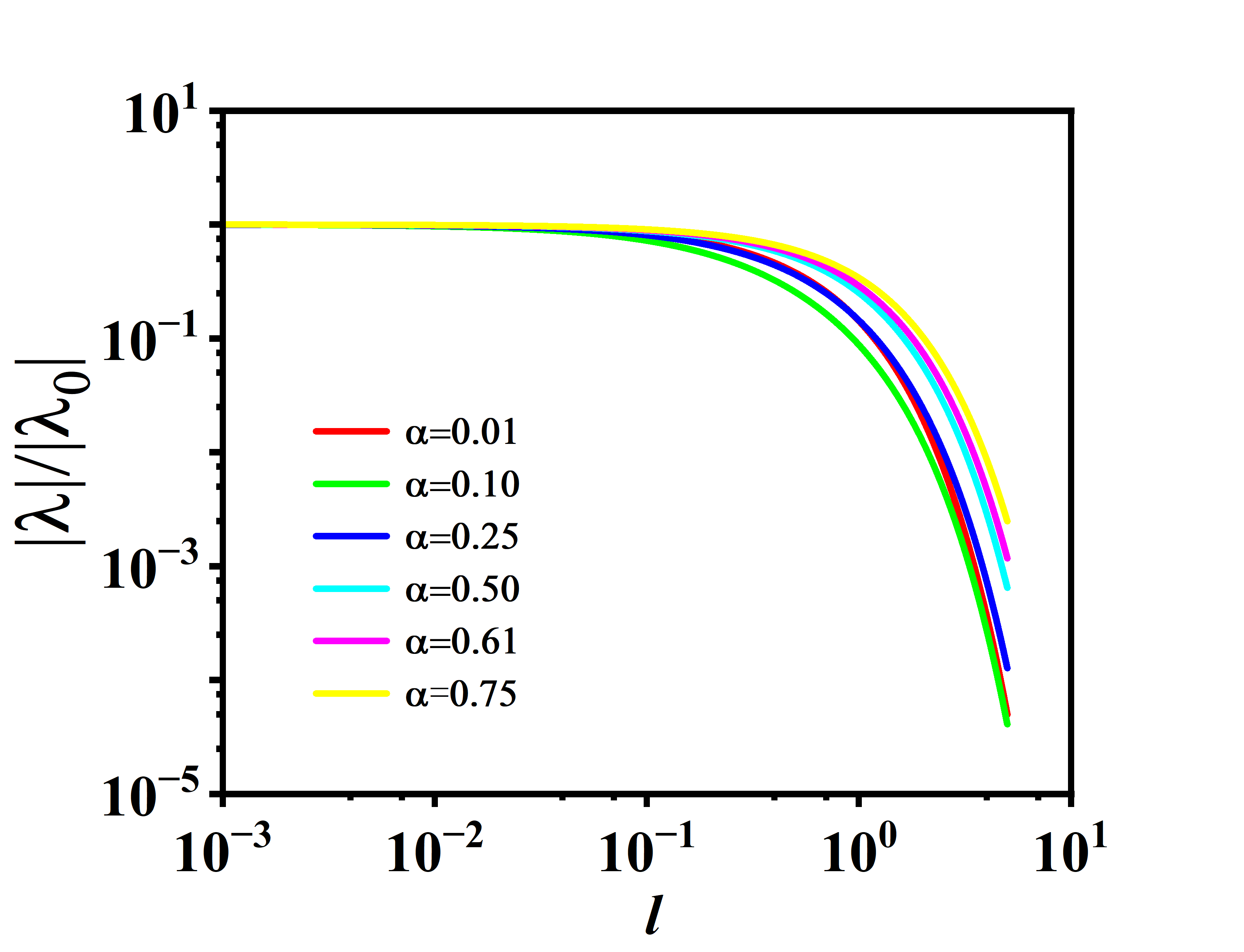}
}\hspace{-1.0cm}
\subfigure[]{
\includegraphics[width=2.45in]{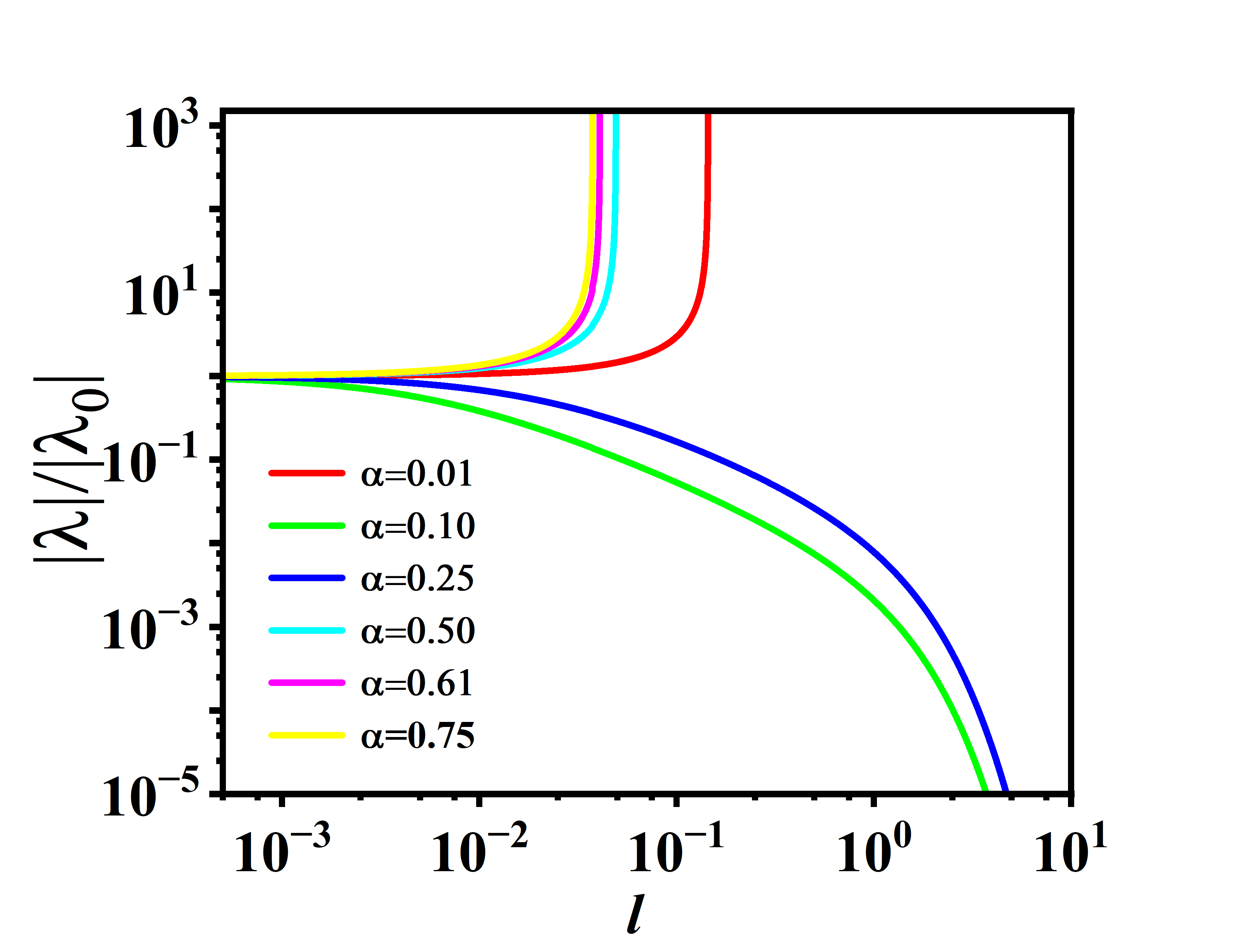}
}\hspace{-0.8cm}
\subfigure[]{
\includegraphics[width=2.45in]{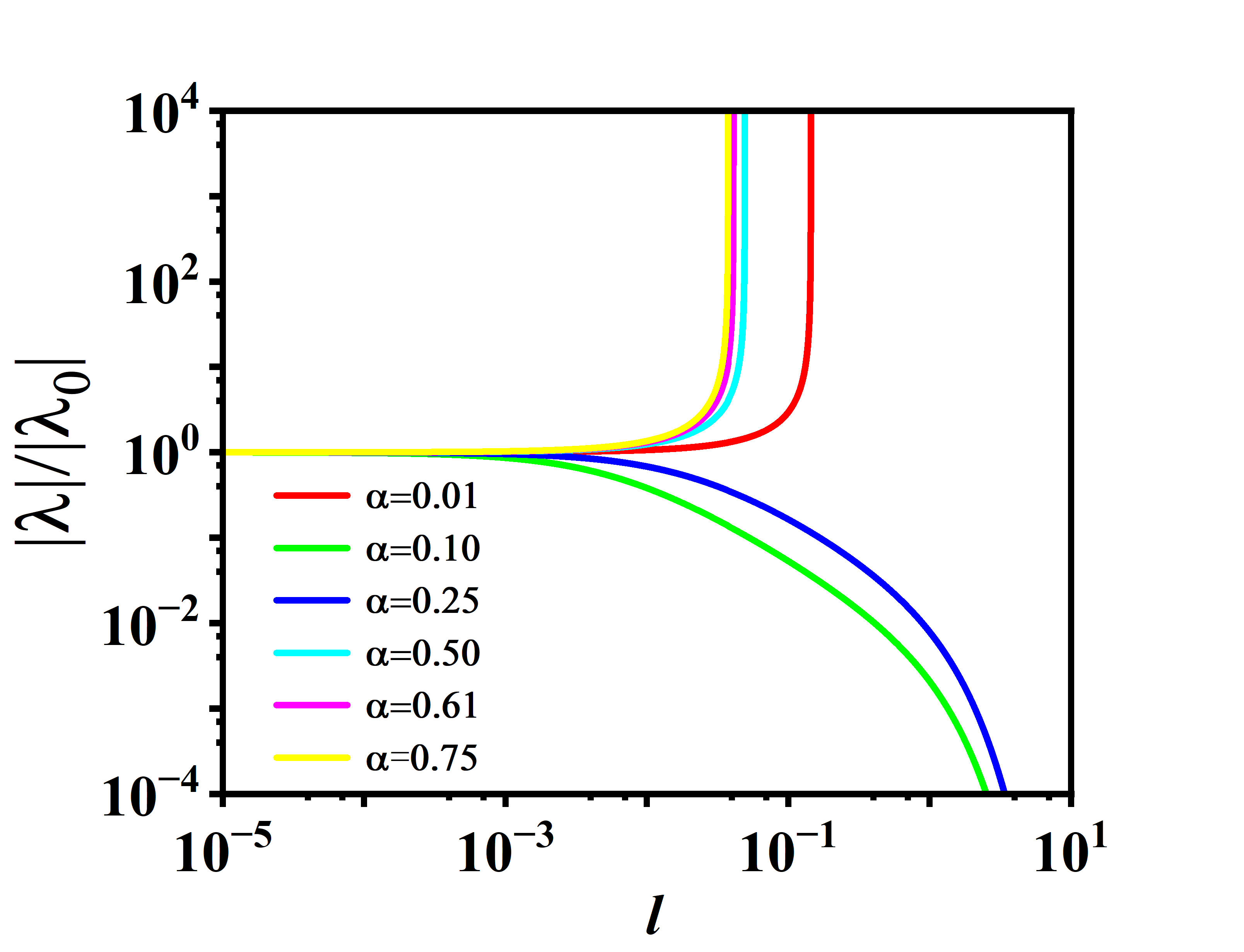}
}
\\
\vspace{-0.2cm}
\caption{(Color online) Energy-dependent evolution of the relative interaction strength
$|\lambda|/|\lambda_0|$ at $(Q=0.01, \phi=\pi)$ (from Fig.~\ref{fig_Q_phi}), varying the initial
values: Cooper interaction $|\lambda_0|$ in (a) $10^{-1}$, (b) $10^{-2}$, (c) $10^{-3}$, as well as the
parameter $v_\alpha$ in (d) $10^{-2}$, (e) $10^{-4}$, and (f) $10^{-5}$.}
\label{fig_lambda_1}
\end{figure*}

At first, paralleling Fig.~\ref{fig_Q_phi}, let us focus on the influence of fractional dispersion index.
By selecting $(Q=0.01, \phi=\pi/4)$, where $|\lambda_c|$ is relatively small, and varying the dispersion index $\alpha$,
we notice from Fig.~\ref{fig_lambda}(a) that the $\lambda$ always goes towards certain instability as long as
$|\lambda_0|$ exceeds the critical value depicted in Fig.~\ref{fig_Q_phi}(a). Besides, Fig.~\ref{fig_lambda}(c)
at $(Q=0.01, \phi=7\pi/4)$ exhibits analogous divergent behavior to Fig.~\ref{fig_lambda}(a).
As a consequence, the evolution of $\lambda$ is insensitive to the value of $\alpha$ in the situation with small $Q$ and $\phi$,
and a smaller $\phi$ is more helpful to the emergence of the Cooper instability.
In comparison, when $Q$ and $\phi$ are properly chosen, the parameter $\alpha$ can impose a significant influence.
As depicted in Fig.~\ref{fig_lambda}(b) with $(Q,\phi)=(0.01,\pi)$ and Fig.~\ref{fig_lambda}(d) with
$(Q,\phi)=(5\times10^{-3},\pi)$, we consider the intermediate region in Fig.~\ref{fig_Q_phi} while maintaining the same
value of the initial Cooper strength $\lambda_0$. From Fig.~\ref{fig_lambda}(b) and Fig.~\ref{fig_lambda}(d),
it is evident that $\lambda$ diverges and Cooper instability emerges for $\alpha\geq 0.50$, whereas for
smaller $\alpha\leq 0.25$, $\lambda$ gradually diminishes and Cooper instability does not occur.
This accordingly exists a critical threshold of fractional exponent for Cooper instability. All these results are
well in agreement with the prior analysis of critical Cooper interaction determined by parameters
$(\alpha, Q, \phi)$ in Fig.~\ref{fig_Q_phi}.

Then, we fix the parameters $(Q, \phi)$ and turn to investigate how the initial values of $\lambda_0$ and $v_\alpha$
affect the Cooper instability, which are demonstrated in Fig.~\ref{fig_lambda_1}. Figs.~\ref{fig_lambda_1}(a)-(c) show
that with $v_\alpha$ fixed, increasing $|\lambda_0|$ is helpful to trigger the Cooper instability. In sharp contrast,
with fixing $\lambda_0$, Fig.~\ref{fig_lambda_1}(d)-(f) present that a smaller $v_\alpha$ enhances the possibility for
the divergence of Cooper interaction. Therefore, this confirms that strong initial attractions and suppressed
fermion kinetics are in cooperatively favour of Cooper instability~\cite{Shankar1994RMP,Sondhi2013PRB,Sondhi2014PRB}.

\begin{table}[h]
    \centering
    \caption{Basic results of  Scenario-A in the clean limit (analogous to Scenario-B under $Q\rightarrow Q'$).
CI denotes Cooper instability, and NO CI indicates that Cooper instability cannot occur.}
    \renewcommand{\arraystretch}{1.5}
    \vspace{+0.2cm}
    \begin{tabular}{c|c|c}
        \hline\hline
         Range of $\alpha$ & Region of ($Q$, $\phi$) & Behavior \\
        \hline
        $0<\alpha<\alpha_{c1}$ & The entire region & CI triggered at $|\lambda_0|>|\lambda_c|$ \\
        \hline
        \multirow{2}{*}{$\alpha_{c1}<\alpha<\alpha_{c2}$} & $(Q,\phi) \in$ \text{Zone-}\uppercase\expandafter{\romannumeral1} & NO CI \\
        \cline{2-3}
                          & $(Q,\phi) \in$ \text{Zone-}\uppercase\expandafter{\romannumeral2} & CI triggered at $|\lambda_0|>|\lambda_c|$ \\
        \hline
        $\alpha_{c2} \le \alpha< 1$ & The entire region & CI triggered at $|\lambda_0|>|\lambda_c|$ \\
        \hline\hline
    \end{tabular}
    \label{conclusion_clean}
\end{table}

\subsection{Results for Scenario-B and clean-limit conclusions}

To proceed, let us provide brief comments on the Scenario-B.
In this circumstance, the RG equation of Cooper interaction~(\ref{eq:RG_theory_clean}) is recast into
\begin{eqnarray}
\frac{d\lambda}{dl}
 & =  &\lambda\left[(\alpha-2)-\frac{\lambda}{4\pi^2v_\alpha }\left (  \mathcal F_0' + v_\alpha\mathcal F_1'    \right) \right ],
\end{eqnarray}
where $\mathcal F_0'$ and $\mathcal F_1'$ are designated in Eqs.~(\ref{Eq_F_0_prime})-(\ref{Eq_F_1_prime}). Compared with their Scenario-A
counterparts $\mathcal F_0$ and $\mathcal F_1$ in Eq.~(\ref{eq:RG_theory_clean}), it is worth emphasizing that
they possess the same functional dependencies but only replace $Q$ with $Q'$. As a consequence,
Scenario-B shares the analogous critical behavior of Cooper interaction with
Scenario-A presented in Sec.~\ref{Subsec_lambda_c}.

In this sense, we are now in a suitable position to address the primary results at the clean limit.
Both analytical and numerical studies indicate that there exists a critical strength of interaction $\lambda$
for Cooper instability in the 2D FDSMs. For either Scenario-A or Scenario-B, such a critical interaction
denoted by $\lambda_c$ is closely dependent upon the fractional dispersion exponent $\alpha$ and
the transfer momentum magnitude $Q$ as well as its angular orientation
$\phi$, while the fermionic velocity $v_\alpha$ provides a quantitative effect.
Table~\ref{conclusion_clean} summarizes our primary results.
Specifically, the $(Q,\phi)$ parameter space is divided into two distinct regions as displayed in Fig.~\ref{fig_Q_phi_1} and Fig.~\ref{fig_Q_phi}. Zone-I with ($\mathcal F_0 + \mathcal F_1v_\alpha < 0$) prohibits the divergence of interaction, whereas Zone-II allows for Cooper instability when $|\lambda_0| > |\lambda_c|$. In particular, $\alpha$ governs the areas of Zone-I and Zone-II via critical thresholds $\alpha_{c1} \approx 10^{-3}$ and $\alpha_{c2} \approx 0.61$£¬ and its increase is helpful to decrease the $|\lambda_c|$. Additionally, with increasing $Q$, the $|\lambda_c|$ can either increase or decrease depending on the
direction of momentum denoted by $\phi$ as schematically illustrated in Fig.~\ref{fig_lambda}.

\section{Impact of disorder scatterings on Cooper instability--doing}\label{Sec_dis_effects}

In the presence of disorder scatterings, the Cooper interaction strength and disorder parameters as well as
fermion velocity become mutually coupled through the coupled RG flows~(\ref{Eq_RG_v_alpha_Q})-(\ref{Eq_RG_v_Delta3_Q_prime}).
Crucially, disorder scatterings are possible to induce significant deviations from
clean-limit behavior in kinds of semimetals~\cite{Ramakrishnan1985RMP,Lerner0307471,Nersesyan1995NPB,Stauber2005PRB,Wang2011PRB, Mirlin2008RMP,Coleman2015Book,Roy2018PRX,LWZ2019JPC} . Accordingly, it is of particular importance
to investigate how disorder scattering modifies the Cooper instability criteria. Given the analogous critical
behavior exhibited in Scenario-A and Scenario-B, we focus exclusively on Scenario A for this analysis,
with its RG equations formally expressed in Eqs.~(\ref{Eq_RG_v_alpha_Q})-(\ref{Eq_RG_Delta3_alpha_Q}).

\begin{figure}[h]
\centering
\includegraphics[width=3.5in]{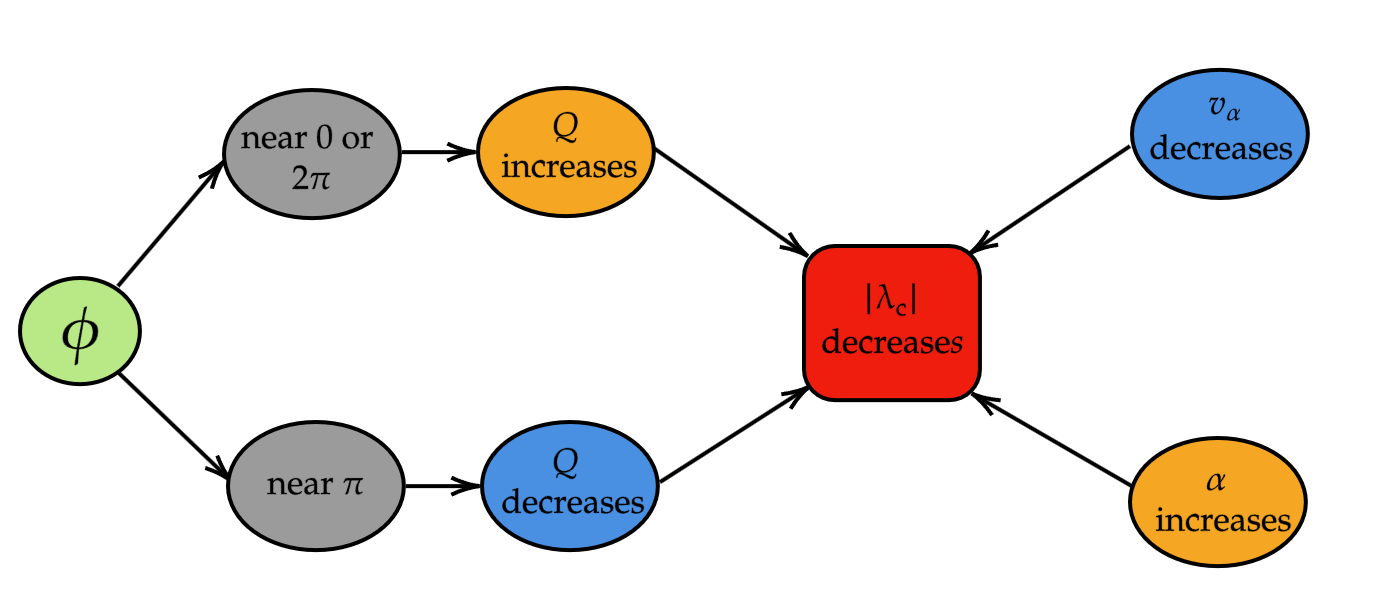}
\\
\vspace{-0.25cm}
\caption{(Color online) Schematic illustration for the influence of parameters $Q$, $\phi$, $\alpha$, and $v_\alpha$
on critical strength of Cooper interaction $|\lambda_c|$ in Zone-\uppercase\expandafter{\romannumeral2}.}
\label{fig_schematic_lambda_c}
\end{figure}


\begin{figure*}[htbp]
\subfigure[]{
\includegraphics[width=2.35in]{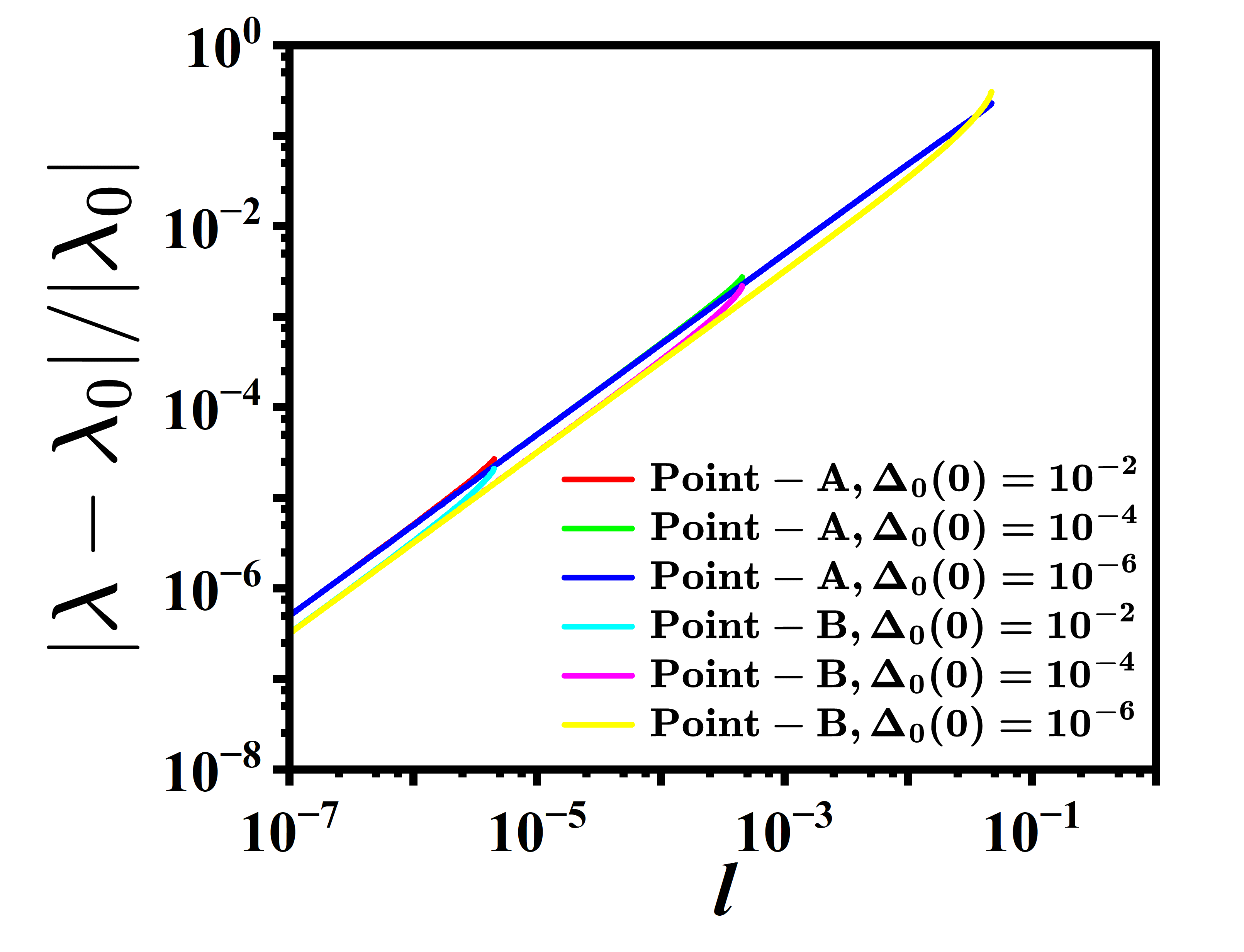}\hspace{-0.6cm}
}
\subfigure[]{
\includegraphics[width=2.35in]{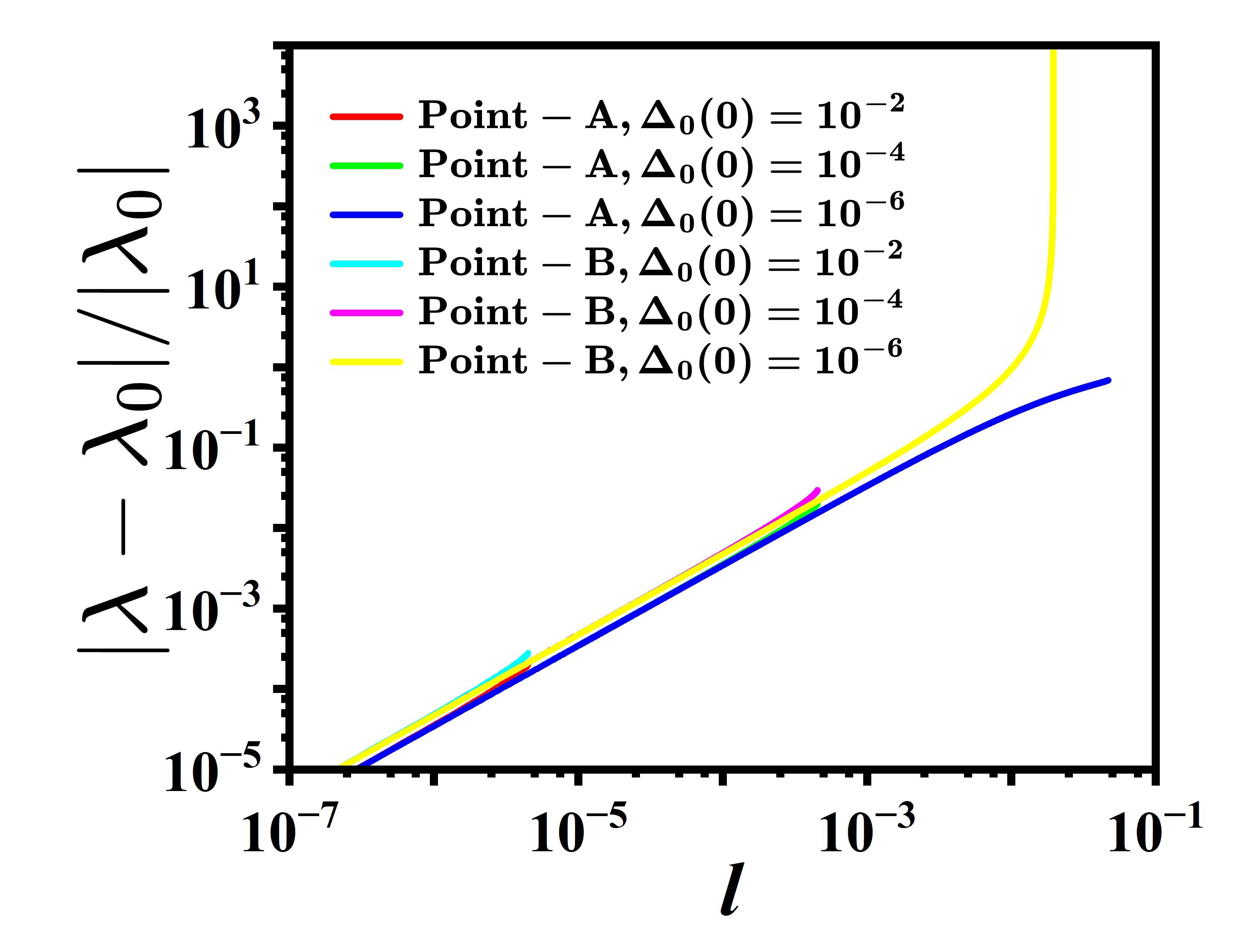}\hspace{-0.6cm}
}
\subfigure[]{
\includegraphics[width=2.35in]{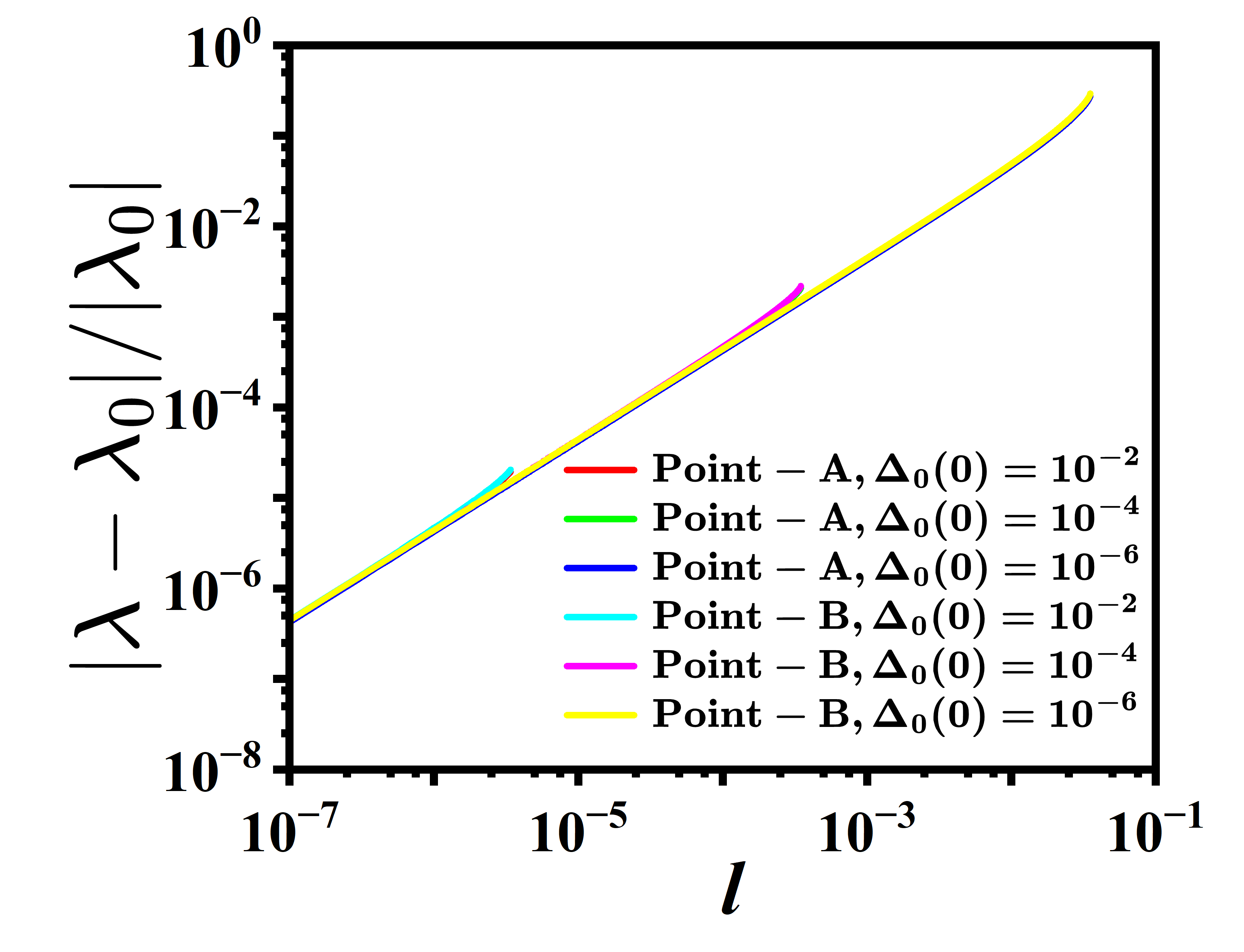}
}
\\
\subfigure[]{
\includegraphics[width=2.35in]{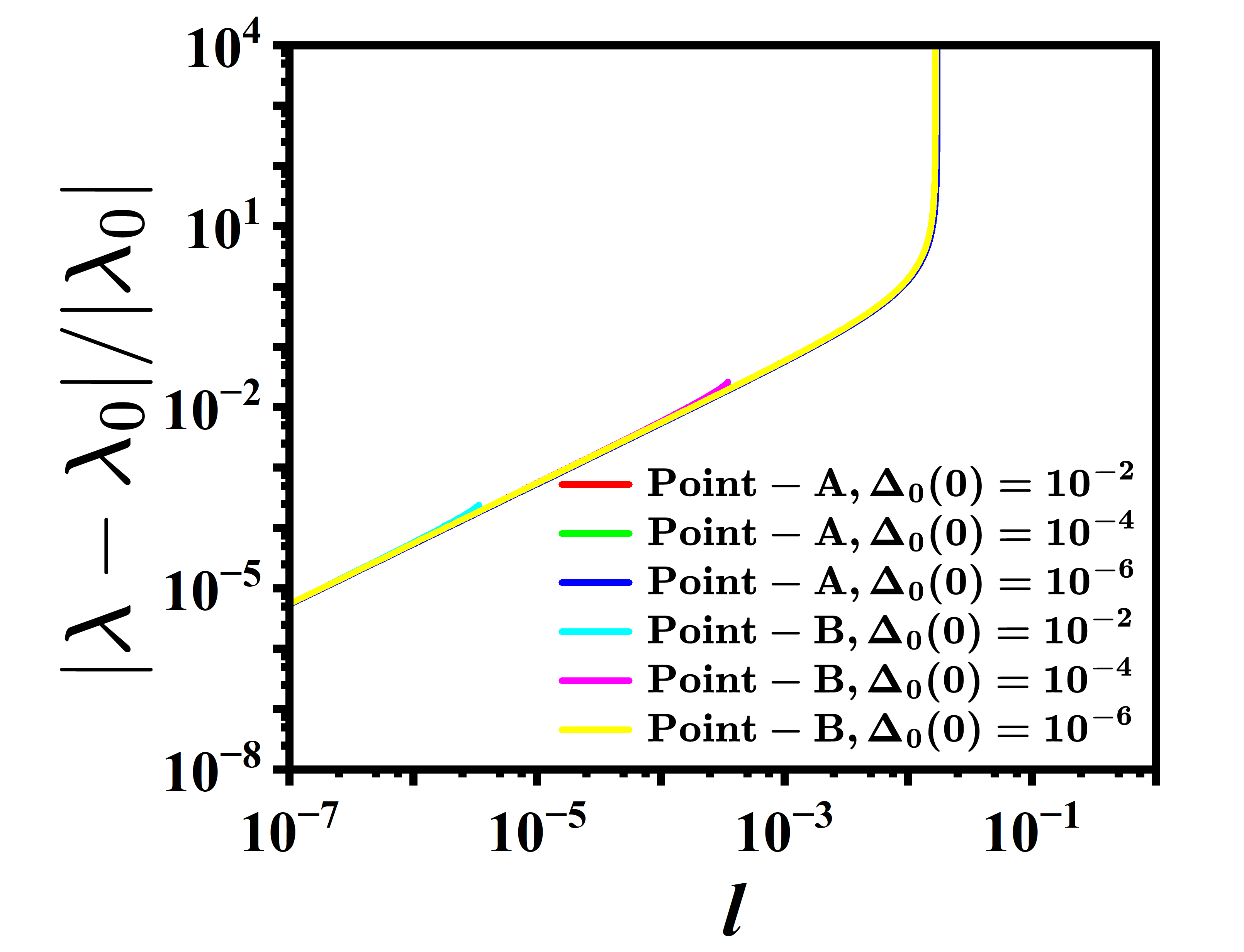}\hspace{-0.6cm}
}
\subfigure[]{
\includegraphics[width=2.35in]{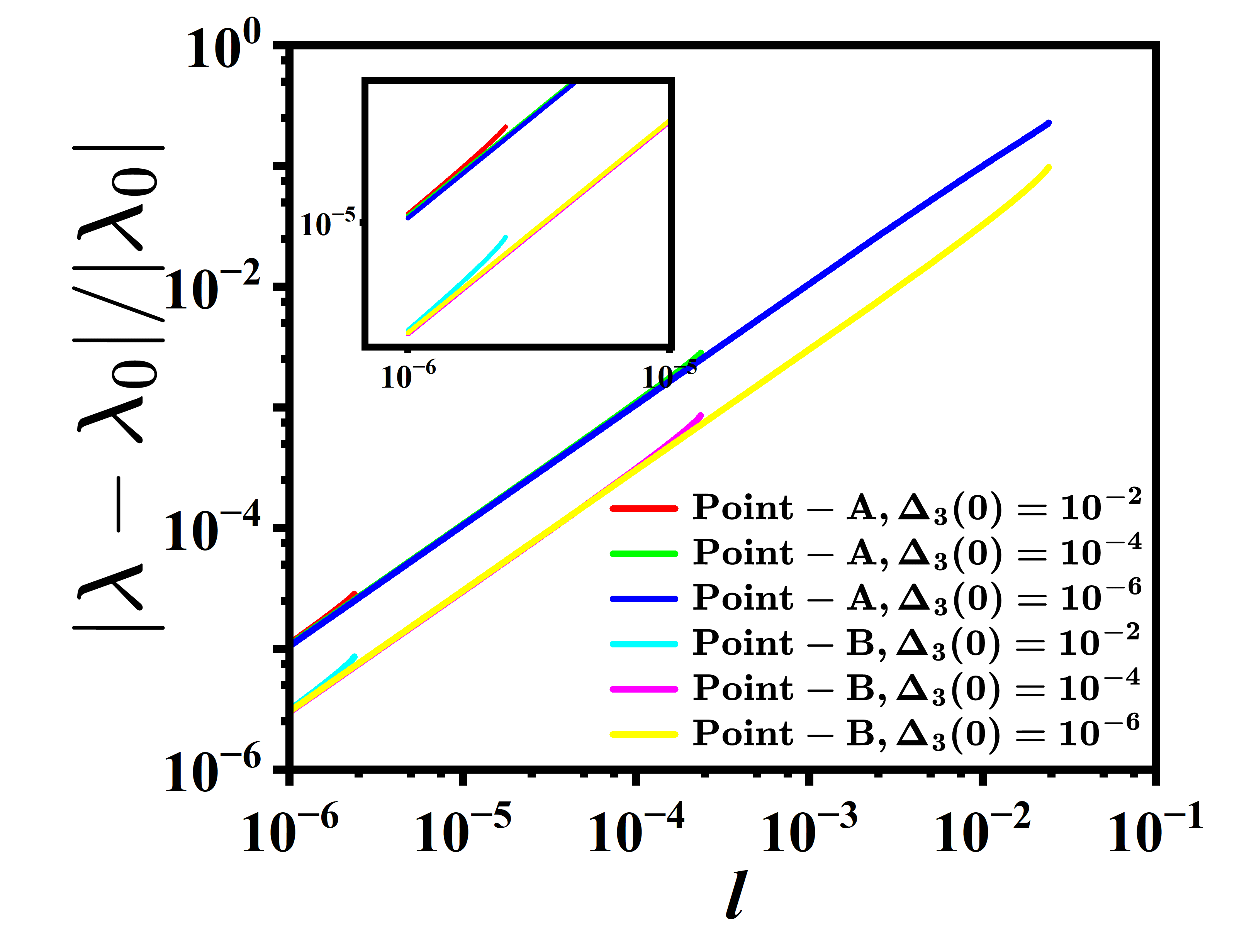}\hspace{-0.6cm}
}
\subfigure[]{
\includegraphics[width=2.35in]{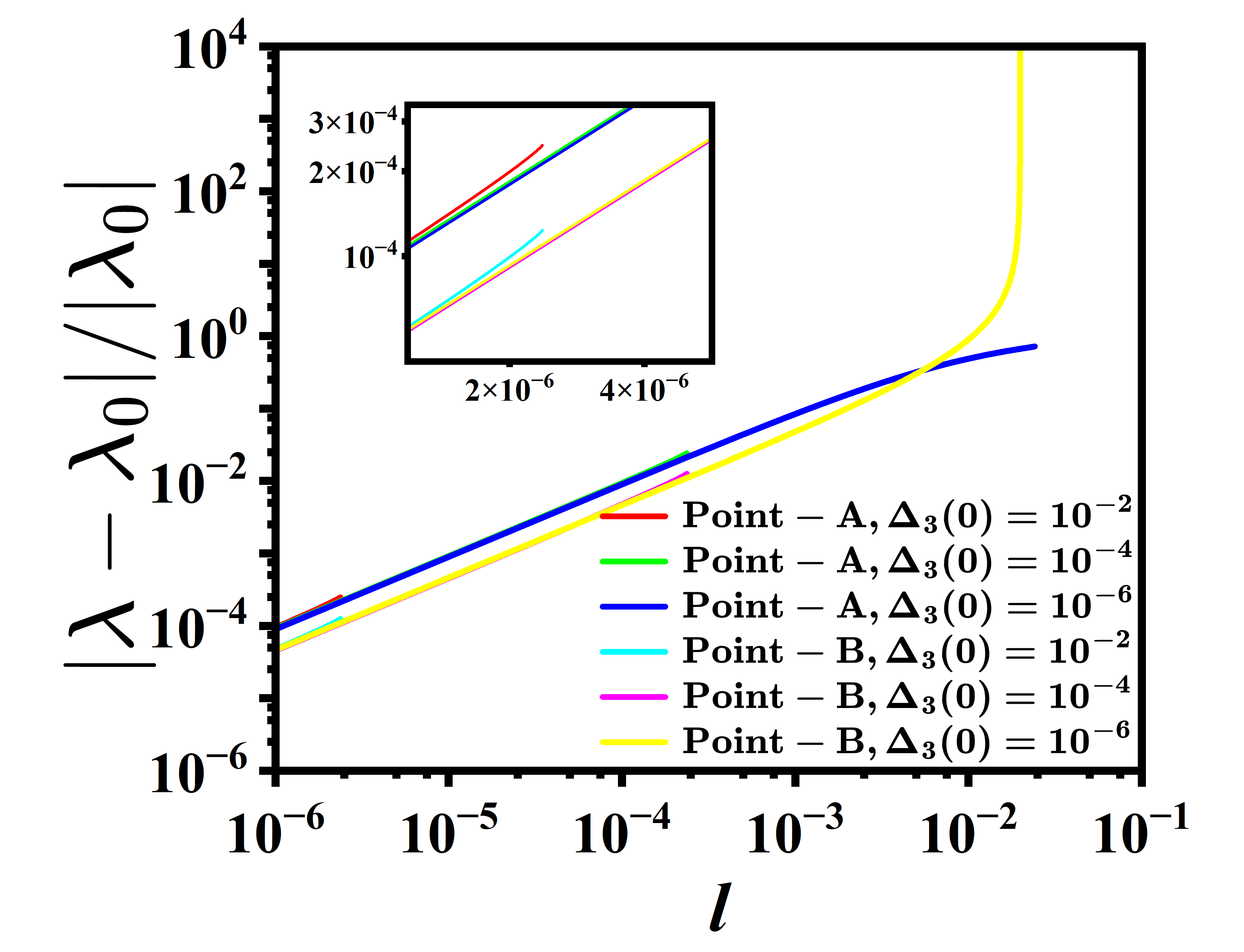}
}
\\
\subfigure[]{
\includegraphics[width=2.35in]{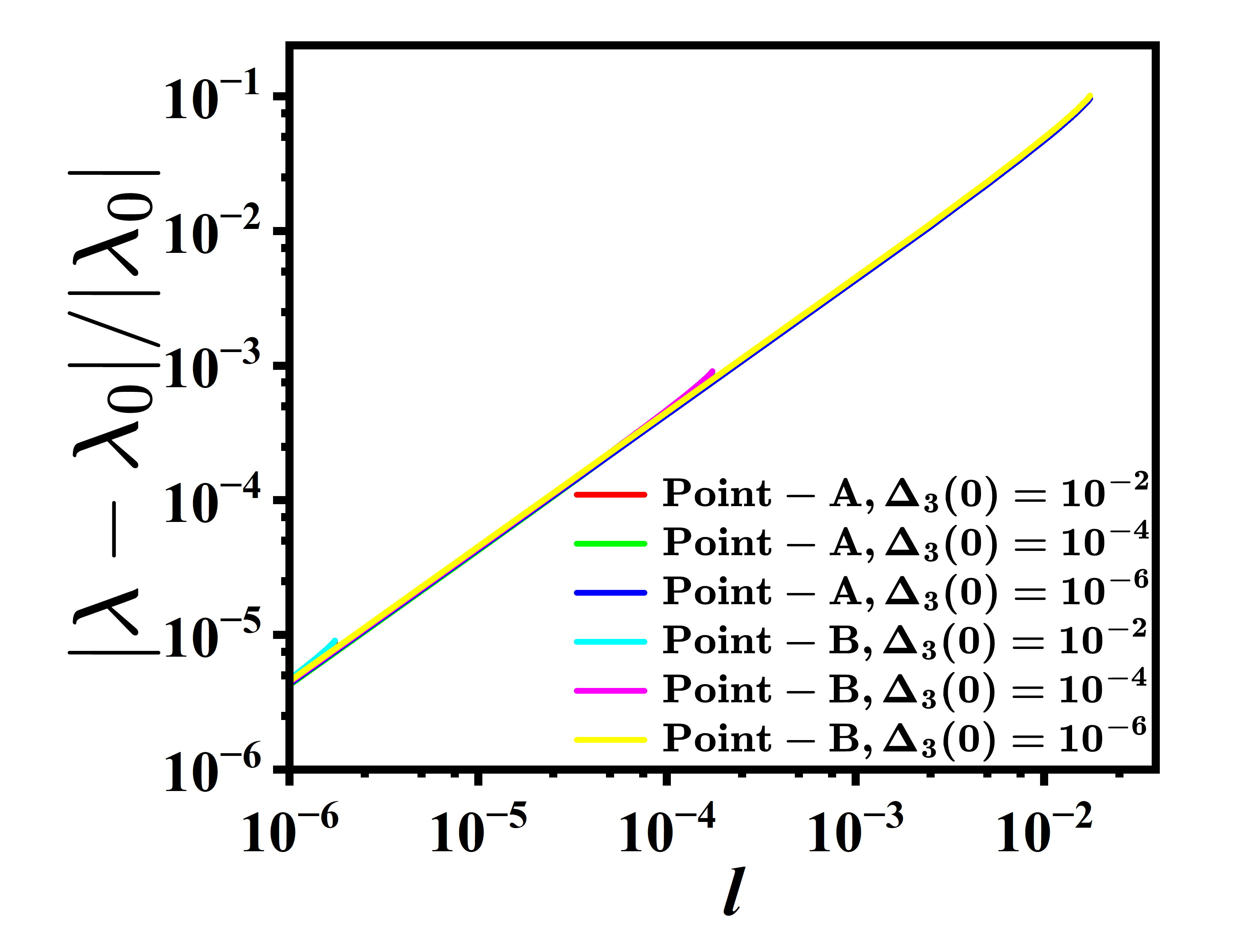}\hspace{0.6cm}
}
\subfigure[]{
\includegraphics[width=2.35in]{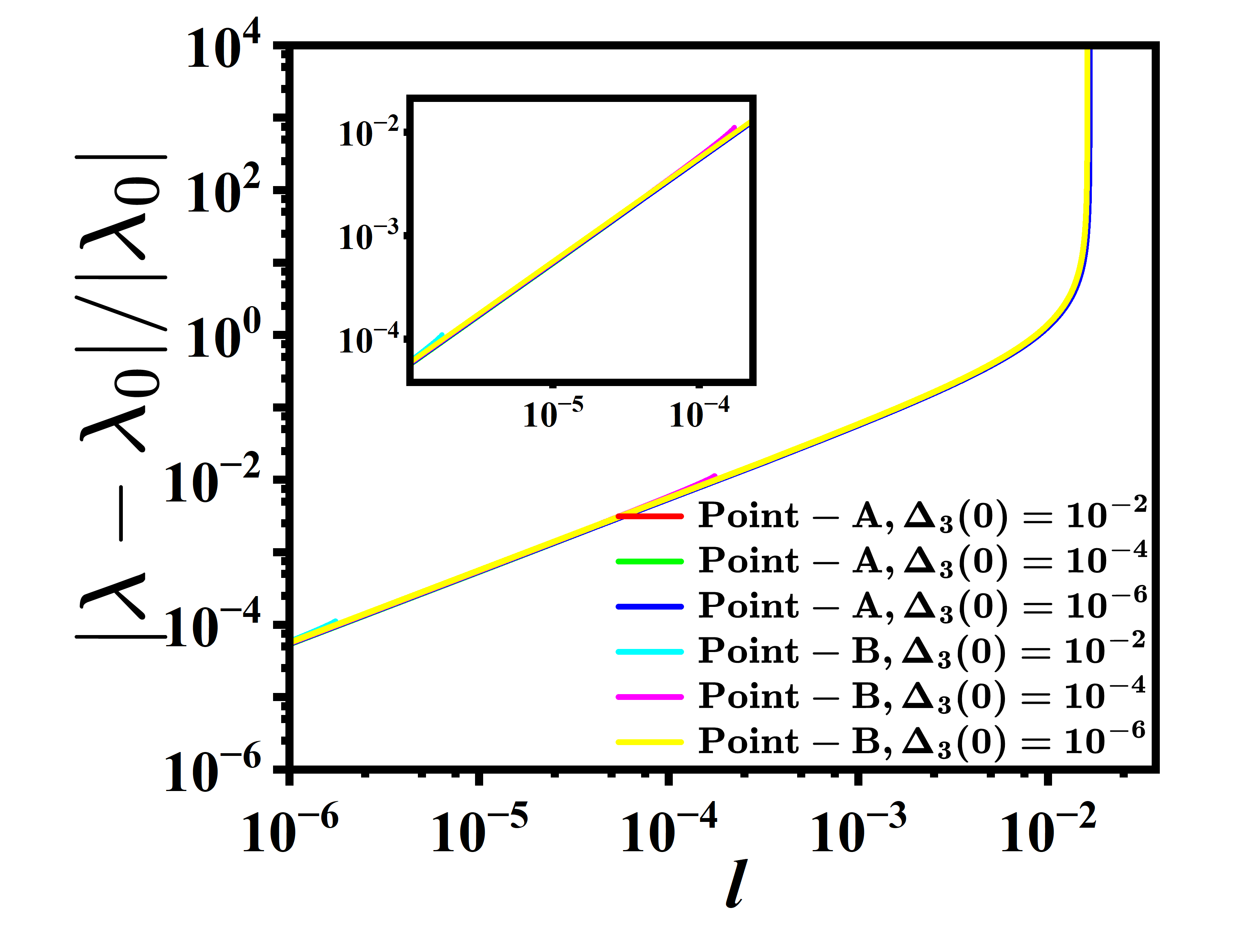}
}
\\
\vspace{-0.35cm}
\caption{(Color online)
Energy-dependent evolutions of the relative interaction strength $\left |\frac{\lambda-\lambda_0}{\lambda_0} \right |$
at Point-A and Point-B for (a)-(b) $\alpha=0.25$ and (c)-(d) $\alpha=0.75$ in the sole presence of disorder $\Delta_0$,
as well as (e)-(f) $\alpha=0.25$ and (g)-(h) $\alpha=0.75$ in the sole presence of disorder $\Delta_3$, respectively.}
\label{fig_delta_0}
\end{figure*}

\subsection{Single type of disorder}\label{Sec_single_disorder}

At first, we consider the presence of sole type of disorder scattering.
In the clean limit, the fractional exponent $\alpha$ plays a critical role in inducing the Cooper instability by
partitioning the $(Q,\phi)$ parameter space into two distinct regions including Zone-\uppercase\expandafter{\romannumeral1} and Zone-\uppercase\expandafter{\romannumeral2} as depicted in Fig.~\ref{fig_Q_phi}. Based on the critical thresholds in Table~\ref{conclusion_clean}, we select two representative values $\alpha = 0.25$ ($\alpha_{c1} < \alpha < \alpha_{c2}$)
and $\alpha = 0.75$ ($\alpha_{c2} \leq \alpha < 1$) to systematically investigate the disorder scattering effects.

Specifically, by analyzing Fig.~\ref{fig_Q_phi}(e) with $\alpha=0.5$ and Fig.~\ref{fig_Q_phi}(h) with $\alpha=0.75$
and taking into account the symmetry of coordinates, we fix the angular orientation at $\phi = \pi$ and choose two distinct momentum
magnitudes $Q = 10^{-2}$ and $Q = 10^{-4}$, which designate two representative points: Point-A $(Q,\phi)=(10^{-2},\pi)$
and Point-B $(Q,\phi)=(10^{-4},\pi)$. At clean limit, Fig.~\ref{fig_Q_phi}(e) shows that Point-A resides in Zone-I
where the condition ($\mathcal F_0 + \mathcal F_1v_\alpha < 0$) forbids the divergence of interaction, while
Point-B lies in Zone-II where Cooper instability emerges when $|\lambda_0|$ exceeds the critical
threshold $|\lambda_c(Q,\phi)|$. In comparison, Fig.~\ref{fig_Q_phi}(h) displays that
both points now occupy Zone-II, enabling the Cooper instability above distinct $|\lambda_c|$ values.

Subsequently, we bring out the controlled disorder perturbations at these four
selected situations Point-A and Point-B at $\alpha=0.5$, as well as Point-A and Point-B
at $\alpha=0.75$ to carefully study how disorder scattering influences the emergence of Cooper instability.

We begin by considering the single presence of random chemical potential denoted by $\Delta_0$.
For $\alpha=0.25$ ($\alpha_{c1} < \alpha < \alpha_{c2}$), Fig.~\ref{fig_delta_0}(a) and Fig.~\ref{fig_delta_0}(b) show
that the presence of $\Delta_0$ at Point-A is incapable of inducing the Cooper instability initially prohibited in the clean limit.
This signals that the properties of Zone-\uppercase\expandafter{\romannumeral1} are insensitive to $\Delta_0$.
In contrast, $\Delta_0$ is manifestly harmful to the Cooper instability at Point-B. As shown in
Fig.~\ref{fig_delta_0}(a), $\Delta_0$ suppresses interaction divergence even with large initial $|\lambda_0|$ values
as long as the initial $\Delta_0$ is adequate. However, $\Delta_0$ fails to prevent the occurrence of Cooper instability if
$\Delta_0$ is very small and $|\lambda_0|$ is sufficiently large such as $\Delta_0=10^{-6}$ and $|\lambda_0|=10^{-1}$ depicted
in Fig.~\ref{fig_delta_0}(b). This implies that a non-zero $\Delta_0$ increases the critical $|\lambda_c|$ required
for Cooper instability in Zone-\uppercase\expandafter{\romannumeral2}. As to $\alpha=0.75$ ($\alpha_{c2} \leq \alpha < 1$), Fig.~\ref{fig_delta_0}(c) demonstrates that $\Delta_0$ hinders the Cooper instability at $|\lambda_0|=10^{-2}$, which
would otherwise occur in the clean limit as shown in Fig.~\ref{fig_Q_phi}(h). However, Fig.~\ref{fig_delta_0}(d)
presents that increasing $|\lambda_0|$ to $10^{-1}$ allows Cooper instability to emerge at both points. This again
confirms that $\Delta_0$ hinders Cooper instability by increasing the critical interaction threshold and cannot
alter the prohibitive nature of Zone-\uppercase\expandafter{\romannumeral1}.

\begin{table}[H]
\centering
\caption{Influence of single presence of disorder on the critical value $|\lambda_c|$ for Cooper instability.
The symbols $\uparrow$ and $\downarrow$ signify the expansion and contraction of
Zone-\uppercase\expandafter{\romannumeral1} and Zone-\uppercase\expandafter{\romannumeral2} which
are designated in Sec.~\ref{Subsec_lambda_c}.}
    \renewcommand{\arraystretch}{2}
    \setlength{\tabcolsep}{6pt}
    \vspace{+0.15cm}
	\begin{tabular}{c|c|c|c|c}
		\hline
        \hline
		Sole disorder &       $\red{\Delta_0}$      &      ${\color{Green} \Delta_1} $      &       ${\color{Green}\Delta_2}$    &     $\red{\Delta_3}$       \\ \hline
		 Area of Zone-\uppercase\expandafter{\romannumeral1}&       $\red{ \uparrow}$      &      ${\color{Green}\downarrow}$       &      ${\color{Green}-}$      &      $\red{ \uparrow}$      \\ \hline
		 Area of Zone-\uppercase\expandafter{\romannumeral2}&       $\red{ \downarrow}$      &       ${\color{Green}\uparrow}$      &      ${\color{Green}-}$      &     $\red{ \downarrow}$       \\
        \hline
        \hline
	\end{tabular}
	\label{tab_disorder}
\end{table}


\begin{figure*}[htbp]
\subfigure[]{
\includegraphics[width=2.35in]{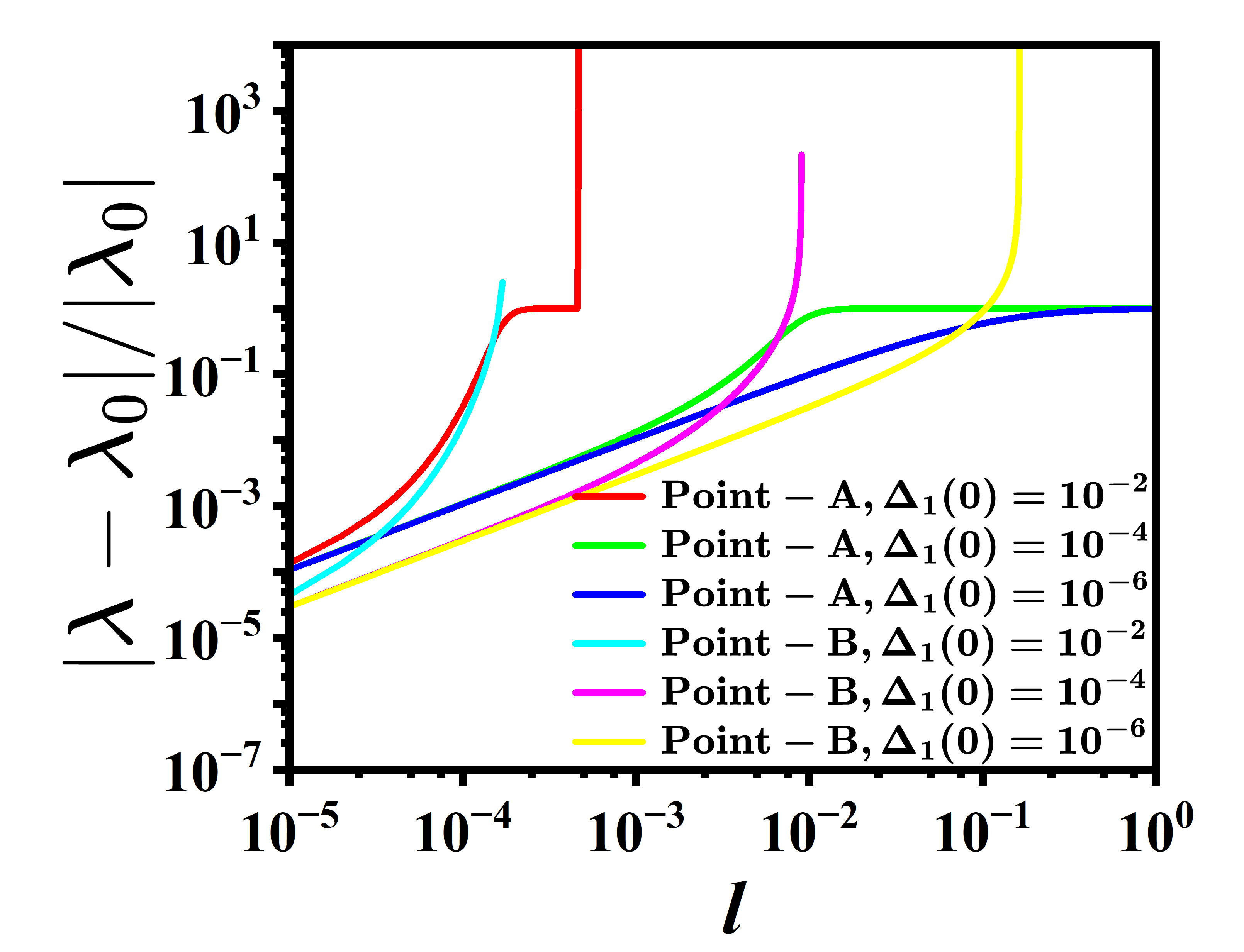}\hspace{-0.6cm}
}
\subfigure[]{
\includegraphics[width=2.35in]{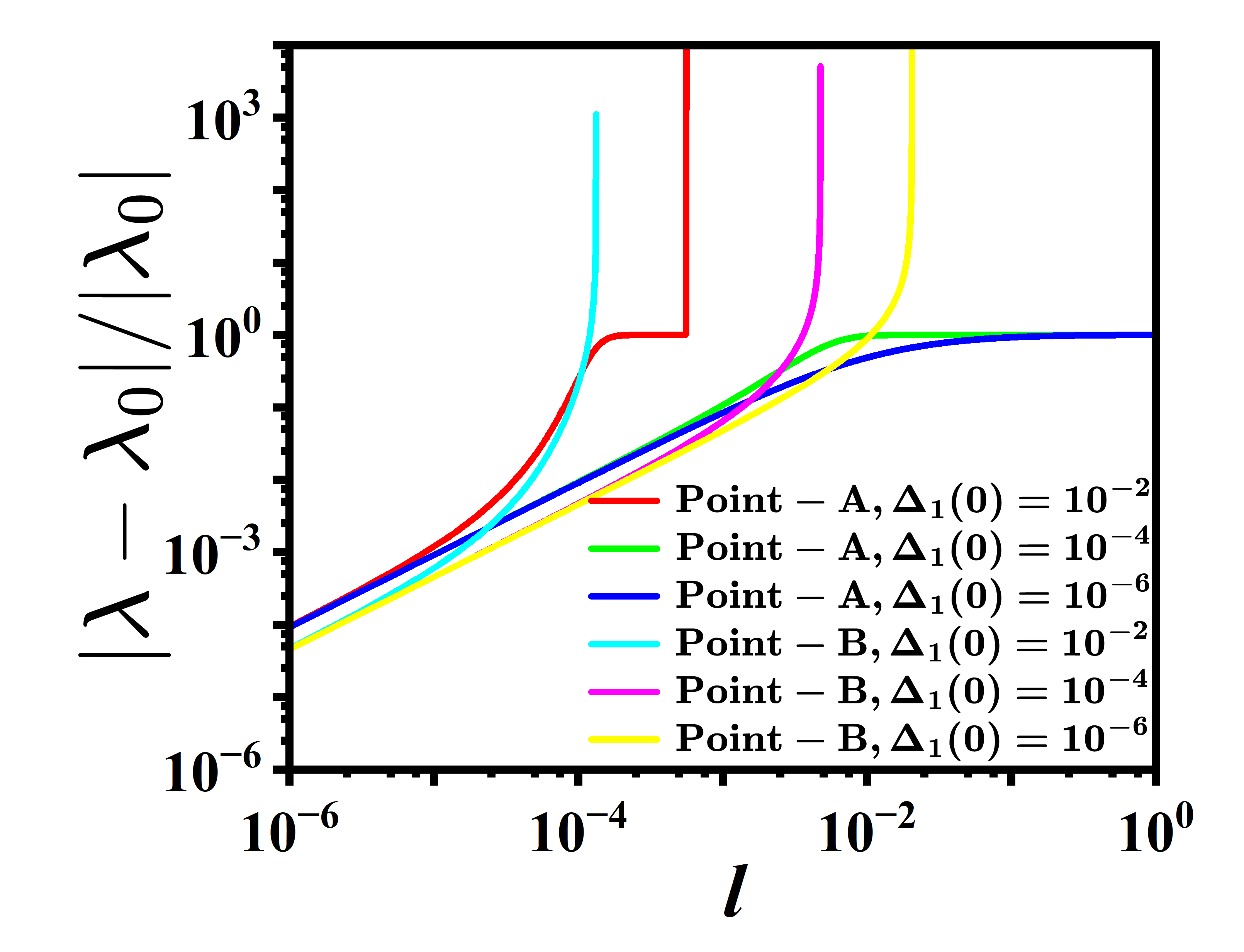}\hspace{-0.6cm}
}
\subfigure[]{
\includegraphics[width=2.35in]{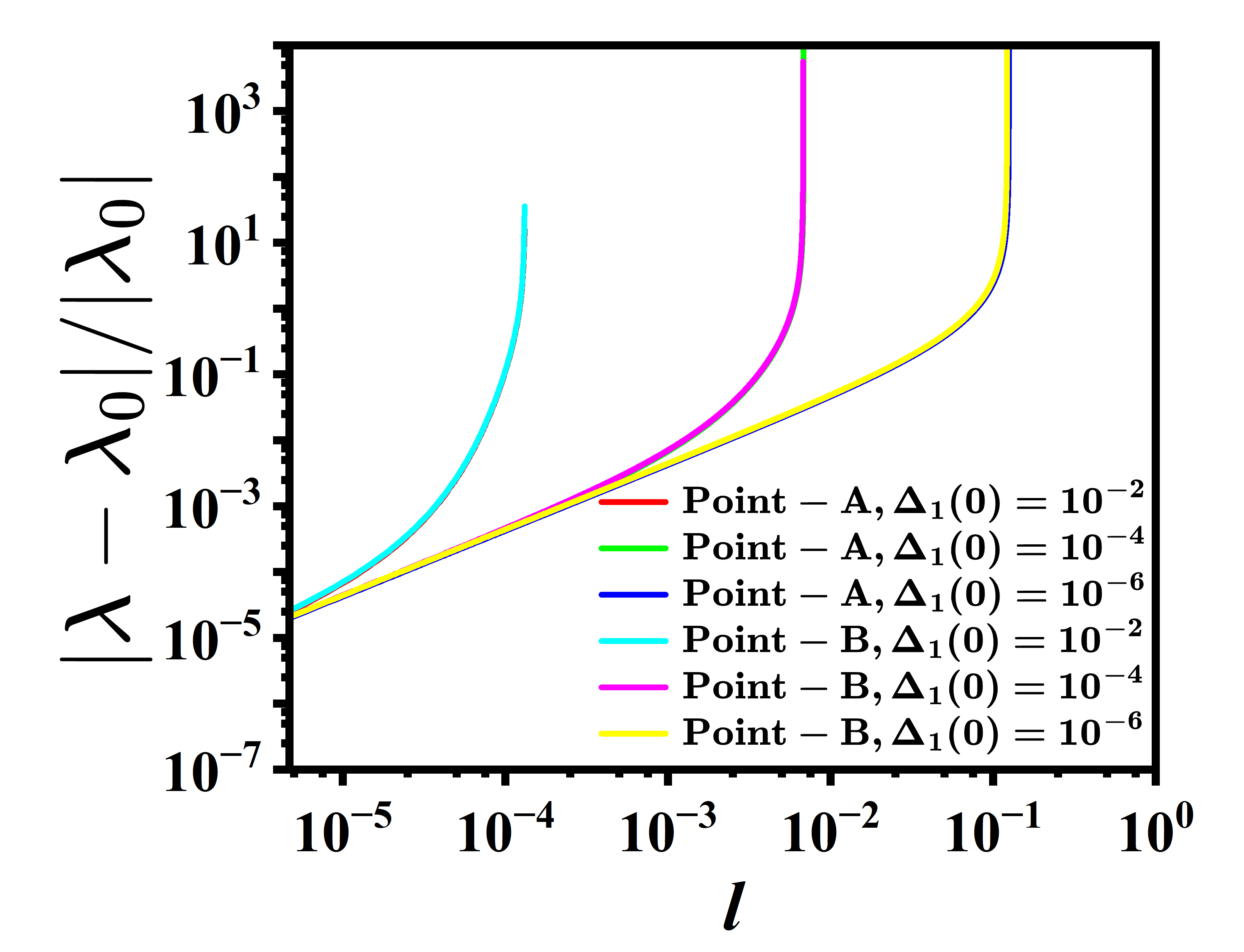}
}
\\
\subfigure[]{
\includegraphics[width=2.35in]{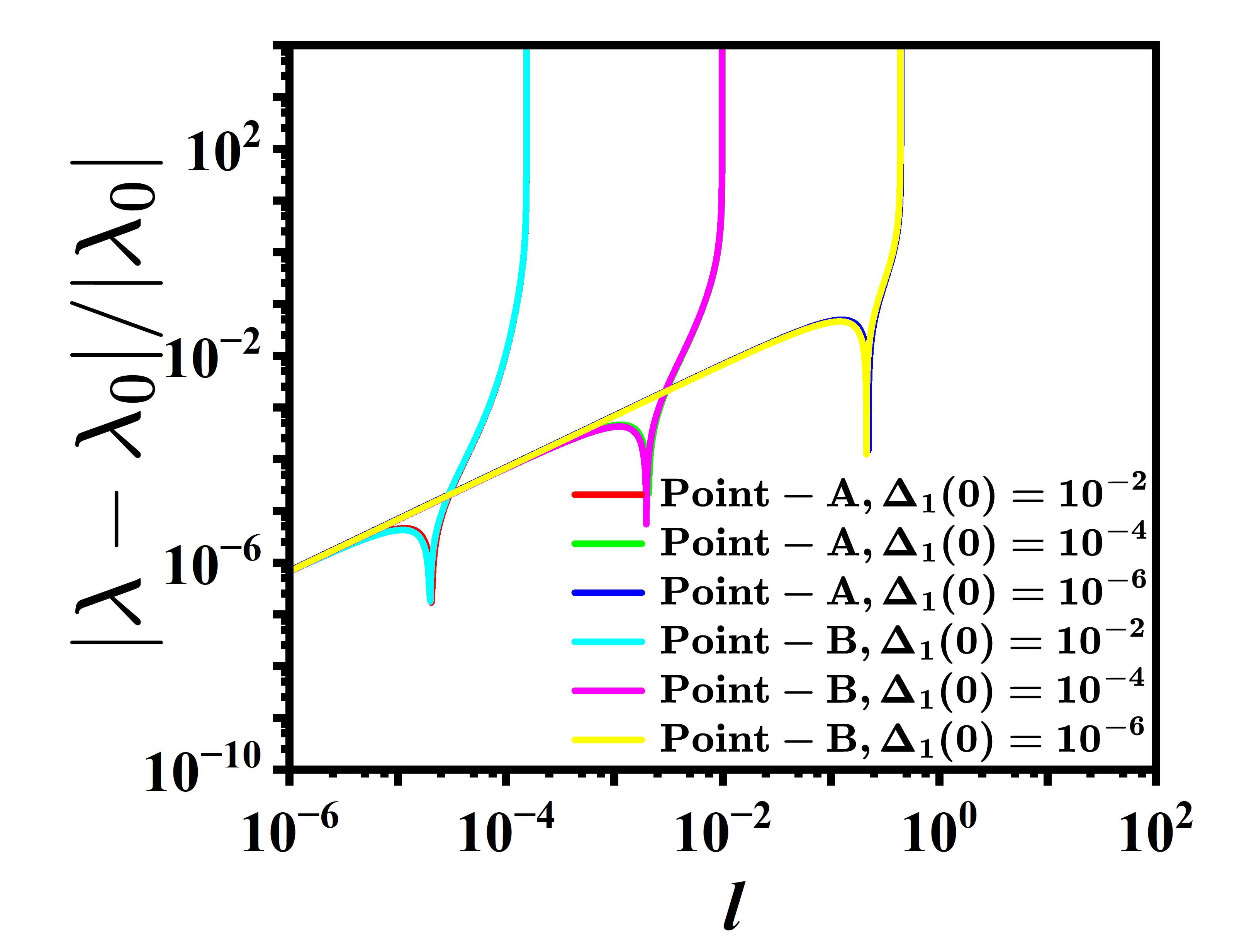}\hspace{-0.6cm}
}
\subfigure[]{
\includegraphics[width=2.35in]{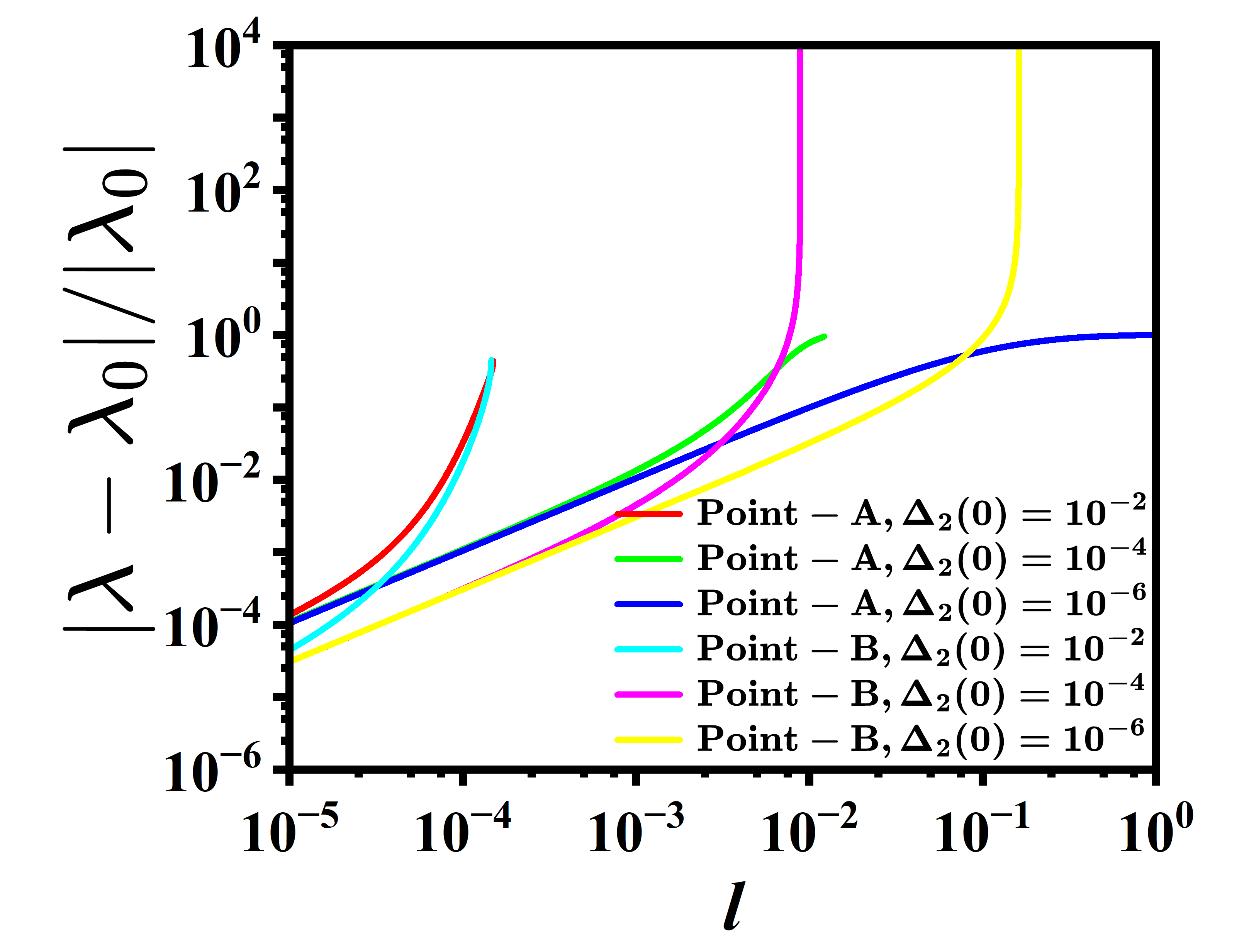}\hspace{-0.6cm}
}
\subfigure[]{
\includegraphics[width=2.35in]{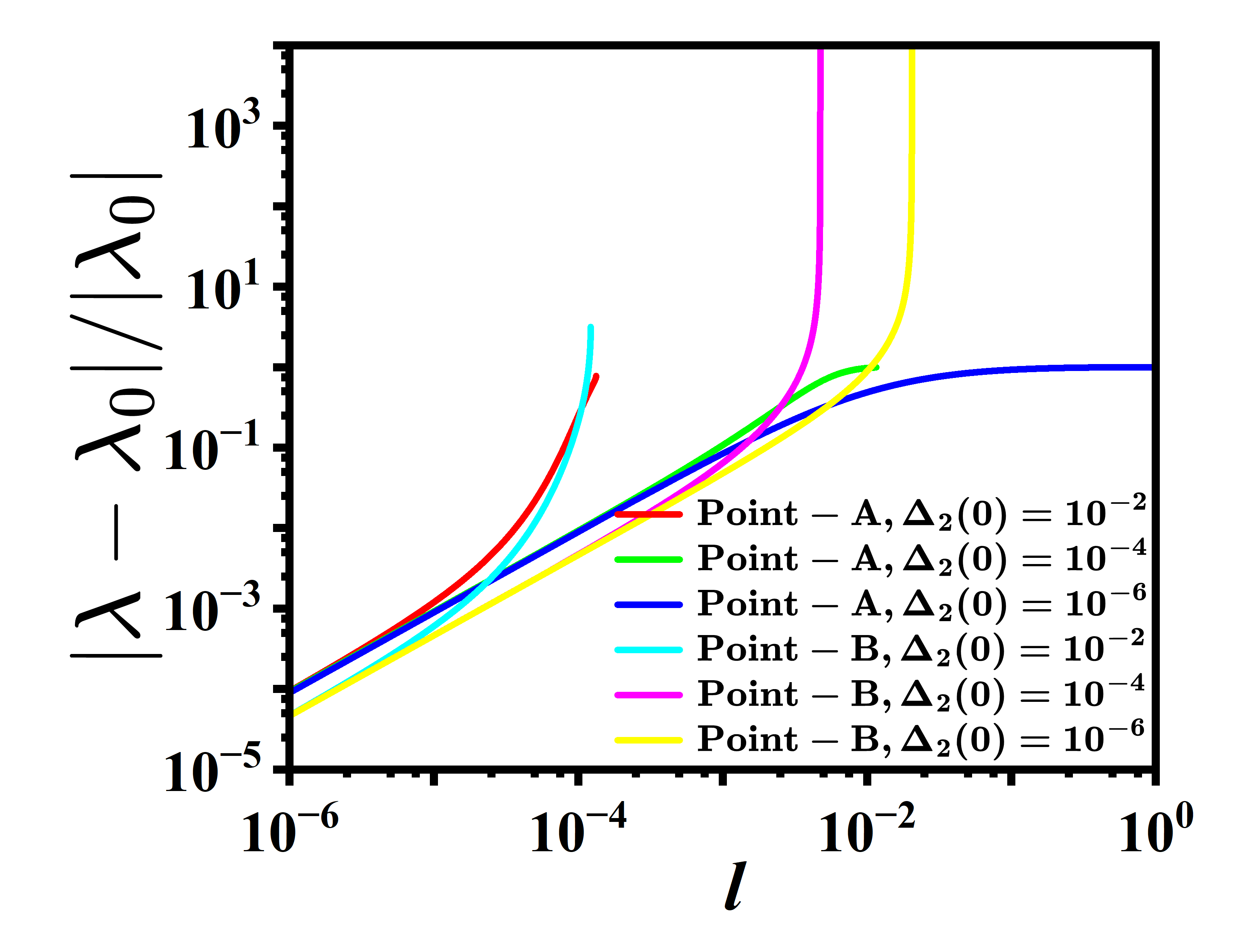}
}
\\
\subfigure[]{
\includegraphics[width=2.35in]{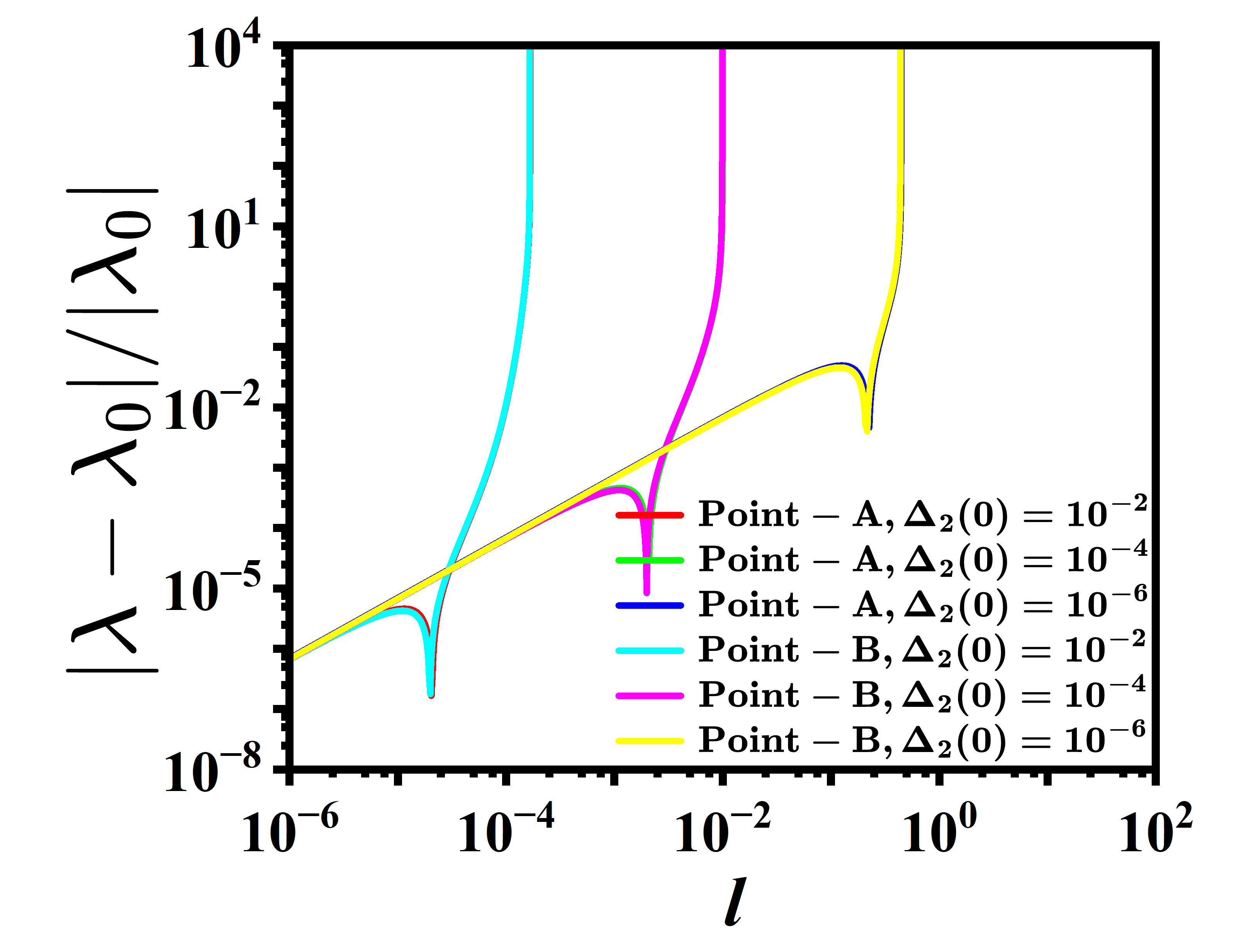}\hspace{0.6cm}
}
\subfigure[]{
\includegraphics[width=2.35in]{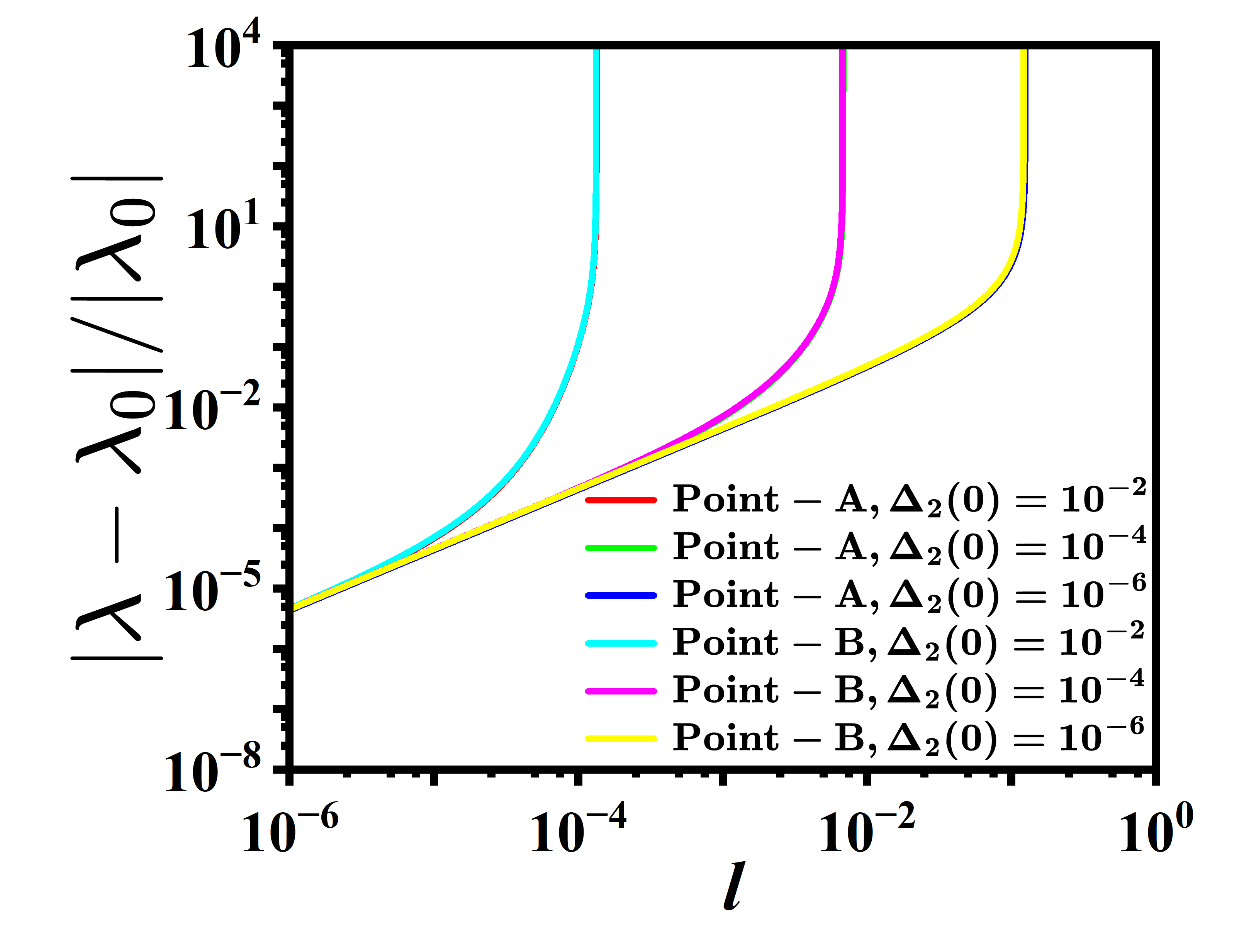}
}
\\
\vspace{-0.25cm}
\caption{(Color online)
Energy-dependent evolutions of the relative interaction strength $\left |\frac{\lambda-\lambda_0}{\lambda_0} \right |$
at Point-A and Point-B for (a)-(b) $\alpha=0.25$ and (c)-(d) $\alpha=0.75$ in the sole presence of disorder $\Delta_1$,
(e)-(f) $\alpha=0.25$ and (g)-(h) $\alpha=0.75$ in the sole presence of disorder $\Delta_2$, respectively.}
\label{fig_delta_1}
\end{figure*}

Then, we shift our focus to the sole presence of random gauge potential, which is characterized by two components, $\Delta_{1}$ and $\Delta_{3}$. Fig~\ref{fig_delta_0}(e) and Fig~\ref{fig_delta_0}(h) present that the effects of $\Delta_3$
component is similar to those of $\Delta_0$ on the Cooper interaction. In this sense, we are going to concentrate
on the influence of the $\Delta_1$ component.

Learning from Fig.~\ref{fig_delta_1}(a) and Fig.~\ref{fig_delta_1}(b) for $\alpha=0.25$, the Cooper interaction goes towards
divergence with variations of initial conditions of $\Delta_1$ at Point-B, which is consistent with the clean-limit results.
As aforementioned in Fig.~\ref{fig_Q_phi}(e) the Cooper instability is forbidden in the clean limit for $\alpha=0.25$ at Point-A in Zone-\uppercase\expandafter{\romannumeral1}.  In the presence $\Delta_1$ presented in Fig.~\ref{fig_delta_1}(a) and Fig.~\ref{fig_delta_1}(b), it is indeed absent of Cooper instability when the initial value of $\Delta_1$ is small but increasing the initial value of $\Delta_1$ to $10^{-2}$, the divergence of Cooper interaction is induced.
Further analysis of Fig.~\ref{fig_delta_1}(c) and (d) shows that
the introduction of disorder $\Delta_1$ with an initial value of $|\lambda_0|=10^{-3}$, which is insufficient to induce the Cooper instability, can clearly drive the divergence of Cooper interaction at either Point-A or Point-B for $\alpha=0.75$.
As a corollary, this indicates that the disorder $\Delta_1$ effectively reduces the critical threshold $|\lambda_c|$ for the
Cooper interaction and thus can induce the Cooper instability with a proper initial value
where it is prohibited at clean limit.


Moreover, we examine the effects of the sole presence of random mass measured by $\Delta_2$.

Fig.~\ref{fig_delta_1}(e)-\ref{fig_delta_1}(h) illustrate that, similar to $\Delta_1$, disorder $\Delta_2$ reduces
the critical coupling strength $|\lambda_c|$ in Zone-\uppercase\expandafter{\romannumeral2}. Nevertheless, it cannot
induce the Cooper instability at Point-A in Zone-\uppercase\expandafter{\romannumeral1}. This sharply contrasts with
the unique ability of $\Delta_1$ to trigger the divergence of the Cooper interaction at Point-A in Zone-\uppercase\expandafter{\romannumeral1}. In this sense, this establishes that while both $\Delta_1$ and
$\Delta_2$ are helpful to reduce $|\lambda_c|$, $\Delta_1$ has a stronger influence on inducing the Cooper instability in Zone-\uppercase\expandafter{\romannumeral1}.

To recapitulate, the effects of single disorders fall into two categories. On one hand, $\Delta_0$ and $\Delta_3$ tend to suppress the divergence of the Cooper interaction but do not alter the regions of Zone-\uppercase\expandafter{\romannumeral1} and Zone-\uppercase\expandafter{\romannumeral2}. On the other hand, $\Delta_1$ and $\Delta_2$ are both effective in reducing the critical threshold ($|\lambda_c|$). Additionally, $\Delta_1$ can modify the regions of Zone-\uppercase\expandafter{\romannumeral2}. Table~\ref{tab_disorder} summarizes the key results for the sole presence of disorder scattering.


\begin{figure*}[htbp]
\subfigure[]{
\includegraphics[width=2.35in]{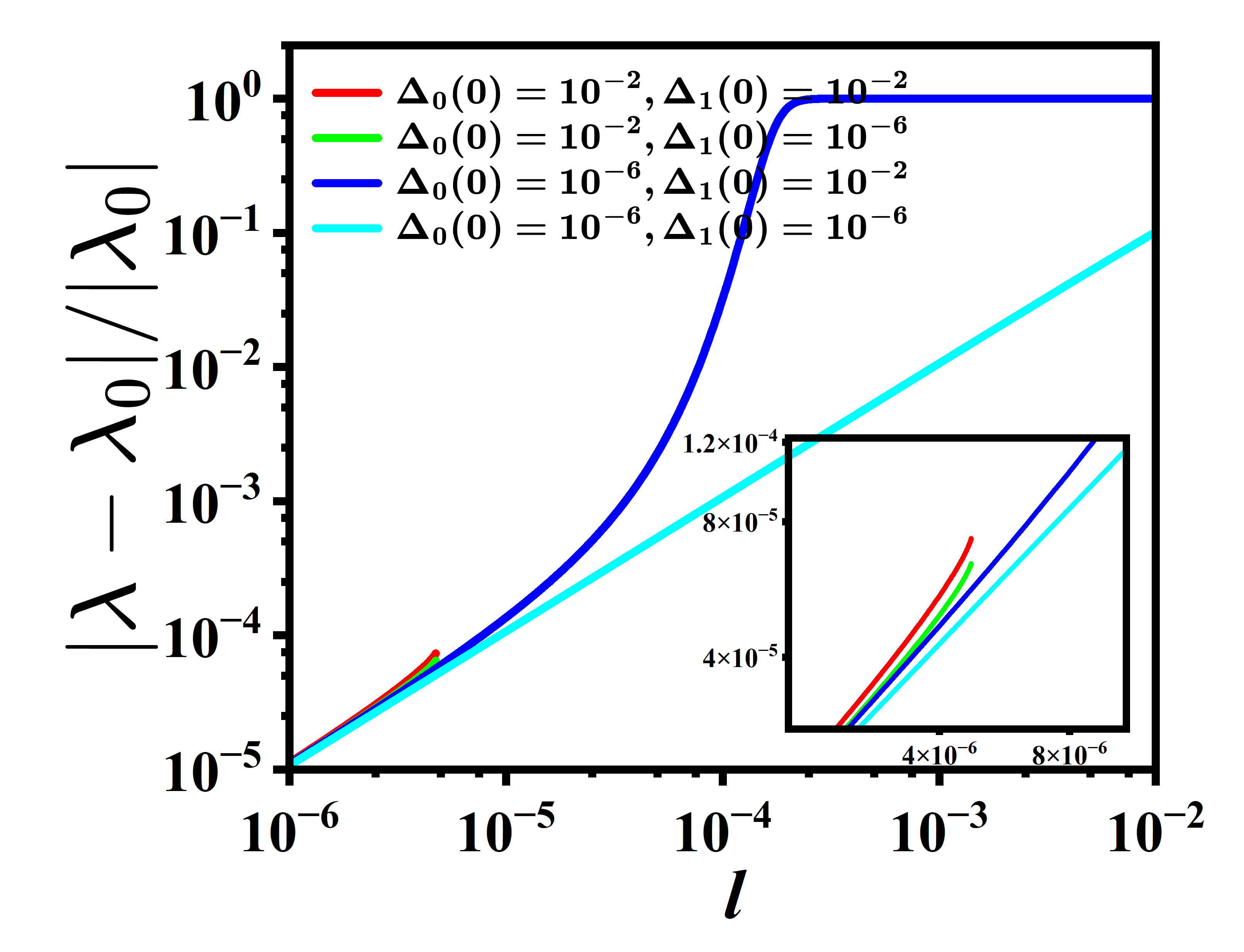}\hspace{-0.6cm}
}
\subfigure[]{
\includegraphics[width=2.35in]{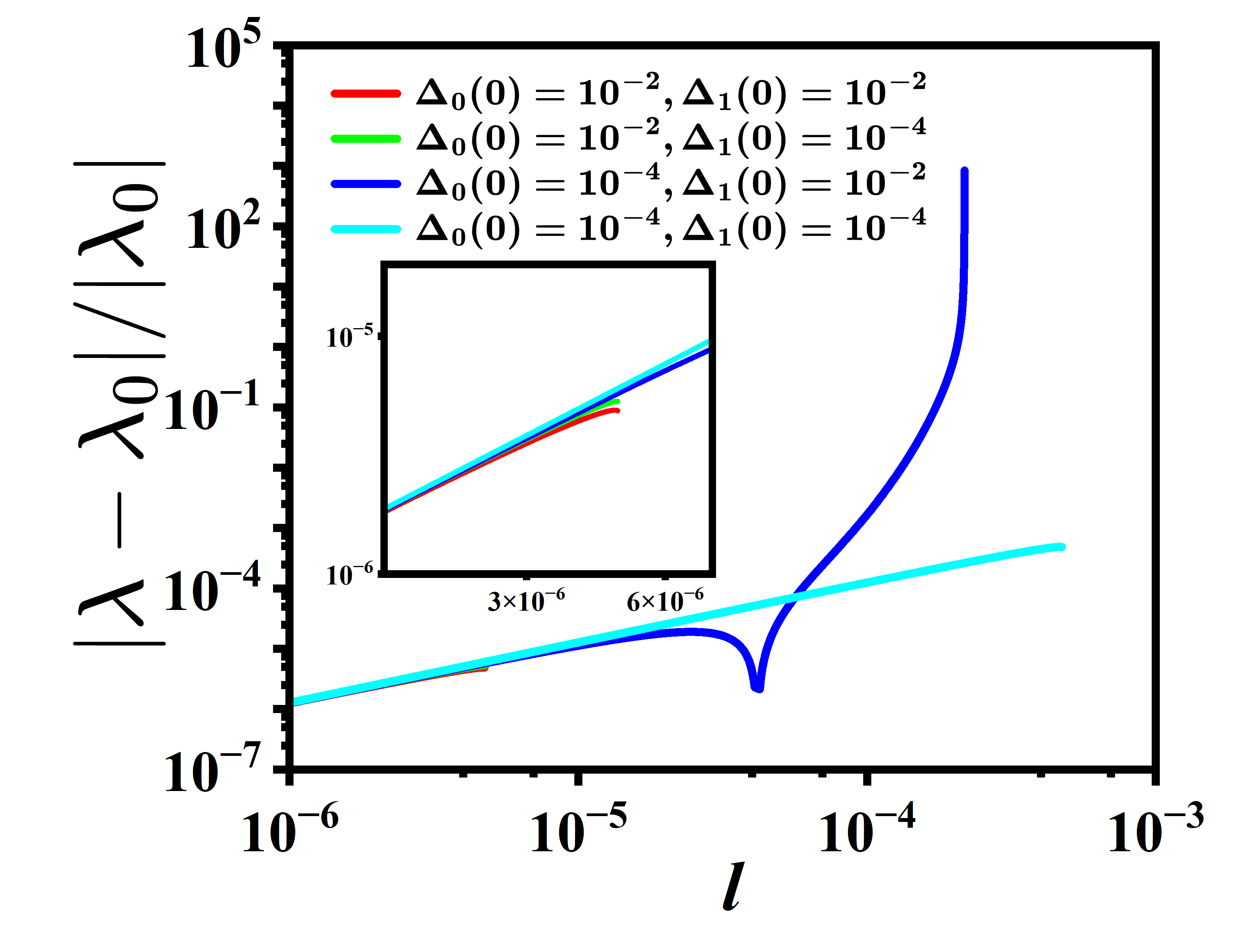}\hspace{-0.6cm}
}
\subfigure[]{
\includegraphics[width=2.35in]{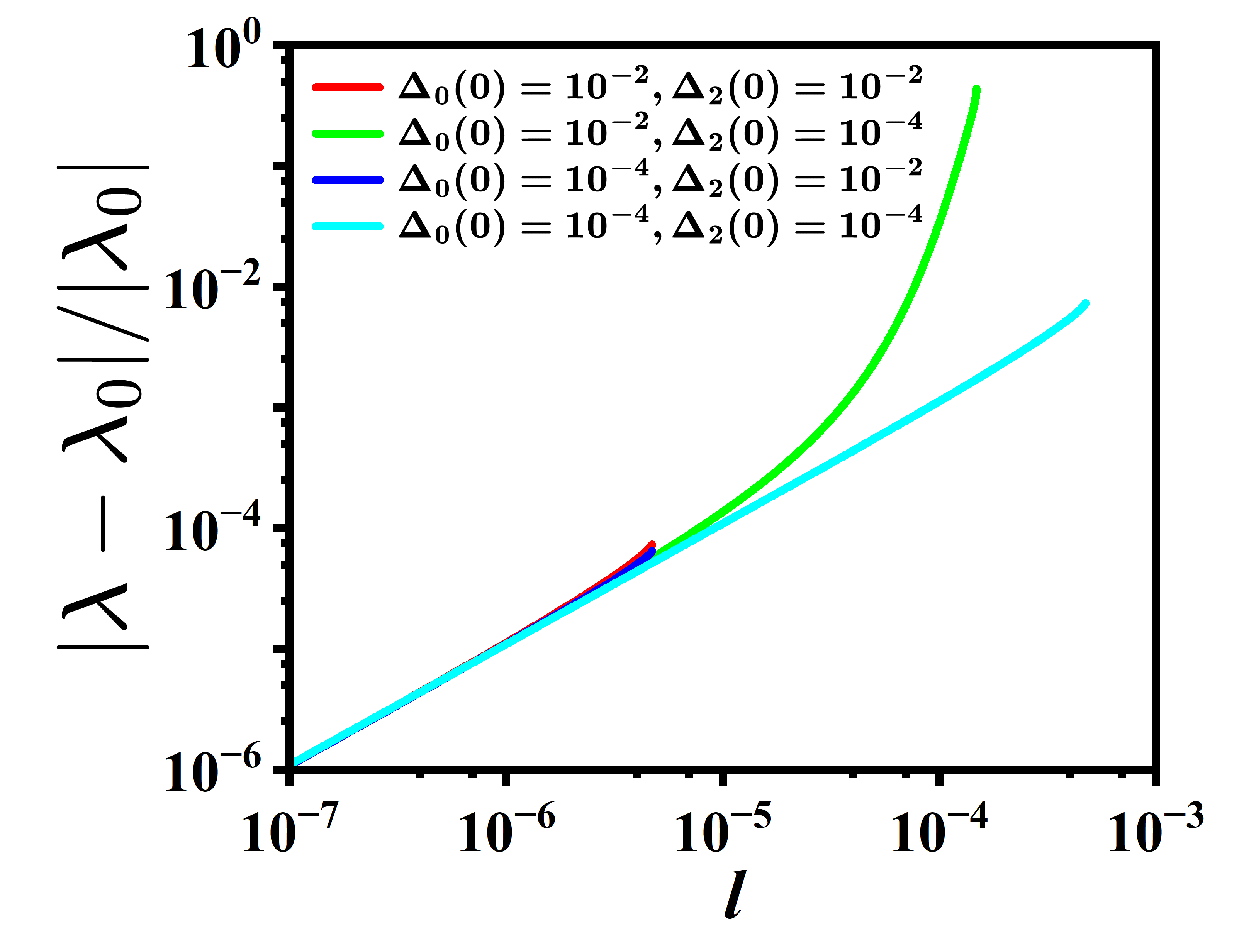}
}
\\
\subfigure[]{
\includegraphics[width=2.35in]{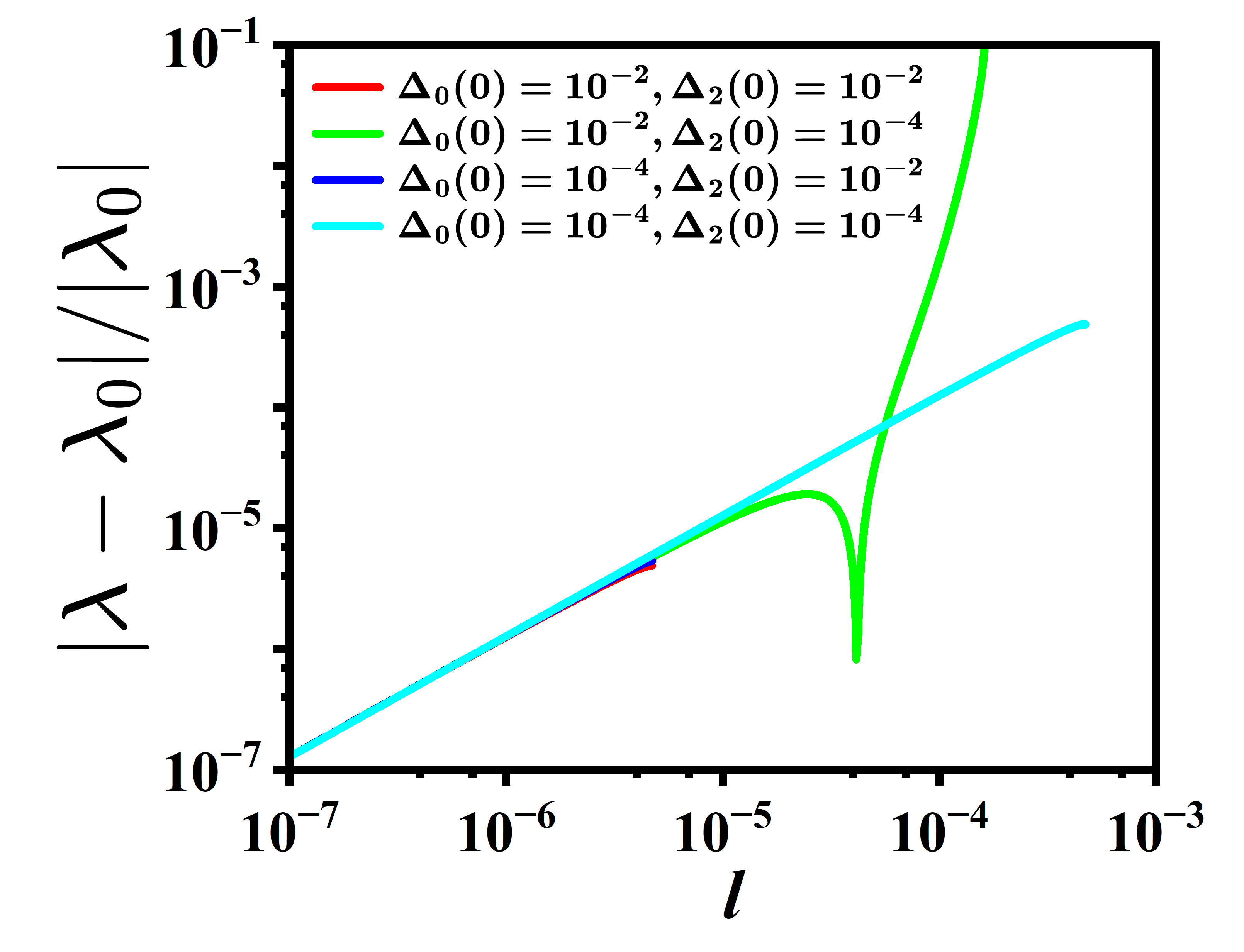}\hspace{-0.6cm}
}
\subfigure[]{
\includegraphics[width=2.35in]{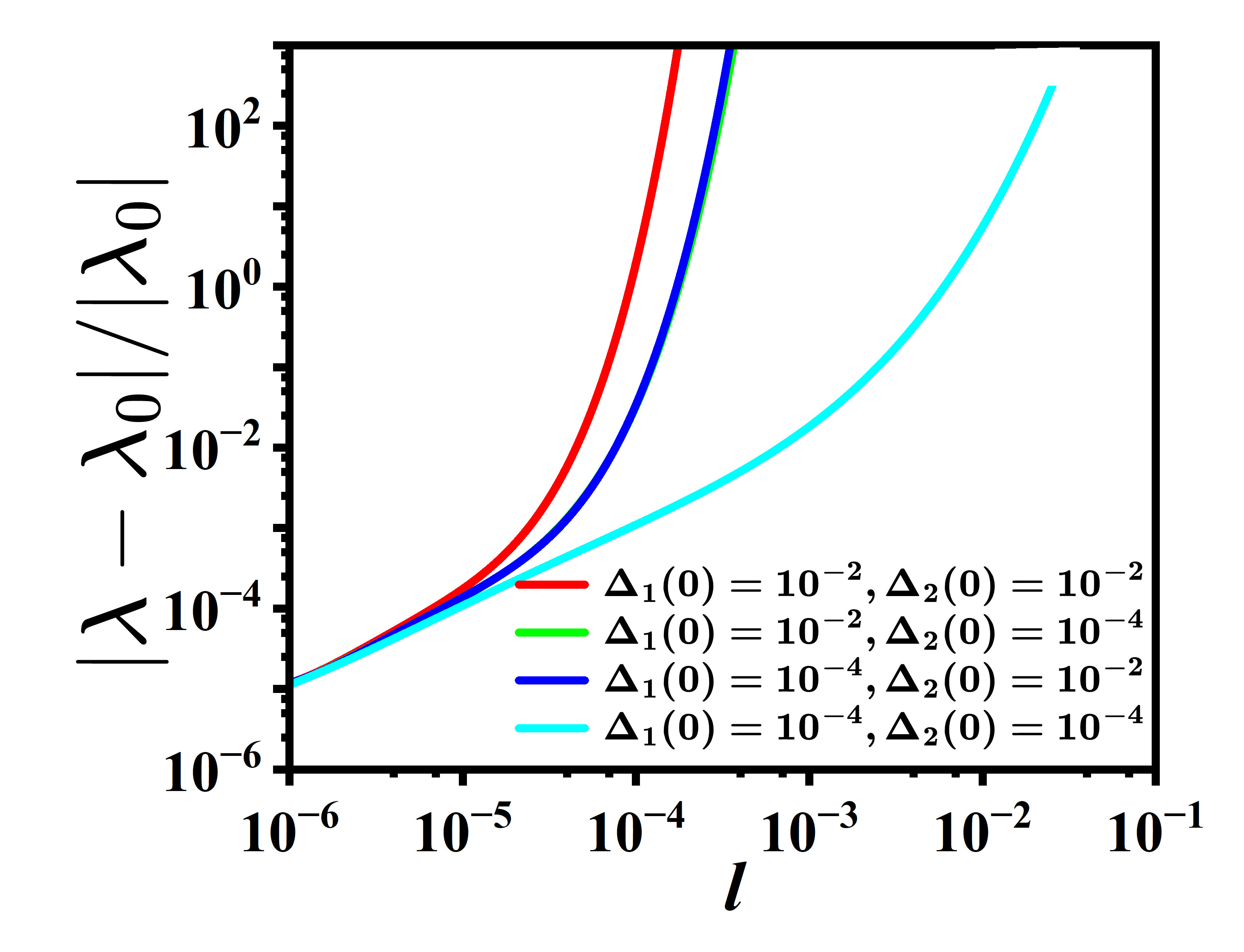}\hspace{-0.6cm}
}
\subfigure[]{
\includegraphics[width=2.35in]{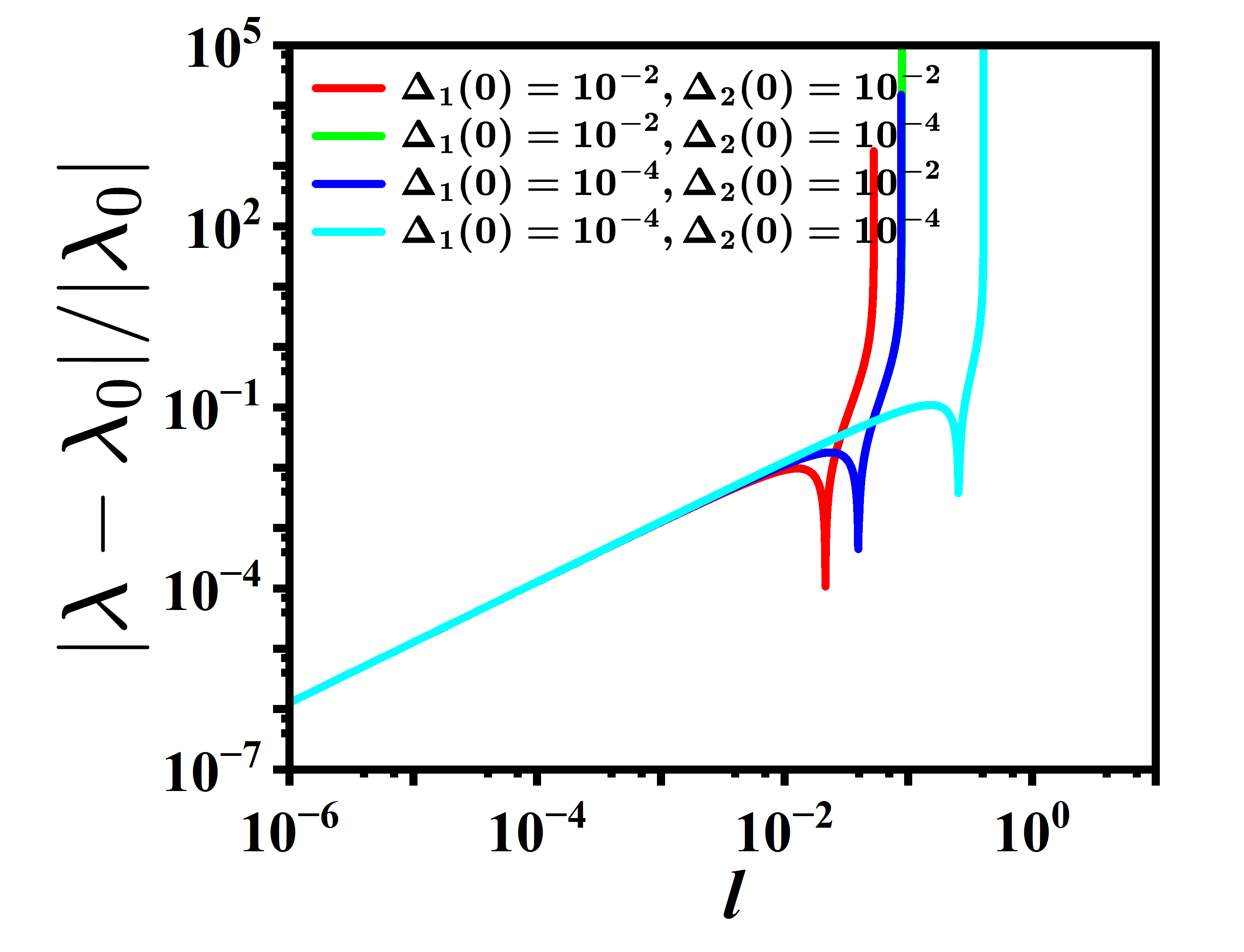}
}
\\
\subfigure[]{
\includegraphics[width=2.35in]{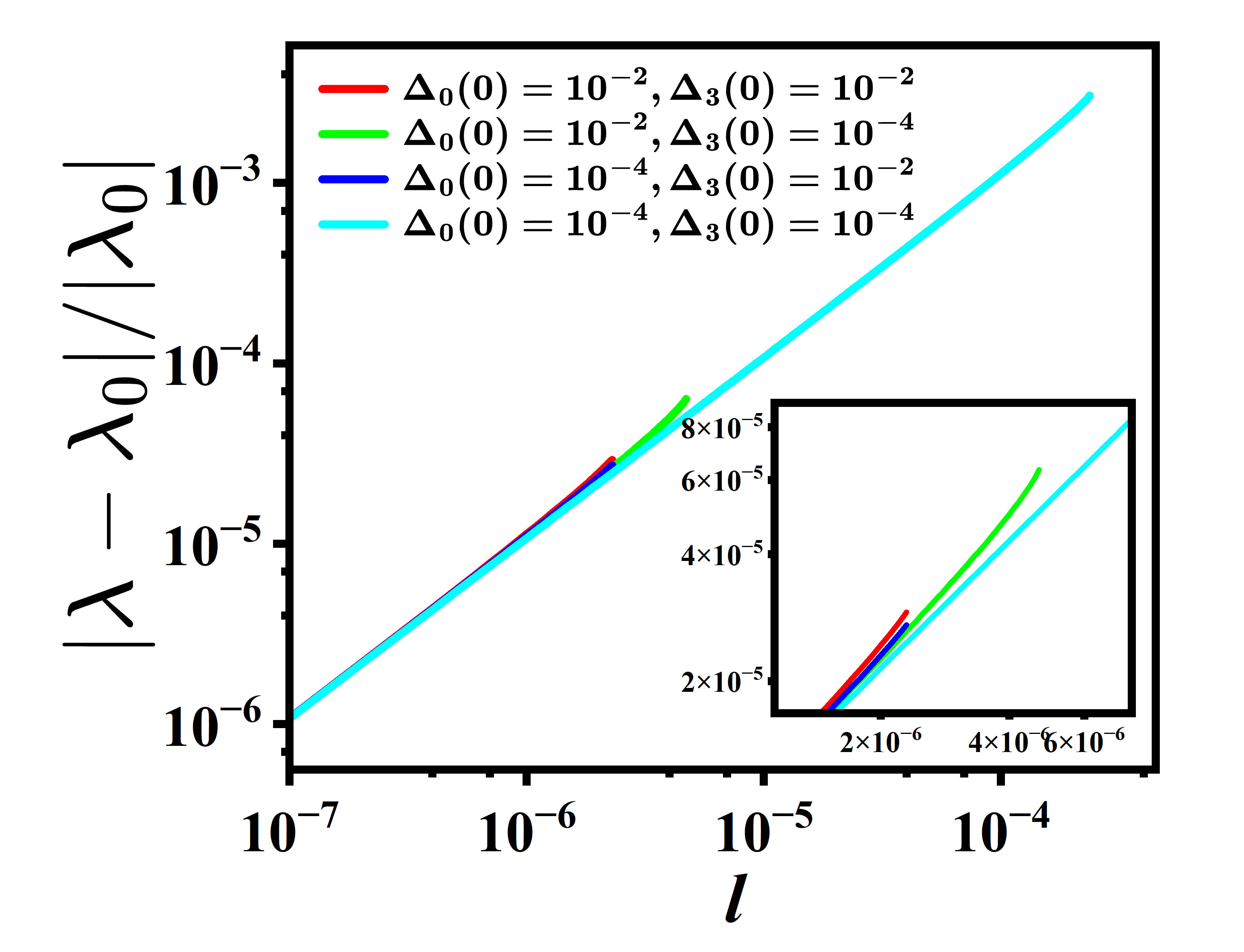}\hspace{0.6cm}
}
\subfigure[]{
\includegraphics[width=2.35in]{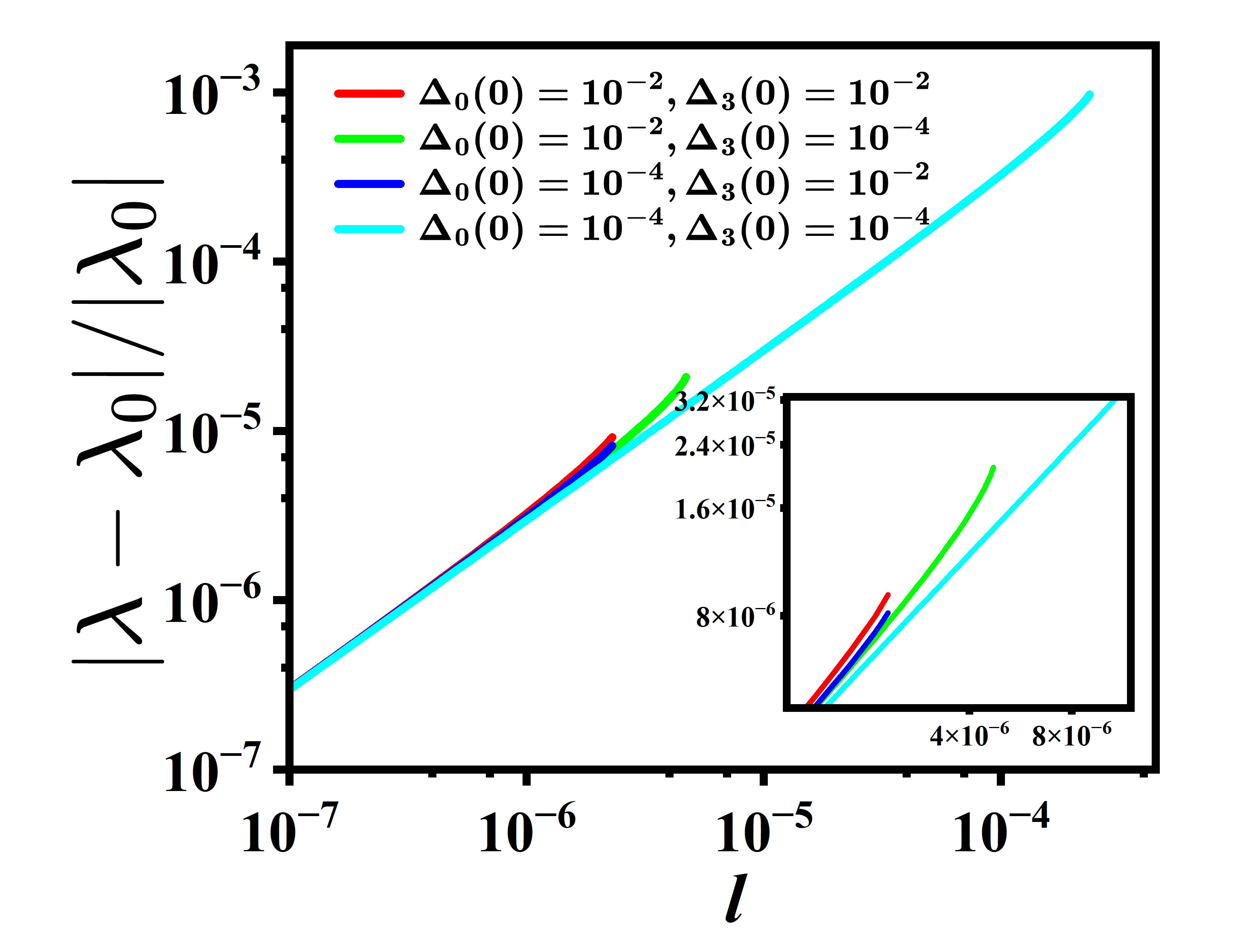}
}
\\
\vspace{-0.2cm}
\caption{(Color online) Energy-dependent evolutions of the relative interaction
strength $\left |\frac{\lambda-\lambda_0}{\lambda_0} \right |$ at $\alpha=0.25$ in the presence of
two kinds of disorders: (a) Point-A ($\Delta_0$, $\Delta_1$), (b) Point-B ($\Delta_0$, $\Delta_1$),
(c) Point-A ($\Delta_0$, $\Delta_2$), (d) Point-B ($\Delta_0$, $\Delta_2$),
(e) Point-A ($\Delta_1$, $\Delta_2$), (f) Point-B ($\Delta_1$, $\Delta_2$),
(g) Point-A ($\Delta_0$, $\Delta_3$), and (h) Point-B ($\Delta_0$, $\Delta_3$).}
\label{fig_two_delta}
\end{figure*}

\subsection{Two types of disorders}\label{Sec_two_disorders}

Next, we study the situation in the presence of two types of disorder scatterings.
By analyzing the effects of sole presence of disorder as presented in Table~\ref{tab_disorder}, it is evident that
the impact of $\Delta_3$ is analogous to that of $\Delta_0$.
We therefore turn attention to the combined influence of two of disorder scatterings from
$\Delta_0$, $\Delta_1$, and $\Delta_2$ on the fate of Cooper interaction.

Specifically, let us consider the simultaneous presence of $\Delta_0$ and $\Delta_1$. At Point-A ($Q=10^{-2}$, $\phi=\pi$),
as shown in Sec.~\ref{Sec_single_disorder}, $\Delta_1$ alone can drive Point-A from Zone-\uppercase\expandafter{\romannumeral1} to Zone-\uppercase\expandafter{\romannumeral2}. However, when both $\Delta_1$ and $\Delta_0$ are present, Fig.~\ref{fig_two_delta}(a) demonstrates that any finite $\Delta_0$ would completely suppresses this transition regardless of the magnitude of $\Delta_1$.
Turning to Point-B ($Q=10^{-2}$, $\phi=\pi$), Fig.~\ref{fig_two_delta}(b) shows that when $\Delta_0$
is very small (such as $\Delta_0=10^{-6}$), the Cooper interaction can go towards divergence with $\Delta_1=10^{-2}$.
However, once $\Delta_0$ exceeds $10^{-4}$, $\Delta_1$ becomes subordinate to $\Delta_0$, and its ability
to lower $|\lambda_c|$ is significantly reduced, keeping $|\lambda_c|$ close to its clean-limit value. This clearly
indicates that $\Delta_0$ generally dominates over $\Delta_1$ when they are present simultaneously.

\begin{table}[H]
\centering
\caption{Influence of simultaneous presence of two kinds of disorders on the critical value $|\lambda_c|$ for Cooper instability.
The symbols $\uparrow$ and $\downarrow$ signify the expansion and contraction of
Zone-\uppercase\expandafter{\romannumeral1} and Zone-\uppercase\expandafter{\romannumeral2} which
are designated in Sec.~\ref{Subsec_lambda_c}.}
    \renewcommand{\arraystretch}{1.5}
    \setlength{\tabcolsep}{4pt}
    \vspace{+0.15cm}
	\begin{tabular}{c|c|c|c|c}
		\hline
        \hline
		Two types of disorder &       $\red{\Delta_0, \Delta_1}$      &      $\red{\Delta_0, \Delta_2}$      &       ${\color{Green}\Delta_1, \Delta_2}$    &     $\red{\Delta_0, \Delta_3}$       \\ \hline
		 Area of Zone-\uppercase\expandafter{\romannumeral1}&       $\red{ \uparrow}$      &      $\red{ \uparrow}$       &      ${\color{Green}\uparrow}$      &      $\red{ \uparrow}$      \\ \hline
		 Area of Zone-\uppercase\expandafter{\romannumeral2}&       $\red{ \downarrow}$      &       $\red{ \downarrow}$      &      ${\color{Green}\downarrow}$      &     $\red{ \downarrow}$       \\
        \hline
        \hline
	\end{tabular}
	\label{tab_two_disorder}
\end{table}

Then, we substitute the disorder $\Delta_2$ for $\Delta_1$ and analyze the combined effect of $\Delta_0$ and $\Delta_2$.
It can be noticed from Fig.~\ref{fig_two_delta}(c) and Fig.~\ref{fig_two_delta}~(d) that this combination
exhibits behavior analogous to that of $\Delta_0$ and $\Delta_1$. In other words, $\Delta_0$ remains dominant
in suppressing the Cooper interaction, while $\Delta_2$ has a significantly weaker effect than $\Delta_1$
under identical conditions and can only slightly reduce the critical value $|\lambda_c|$ when $\Delta_0$ is very small.
Furthermore, Fig.~\ref{fig_two_delta}(g) and Fig.~\ref{fig_two_delta}(h) demonstrate that
the combination of $\Delta_0$ and $\Delta_3$ still
suppresses the Cooper instability and produces behavior similar to that of each disorder acting alone.
In consequence, these corroborate that $\Delta_0$ compete with other types of disorders and
is the leading ingredient in suppressing the Cooper instability regardless of whether it is paired
with $\Delta_1$, $\Delta_2$, or $\Delta_3$.

At the end, we briefly comment on the combined presence of $\Delta_1$ with $\Delta_2$.
In contrast to the single $\Delta_1$ case shown in Figs.~\ref{fig_delta_1}(a)-(b), where Point-A transitions from Zone I to Zone II with $\Delta_1(0)=10^{-2}$, Fig.~\ref{fig_two_delta}(e)-Fig.~\ref{fig_two_delta}(f) demonstrate that, with $\Delta_2$ present, the critical threshold for the divergence of the Cooper interaction is further reduced. This allows the Cooper instability to emerge with a smaller initial value of $\Delta_1$ ($\Delta_1(0) < 10^{-2}$). Accordingly, this indicates that $\Delta_1$ and $\Delta_2$ do not compete but instead cooperate to promote the Cooper instability by lowering the critical value $|\lambda_c|$.
Table~\ref{tab_two_disorder} collects the key results for the presence of two distinct
kinds of disorder scatterings.


\begin{figure*}[htbp]
\subfigure[]{
\includegraphics[width=2.35in]{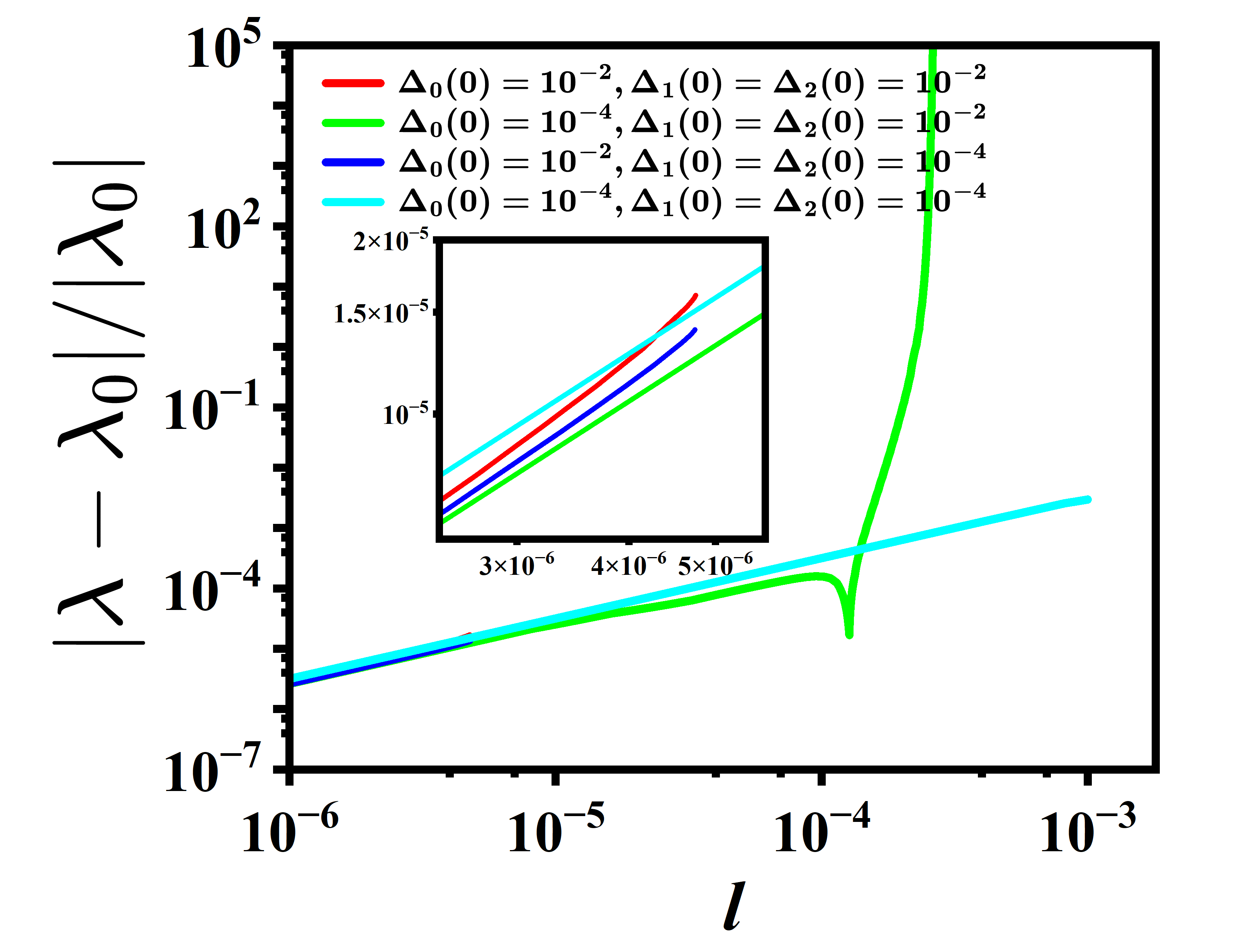}\hspace{-0.6cm}
}
\subfigure[]{
\includegraphics[width=2.35in]{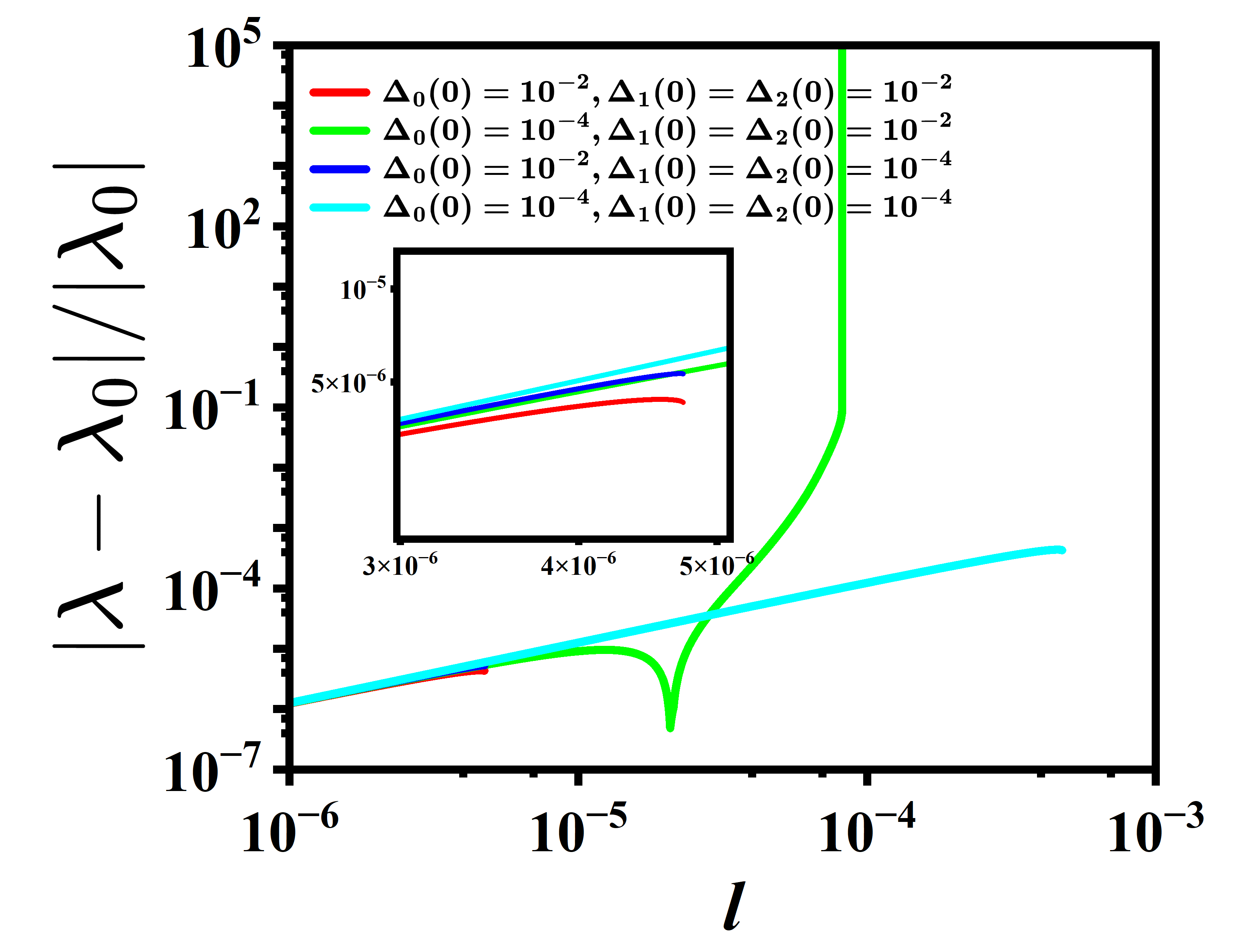}\hspace{-0.6cm}
}
\subfigure[]{
\includegraphics[width=2.35in]{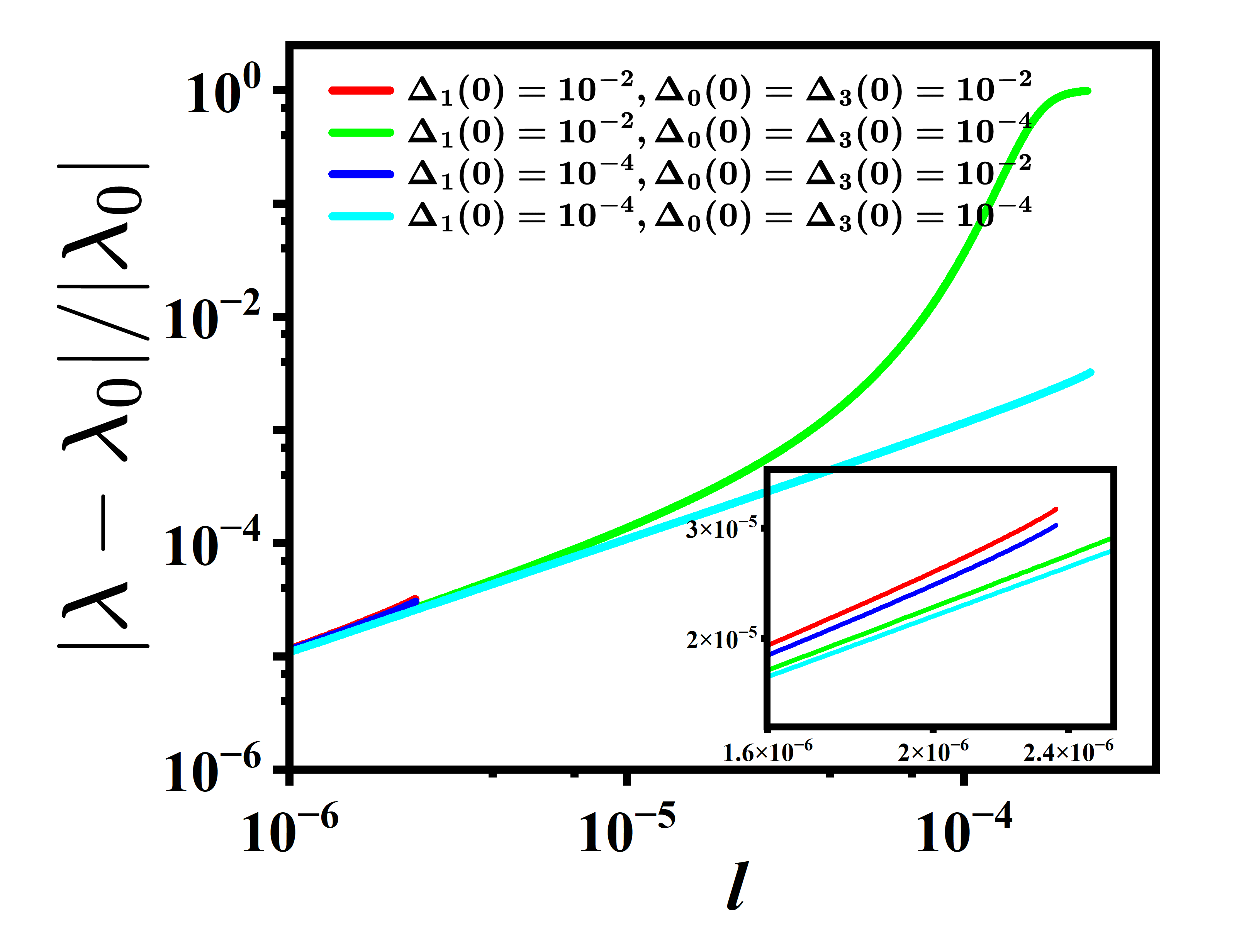}
}
\\
\subfigure[]{
\includegraphics[width=2.35in]{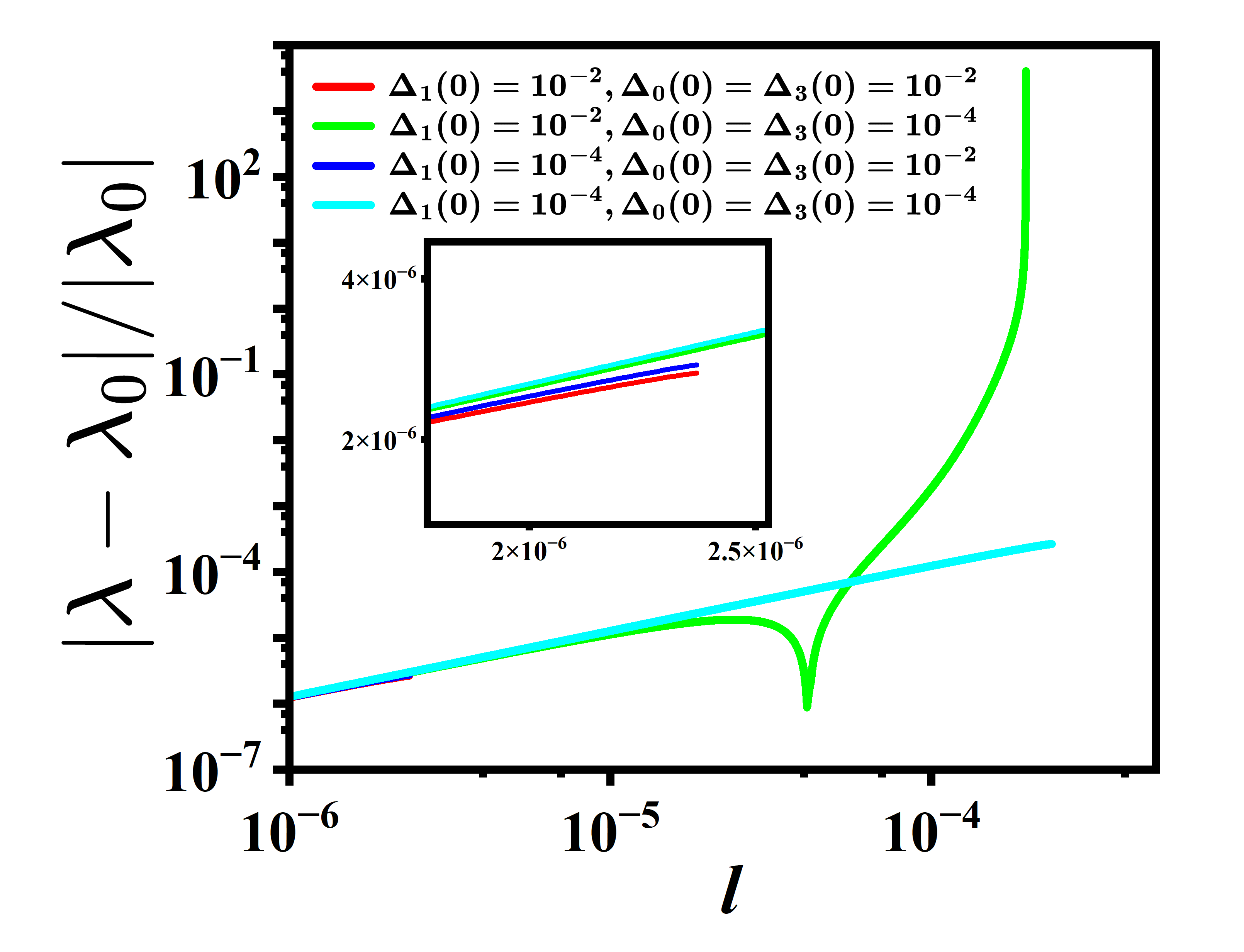}\hspace{-0.6cm}
}
\subfigure[]{
\includegraphics[width=2.35in]{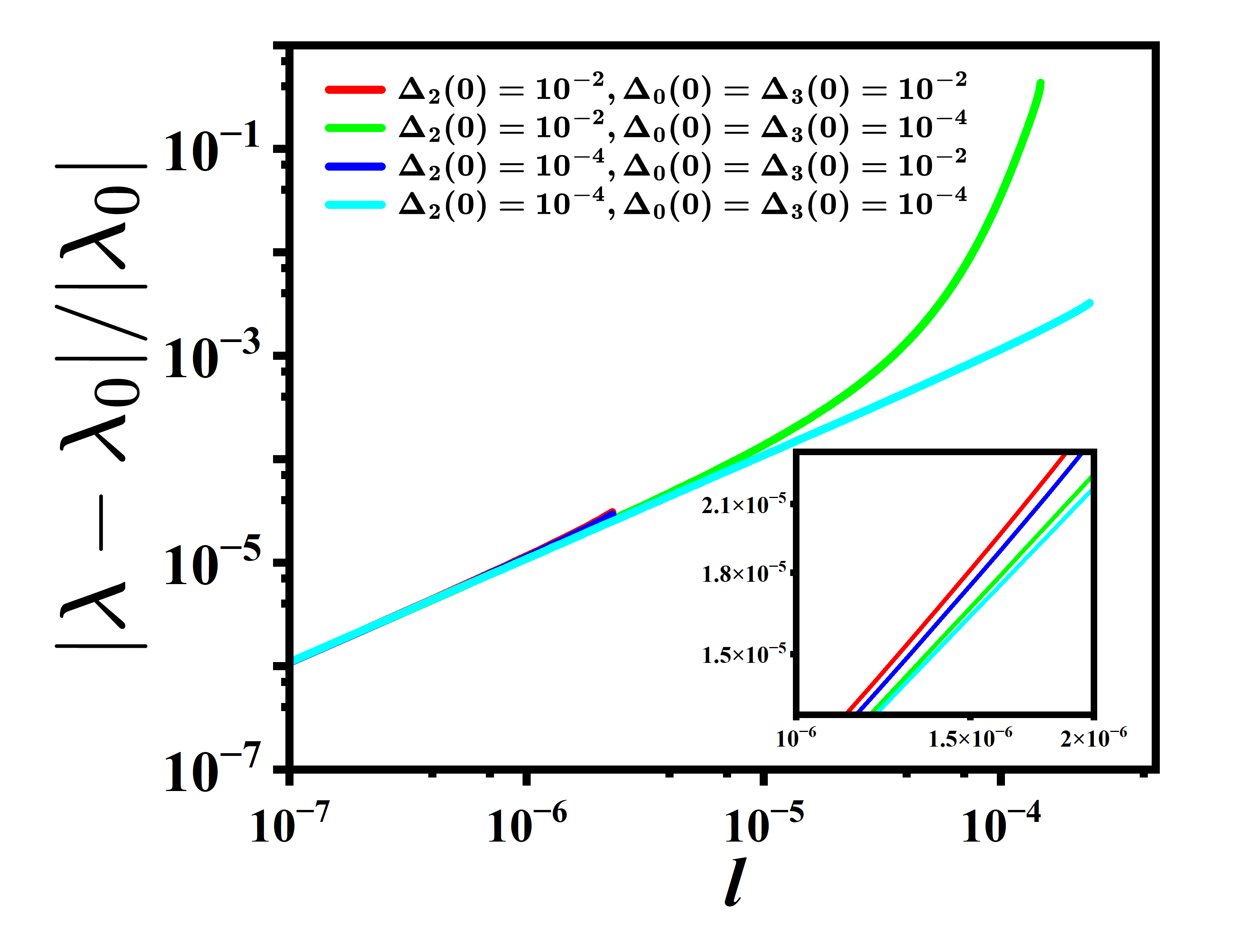}\hspace{-0.6cm}
}
\subfigure[]{
\includegraphics[width=2.35in]{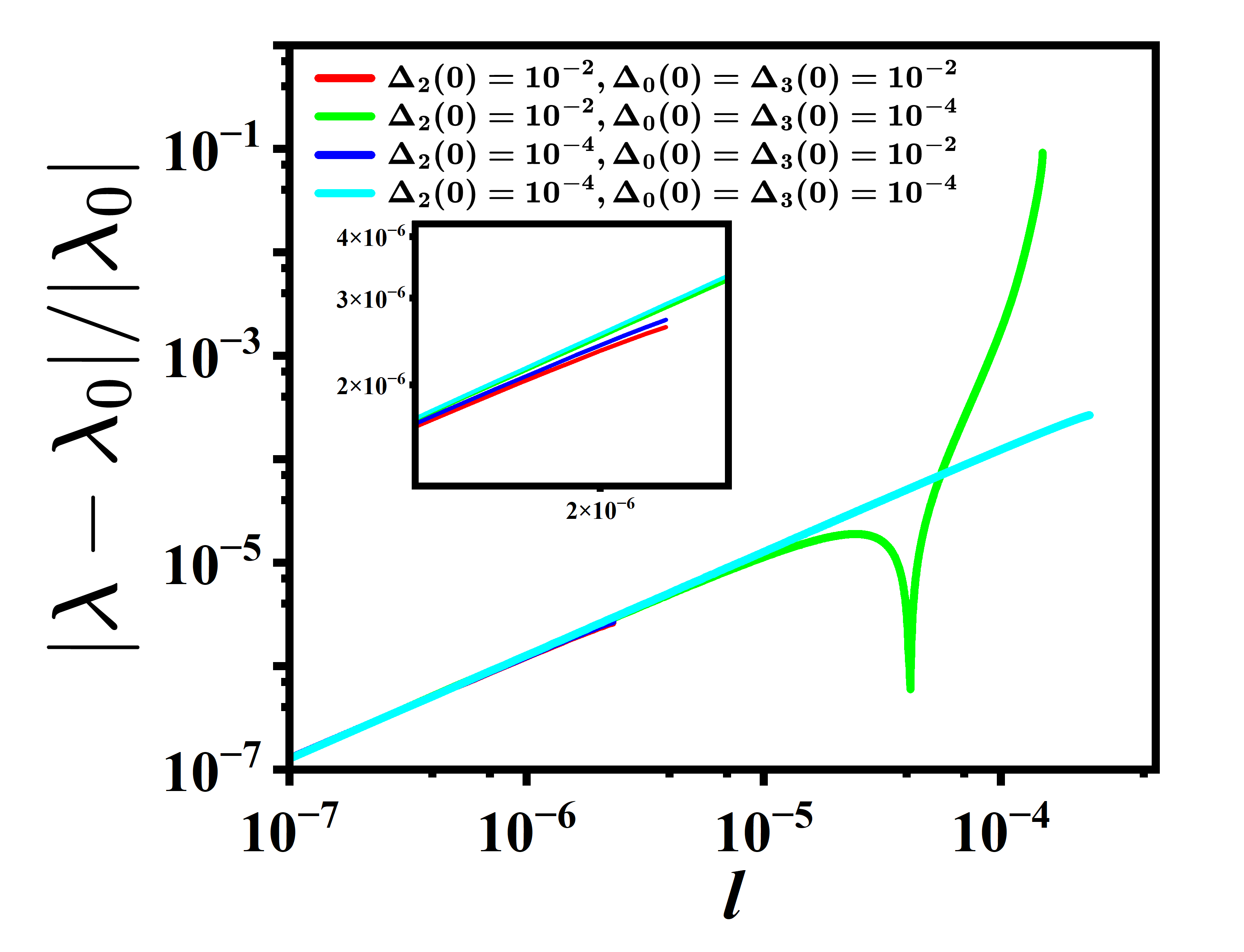}
}
\\
\subfigure[]{
\includegraphics[width=2.35in]{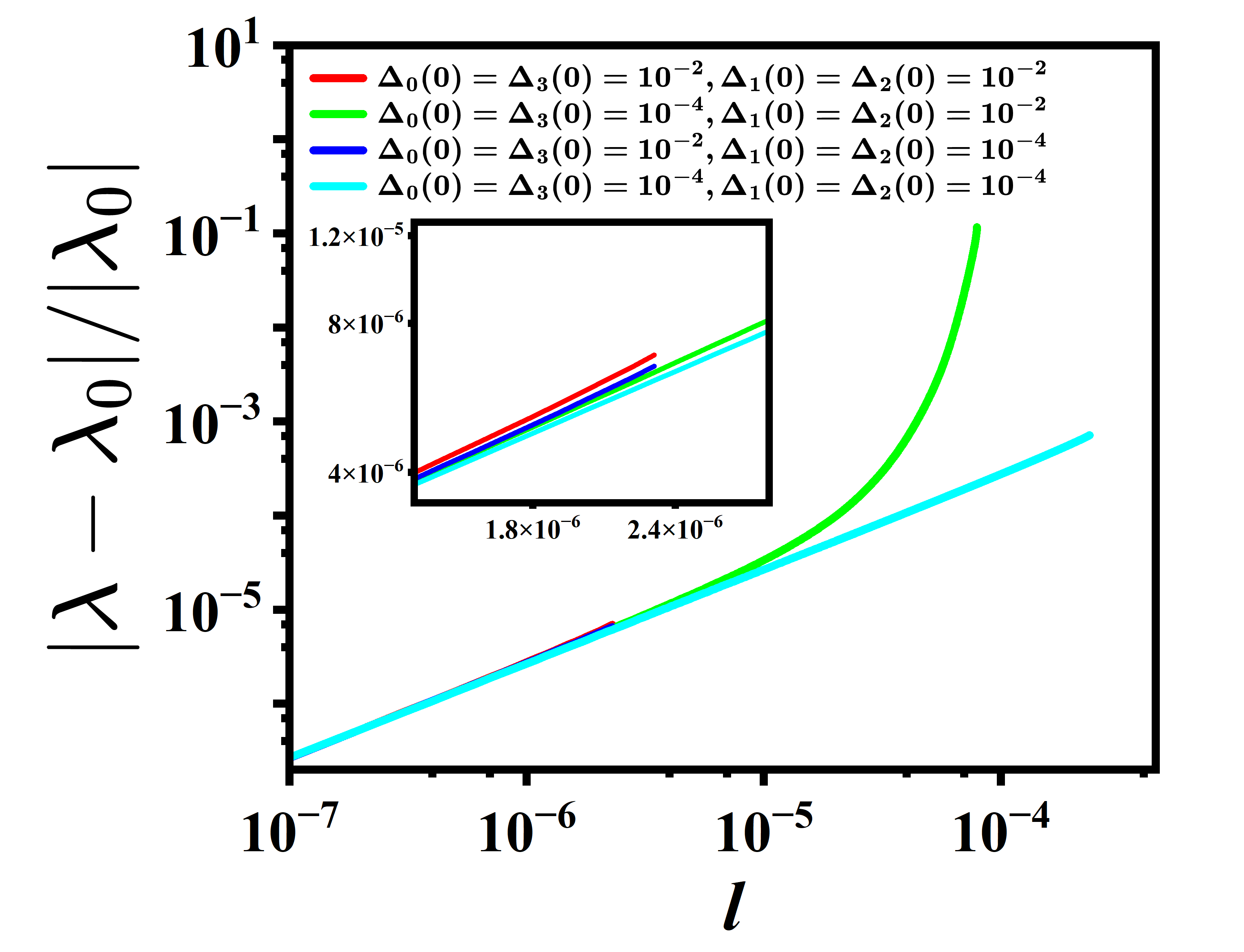}\hspace{0.6cm}
}
\subfigure[]{
\includegraphics[width=2.35in]{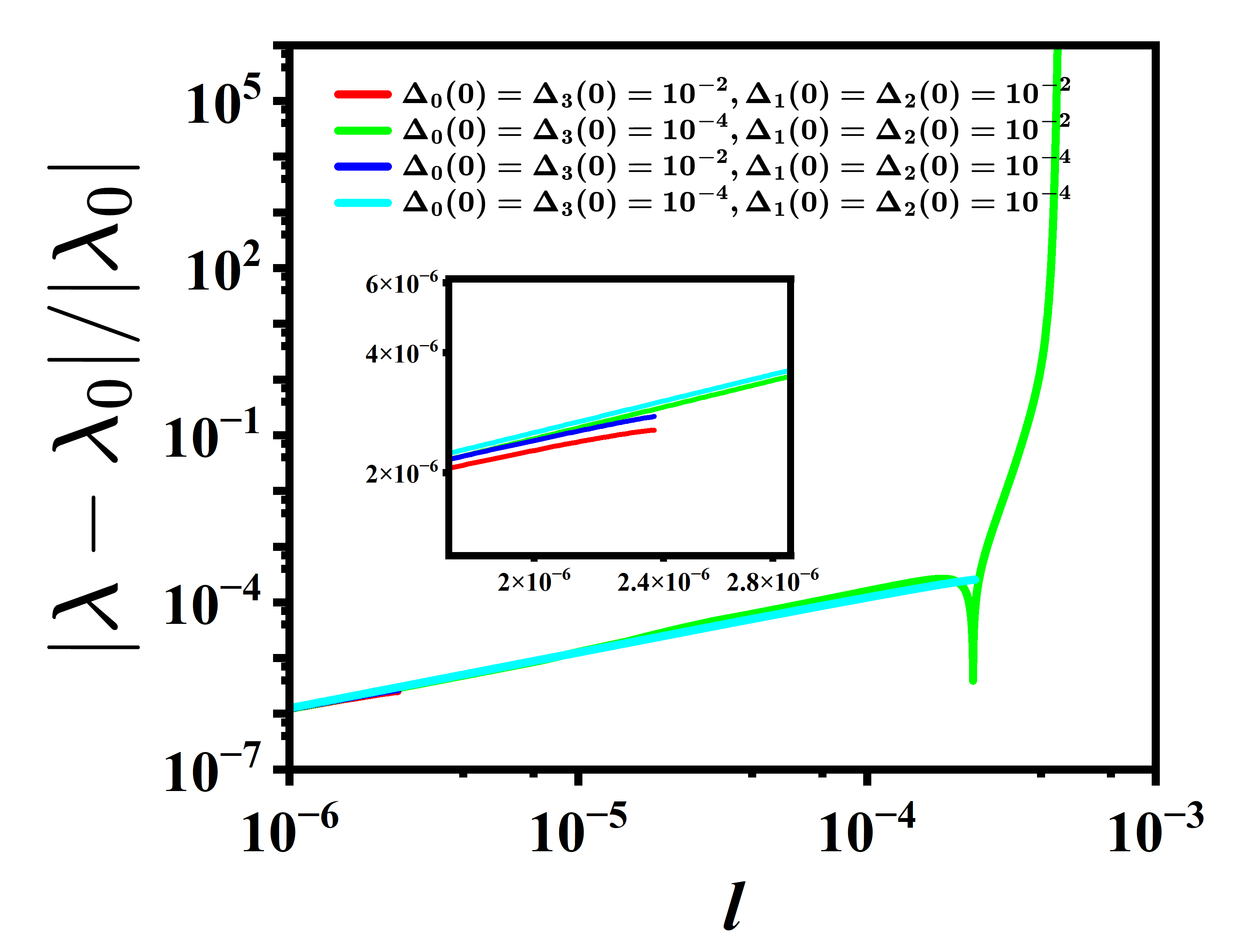}
}
\\
\vspace{-0.2cm}
\caption{(Color online) Energy-dependent evolutions of the relative interaction
strength $\left |\frac{\lambda-\lambda_0}{\lambda_0} \right |$ at $\alpha=0.25$ in the presence of
multiple kinds of disorders: (a) Point-A ($\Delta_0$, $\Delta_1$, $\Delta_2$), (b) Point-B ($\Delta_0$, $\Delta_1$, $\Delta_2$),
(c) Point-A ($\Delta_0$, $\Delta_1$, $\Delta_3$), (d) Point-B ($\Delta_0$, $\Delta_1$, $\Delta_3$),
(e) Point-A ($\Delta_0$, $\Delta_2$, $\Delta_3$), (f) Point-B ($\Delta_0$, $\Delta_2$, $\Delta_3$),
(g) Point-A ($\Delta_0$, $\Delta_1$, $\Delta_2$, $\Delta_3$), and (h) Point-B ($\Delta_0$, $\Delta_1$, $\Delta_2$, $\Delta_3$).}
\label{fig_multiple_delta}
\end{figure*}

\begin{table}[H]
\centering
\caption{Influence of presence of multiple sorts of disorders on the critical value $|\lambda_c|$ for Cooper instability.
The symbols $\uparrow$ and $\downarrow$ signify the expansion and contraction of
Zone-\uppercase\expandafter{\romannumeral1} and Zone-\uppercase\expandafter{\romannumeral2} which
are designated in Sec.~\ref{Subsec_lambda_c}.}
	\renewcommand{\arraystretch}{1.5}
	\setlength{\tabcolsep}{6pt}
    \vspace{+0.15cm}
	\begin{tabular}{c|c|c}
		\hline
		\hline
		$\Delta$ combinations & Type of ($Q$, $\phi$) region & Area change \\
		\hline
		\multirow{2}{*}{${\color{Green}\Delta_0, \Delta_1, \Delta_2}$}
		& Zone-\uppercase\expandafter{\romannumeral1} & ${\color{Green}\uparrow}$ \\
		\cline{2-3}
		& Zone-\uppercase\expandafter{\romannumeral2} & ${\color{Green}\downarrow}$ \\
		\hline
		\multirow{2}{*}{$\red{\Delta_0, \Delta_1, \Delta_3}$}
		& Zone-\uppercase\expandafter{\romannumeral1} & $\red{\uparrow}$ \\
		\cline{2-3}
		& Zone-\uppercase\expandafter{\romannumeral2} & $\red{\downarrow}$ \\
		\hline
		\multirow{2}{*}{$\red{\Delta_0, \Delta_2, \Delta_3}$}
		& Zone-\uppercase\expandafter{\romannumeral1} & $\red{\uparrow}$ \\
		\cline{2-3}
		& Zone-\uppercase\expandafter{\romannumeral2} & $\red{\downarrow}$ \\
		\hline
		\multirow{2}{*}{$\red{\Delta_0, \Delta_1, \Delta_2, \Delta_3}$}
		& Zone-\uppercase\expandafter{\romannumeral1} & $\red{-}$ \\
		\cline{2-3}
		& Zone-\uppercase\expandafter{\romannumeral2} & $\red{-}$ \\
		\hline
		\hline
	\end{tabular}
	\label{tab_three_disorder}
\end{table}

\subsection{Multiple sorts of disorders}

At last, we now investigate the circumstance with multiple types of disorders.
As detailed in Sec.~\ref{Sec_single_disorder} and Sec.~\ref{Sec_two_disorders},
all disorders can be clustered into two distinct categories:
a suppressing combination of $\Delta_0$ and $\Delta_3$, and an enhancing combination of $\Delta_1$
and $\Delta_2$. To examine the effects of multiple disorders, an effective approach is to introduce
a promotive disorder to a suppressive pair, or vice versa.

Reading from Fig.~\ref{fig_multiple_delta}(a) and Fig.~\ref{fig_multiple_delta}(b), it is clear that introducing a suppressive $\Delta_0$ to the enhancing combination of $\Delta_1$ and $\Delta_2$ significantly reduces the critical value of Cooper interaction.
This suppression sabotages the possibility to alter the areas of Zone-\uppercase\expandafter{\romannumeral1} and Zone-\uppercase\expandafter{\romannumeral2}. In contrast, Figs.~\ref{fig_multiple_delta}(c)-(f) illustrate that the impact of introducing
promotive disorders $\Delta_1$  or $\Delta_2$ into a suppressive combination of $\Delta_0$ and $\Delta_3$.
As to Point-A, neither $\Delta_1$ nor $\Delta_2$ is able to induce a transition from Zone-\uppercase\expandafter{\romannumeral1} to Zone-\uppercase\expandafter{\romannumeral2}, thus preventing the emergence of Cooper interaction divergence. With
respect to Point-B, Fig.~\ref{fig_multiple_delta}(e) shows that the critical Cooper interaction for divergence
is insusceptible to the introduction of $\Delta_2$. However, the addition of $\Delta_1$ is in favor
of reducing the critical value $|\lambda_c|$ even in the suppressive combination of $\Delta_0$ and $\Delta_3$
as shown in Fig.~\ref{fig_multiple_delta}(d). This is well in agreement with
the sole presence of $\Delta_1$ which is helpful to promote the Cooper instability.

For completeness, we verify the fates of Cooper interaction in the presence of all kinds of disorders as depicted in
Fig.~\ref{fig_multiple_delta}(g) and Fig.~\ref{fig_multiple_delta}(h). It can be found that the tendencies of Cooper
interactions are analogous to the circumstance with adding $\Delta_1$ or $\Delta_2$ to the
$\Delta_0$-$\Delta_3$ suppressive combination. To wrap up, we summarize the basic results in
Table~\ref{tab_three_disorder} for the presence of multiple kinds of disorder scatterings.

\section{Summary}\label{Sec_summary}

In summary, this work investigates the fate of Cooper instability in the low-energy regime of 2D FDSMs under the influence of attractive fermion-fermion interaction and disorder scatterings. To treat these competing physical ingredients on an equal footing, we employ a Wilsonian momentum-shell RG analysis~\cite{Wilson1975RMP,Polchinski9210046,Shankar1994RMP}, which yields the energy-dependent flow equations for all coupling strengths. Employing these RG equations~(\ref{Eq_RG_v_alpha_Q})-(\ref{Eq_RG_v_Delta3_Q_prime}),
we systematically examine the low-energy fate of Cooper pairing in the clean limit and the presence of disorder scatterings, with particular emphasis on its dependence on physical parameters and initial conditions.

Our analysis considers two complementary scenarios, namely Scenario-A ($Q \neq 0, Q' = 0$) and
Scenario-B ($Q = 0, Q' \neq 0$) that correspond to distinct configurations of the transfer momenta in the
ZS and $\mathrm{ZS}'$ channels~\cite{Shankar1994RMP}. As Scenario-B shares the basic similar results with
Scenario-A due to an underlying exchange symmetry, we focus primarily on Scenario-A. We begin by examining
Cooper instability in the clean limit using a combination of analytical and numerical methods. At tree level,
the pairing strength $\lambda$ flows to zero with decreasing energy scale, indicating the absence of instability.
However, upon including one-loop corrections, we identify a critical threshold of Cooper-pairing interaction
denoted by $\lambda_c$. This implies that Cooper instability emerges only when the initial interaction
strength $|\lambda_0|$ exceeds $|\lambda_c|$. Numerical results reveal that $\lambda_c$ depends sensitively on
the fractional dispersion exponent $\alpha$ and the transfer momentum magnitude $Q$ as well as its angular
orientation $\phi$, which are summarized in Table~\ref{conclusion_clean}. By contrast, the fermionic
velocity $v_\alpha$ only modifies $\lambda_c$ quantitatively, without altering qualitative behavior as shown
in Fig.~\ref{fig_Q_phi}. Specifically, the $(Q, \phi)$ parameter space divides into two distinct regions illustrated
in Fig.~\ref{fig_Q_phi_1}: Zone-\uppercase\expandafter{\romannumeral1}, where $|\lambda_c|$ diverges and Cooper
instability is suppressed, and Zone-\uppercase\expandafter{\romannumeral2}, where $|\lambda_c|$ remains finite
and instability is allowed. Such two zones coexist and compete with each other when $\alpha$ lies between two
critical values $\alpha_{c1}$ and $\alpha_{c2}$, but instead Zone-\uppercase\expandafter{\romannumeral2} completely
dominates over Zone-\uppercase\expandafter{\romannumeral1} at $\alpha<\alpha_{c1}$ or $\alpha>\alpha_{c2}$.
Additionally, we notice from Fig.~\ref{fig_Q_phi} and Fig.~\ref{fig_lambda_1}
that higher values of $\alpha$ and lower values of $v_\alpha$ reduce the critical interaction
$|\lambda_c|$, thereby enhancing the opportunity for Cooper instability.

Subsequently, moving beyond the idealized clean limit, we systematically investigate how disorder impacts
Cooper instability. Three distinct types of disorders dubbed by $\Delta_0$, $\Delta_{1,3}$, and $\Delta_2$~\cite{Nersesyan1995NPB,Stauber2005PRB,Wang2011PRB, Mirlin2008RMP,Coleman2015Book,Roy2018PRX}
are taken into account. In the presence of single type of disorder, we find that either $\Delta_0$ or $\Delta_3$
not only increases the critical value of $|\lambda_c|$ but also alters the regions of Zone-\uppercase\expandafter{\romannumeral1} and Zone-\uppercase\expandafter{\romannumeral2}, indicating that they suppress the Cooper instability. In contrast,
either $\Delta_1$ or $\Delta_2$ reduces the critical value $|\lambda_c|$ and hence promotes the Cooper instability
as provided in Table~\ref{tab_disorder}. When two or more types of disorders coexist, their competing effects
lead to a more intricate scenario. Our analysis shows that the simultaneous presence of both promotive
disorders ($\Delta_{1}$ and $\Delta_{2}$) together with a single suppressive
disorder ($\Delta_0$ or $\Delta_3$) is in favor of Cooper instability. Notably, $\Delta_1$ exhibits a stronger
promotive effect than $\Delta_2$. However, in configurations involving all disorder types, the combined suppressive
influence of $\Delta_0$ and $\Delta_3$ dominates over the promotive effects of $\Delta_1$ and $\Delta_2$. The
basic results are summarized in Table~\ref{tab_two_disorder} and Table~\ref{tab_three_disorder}.

These results elucidate distinctive properties of 2D FDSMs, governed by the competition between Cooper pairing and multiple disorder types, and reveal their intricate relationship with superconductivity through the emergence of Cooper instability. We anticipate that our results would be helpful for further studies of low-energy quantum criticality not only in 2D FDSMs, but also in a broader class of Dirac-type materials with anisotropic or fractional band dispersions.

\section*{ACKNOWLEDGEMENTS}

We thank Wen Liu and Wen-Hao Bian for the helpful discussions.
J.W. is supported  by Tianjin Natural Science Foundation Project (25JCYBJC01640).

\appendix

\section{Related coefficients}\label{Appendix_Related coefficients}

All the coefficients introduced in Sec.~\ref{Sec_RG_analysis} are designated as follows,
\begin{widetext}
\begin{eqnarray}
\mathcal F_0 & \equiv & \!\!\int_{0}^{2\pi } \!\!\mathrm{d}\theta\left [ \frac{3U}{8 P^{\frac{3}{2} }} +\frac{3\alpha (2UL+JP)}{8 P} -\frac{9\alpha (1+2P^{\frac{1}{2}})LU}{16 P^{\frac{5}{2} }}    +\frac{3\alpha UL}{8 P^{\frac{5}{2} }} -\frac{\left ( 2P-U \right )}{8 P^{\frac{3}{2}}}  \right .\nonumber \\
&&\left .-\frac{\alpha \left ( 2L-J \right ) }{8 P^{\frac{3}{2}}} +\frac{3\alpha L \left ( 2P-U \right ) }{16 P^{\frac{5}{2}}}- \frac{5 }{8 P^{\frac{1}{2}}}  +\frac{5\alpha L  }{16 P^{\frac{3}{2}}}\right]+2\mathcal F_3-2\mathcal F_4,\label{Eq_F_0}\\
\mathcal F_1 & \equiv & \!\!\int_{0}^{2\pi } \!\!\mathrm{d}\theta\left[\frac{2U+P}{8P^{\frac{1}{2} }} +\frac{\alpha(UL+PJ+PL)}{4P^{\frac{3}{2} }}-\frac{\alpha L(2U+P)}{16 P^{\frac{3}{2}} }\right ]\!\!,\label{Eq_F_1}\\
\mathcal F_0' & \equiv & \!\!\int_{0}^{2\pi }\!\! \mathrm{d}\theta\left [ \frac{3U}{8 P^{\frac{3}{2} }} +\frac{3\alpha (2UL'+J'P)}{8 P} -\frac{9\alpha (1+2P^{\frac{1}{2}})L'U}{16 P^{\frac{5}{2} }}    +\frac{3\alpha UL'}{8 P^{\frac{5}{2} }} -\frac{\left ( 2P-U \right )}{8 P^{\frac{3}{2}}}  \right .\nonumber \\
&&\left .-\frac{\alpha \left ( 2L'-J' \right ) }{8 P^{\frac{3}{2}}} +\frac{3\alpha L' \left ( 2P-U \right ) }{16 P^{\frac{5}{2}}}- \frac{5 }{8 P^{\frac{1}{2}}}  +\frac{5\alpha L'  }{16 P^{\frac{3}{2}}}\right]+2\mathcal F_3-2\mathcal F_4,\label{Eq_F_0_prime}\\
\mathcal F_1' & \equiv & \!\!\int_{0}^{2\pi } \!\!\mathrm{d}\theta\left[\frac{2U+P}{8P^{\frac{1}{2} }} +\frac{\alpha(UL'+PJ'+PL')}{4P^{\frac{3}{2} }}-\frac{\alpha L'(2U+P)}{16 P^{\frac{3}{2}} }\right ]\!\!,\label{Eq_F_1_prime}\\
\mathcal F_2&\equiv&\!\!\int_{0 }^{2\pi  } \!\!\!\mathrm{d}\theta \frac{ \left | \cos \theta \right |^{2\alpha }-\left | \sin \theta \right |^{2\alpha } }{\left ( \left | \cos \theta \right |^{2\alpha }+\left | \sin \theta \right |^{2\alpha } \right )^{2 }},
\mathcal F_3\equiv \!\!\int_{0 }^{2\pi  } \!\!\!\mathrm{d}\theta \frac{ \left | \sin \theta  \right |^{2\alpha } }{ \left ( \left | \cos \theta \right |^{2\alpha }+\left | \sin \theta \right |^{2\alpha } \right )^{\frac{3}{2} }},
\mathcal F_4\equiv \!\!\int_{0 }^{2\pi  } \!\!\!\!\mathrm{d}\theta \frac{ \left | \cos \theta \right |^{2\alpha } }{ \left ( \left | \cos \theta \right |^{2\alpha }+\left | \sin \theta \right |^{2\alpha } \right )^{\frac{3}{2}  }},\\
\mathcal F_{5}&\equiv& \!\! \int_{0 }^{2\pi  } \!\!\!\mathrm{d}\theta \frac{1 }{\left ( \left | \cos \theta \right |^{2\alpha }+\left | \sin \theta \right |^{2\alpha } \right )},
\mathcal F_6\equiv\!\! \int_{0 }^{2\pi  } \!\!\!\mathrm{d}\theta \frac{ \left | \sin \theta \right |^{2\alpha } }{ \left ( \left | \cos \theta \right |^{2\alpha }+\left | \sin \theta \right |^{2\alpha } \right )^{\frac{5}{2} }},
\mathcal F_8\equiv\!\!\int_{0 }^{2\pi  } \!\!\!\!\mathrm{d}\theta \frac{ 1 }{ \left ( \left | \cos \theta \right |^{2\alpha }+\left | \sin \theta \right |^{2\alpha } \right )^{2 }},\\
\mathcal{F}_{ 7 } &\equiv&\!\! \int_{0}^{2\pi } \!\!\!\!\mathrm{d}\theta\!\!\left [ - \frac{3}{8v_\alpha^3 P^{\frac{3}{2} }}-\frac{3\alpha L }{2v_\alpha^5 P^{\frac{5}{2} }}- \frac{9\alpha (1+2P^{\frac{1}{2}})L+6L}{16v_\alpha^3 P^{\frac{5}{2} }}+\frac{U-4P+\alpha (J-4L)}{4v_\alpha^2 P^{\frac{3}{2}} }-\frac{\alpha L(3+2P)(U-4P)}{8v_\alpha^2 P^{3} } \right ]\!\!,\\
\mathcal{F}_{ 7 }'&\equiv&\!\! \int_{0}^{2\pi }\!\!\! \mathrm{d}\theta\!\left [\!\frac{U-4P+\alpha (J'-4L')}{4v_\alpha^2 P^{\frac{3}{2}} } - \frac{3}{8v_\alpha^3 P^{\frac{3}{2} }}-\frac{3\alpha L' }{2v_\alpha^5 P^{\frac{5}{2} }}- \frac{9\alpha (1+2P^{\frac{1}{2}})L'+6L'}{16v_\alpha^3 P^{\frac{5}{2} }}-\frac{\alpha L'(3+2P)(U-4P)}{8v_\alpha^2 P^{3} } \!\right ]\!\!,\\
\mathcal F_9&\equiv&\!\! \int_{0 }^{2\pi  } \!\!\!\!\mathrm{d}\theta \frac{ \left | \cos \theta \right |^{2\alpha } }{ \left ( \left | \cos \theta \right |^{2\alpha }+\left | \sin \theta \right |^{2\alpha } \right )^{\frac{5}{2}  }},\hspace{1.0cm}
\mathcal F_{10}\equiv\!\! \int_{0 }^{2\pi  } \!\!\!\mathrm{d}\theta \frac{3\left | \sin \theta \right |^{2\alpha }- \left | \cos \theta \right |^{2\alpha } }{ \left ( \left | \cos \theta \right |^{2\alpha }+\left | \sin \theta \right |^{2\alpha } \right )^{3  }},
\\
\mathcal F_{11}&\equiv&\!\! \int_{0 }^{2\pi  } \!\!\!\!\mathrm{d}\theta \frac{ 1 }{\left ( \left | \cos \theta \right |^{2\alpha }+\left | \sin \theta \right |^{2\alpha } \right )^{3}},\hspace{1.0cm}
\mathcal F_{12} \equiv\!\! \int_{0}^{2\pi}\!\!\!\!\mathrm{d}\theta \frac{1}{\left ( \left | \cos \theta  \right |^{2\alpha } +\left | \sin \theta  \right |^{2\alpha } \right ) ^{\frac{5}{2} }},
\\
\mathcal F_{13} &\equiv&\!\! \int_{0}^{2\pi}\!\!\!\!\mathrm{d}\theta \frac{1}{\left ( \left | \cos \theta  \right |^{2\alpha } +\left | \sin \theta  \right |^{2\alpha } \right ) ^{\frac{7}{2} }},\hspace{1.0cm}
\mathcal F_{14} \equiv\!\!\int_{0}^{2\pi}\!\!\!\!\mathrm{d}\theta \frac{ \left | \cos \theta  \right |^{2\alpha }}{\left ( \left | \cos \theta  \right |^{2\alpha } +\left | \sin \theta  \right |^{2\alpha } \right ) ^{\frac{7}{2} }},
\\
\mathcal F_{15} &\equiv&\!\! \int_{0}^{2\pi}\!\!\!\mathrm{d}\theta \frac{1}{\left ( \left | \cos \theta  \right |^{2\alpha } +\left | \sin \theta  \right |^{2\alpha } \right ) ^{\frac{9}{2} }},\hspace{1.0cm}
\mathcal F_{16} \equiv\!\!\int_{0}^{2\pi}\!\!\!\!\mathrm{d}\theta \frac{ \left | \cos \theta  \right |^{2\alpha }}{\left ( \left | \cos \theta  \right |^{2\alpha } +\left | \sin \theta  \right |^{2\alpha } \right ) ^{\frac{9}{2} }},
\end{eqnarray}
\end{widetext}
with
\begin{eqnarray}
P \!&\equiv&\!\left | \cos \theta  \right |^{2\alpha } +\left | \sin \theta  \right |^{2\alpha },\\
U \!&\equiv&\!-\left | \cos\theta \right |^{2\alpha }+ \left |\sin\theta \right |^{2\alpha },\\
J\!&\equiv&\!\left | \sin\theta  \right |^{2\alpha-1}\!Q\!\sin\phi -\left | \cos\theta  \right |^{2\alpha-1}\!Q\!\cos\phi,\\
L\!&\equiv&\!\left | \sin\theta  \right |^{2\alpha-2}\!Q\!\sin\phi +\left | \cos\theta  \right |^{2\alpha-2}\!Q\!\cos\phi,\\
J'\!&\equiv&\!\left | \sin\theta  \right |^{2\alpha-1}\!Q'\!\sin\phi' -\left | \cos\theta  \right |^{2\alpha-1}\!Q'\!\cos\phi',\\
L'\!&\equiv&\!\left | \sin\theta  \right |^{2\alpha-2}\!Q'\!\sin\phi' +\left | \cos\theta  \right |^{2\alpha-2}\!Q'\!\cos\phi'.
\end{eqnarray}
Hereby, the finite transfer momentum is parameterized as $\mathbf{Q} = Q(\cos\phi, \sin\phi)$
and $\mathbf{Q}' = Q'(\cos\phi', \sin\phi')$,
where $Q/Q'$ and $\phi/\phi'$ denote the magnitude and angular orientation of transfer momentum, respectively.



\end{document}